\pdfoutput=1
\documentclass[%
 reprint,
 amsmath,amssymb,
 aps,
floatfix,
]{revtex4-2}
\usepackage{acronym}
\usepackage{graphicx}% Include figure files
\usepackage{dcolumn}% Align table columns on decimal point
\usepackage{bm}% bold math
\usepackage{hyperref}% add hypertext capabilities
\usepackage{cleveref}
\usepackage{xcolor}
\usepackage{url}
\usepackage{booktabs}
\usepackage[acronym]{glossaries}
\usepackage{todonotes}
\usepackage{bm}
\usepackage{physics}
\usepackage{float}
\usepackage[export]{adjustbox} % for fractional cropping
\usepackage{subfigure} % provides \subfigure and subfigure counter

\makeglossaries

\newcommand{\cropTopThirty}[2][]{%
  \begin{adjustbox}{trim=0 0 0 {0.18\totalheight},clip}%
    \includegraphics[#1]{#2}%
  \end{adjustbox}%
}

\newcommand{\onlyTopThirty}[2][]{%
  \begin{adjustbox}{trim=0 {0.82\totalheight} 0 0,clip}%
    \includegraphics[#1]{#2}%
  \end{adjustbox}%
}

\newacronym{vqe}{VQE}{Variational Quantum Eigensolver}
\newacronym{cas}{CAS}{Complete Active Space}
\newacronym{bfgs}{BFGS}{Broyden-Fletcher-Goldfarb-Shanno method}
\newacronym{slsqp}{SLSQP}{Sequential Least Squares Quadratic Programming}
\newacronym{nm}{NM}{Nelder-Mead Simplex method}
\newacronym{pm}{PM}{Powell's method}
\newacronym{cobyla}{COBYLA}{Constrained Optimization BY Linear Approximations}
\newacronym{isoma}{iSOMA}{Improved Self-Organizing Migrating Algorithm}
\newacronym{sa-oo-vqe}{SA-OO-VQE}{State-Averaged Orbital Optimized Variational Quantum Eigensolver}
\newacronym{manova}{MANOVA}{Multivariate Analysis of Variance}
\newacronym{permanova}{PERMANOVA}{Permutational Multivariate Analysis of Variance}
\newacronym{permdisp}{PERMDISP}{Permutational Analysis of Multivariate Dispersions}
\newacronym{anova}{ANOVA}{Analysis of variance}
\newacronym{nisq}{NISQ}{Noisy Intermediate-Scale Quantum}
\newacronym{vqa}{VQA}{Variational Quantum Algorithm}
\newacronym{casscf}{CASSCF}{Complete Active Space Self-Consistent Field}
\begin{document}

\preprint{APS/123-QED}

\title{Statistical Benchmarking of Optimization Methods for Variational Quantum Eigensolver under Quantum Noise}% Force line breaks with \\
\newcommand{\bruno}[1]{\textcolor{blue}{Bruno: #1}}
\newcommand{\bmkappa}{{\bm \kappa}}
\newcommand{\bmtheta}{{\bm \theta}}

\author{Silvie Ill\'{e}sov\'{a}}
\email{silvie.illesova@seznam.cz}%
\affiliation{IT4Innovations National Supercomputing Center, VSB - Technical University of Ostrava, 708 00 Ostrava, Czech Republic}
\affiliation{Gran Sasso Science Institute, L'Aquila, Italy}%

\author{Tom\'{a}\v{s} Bezd\v{e}k}%
\affiliation{Department of Mathematics, TUM School of Computation, Information and Technology, Technical University of Munich, Boltzmannstraße 3, D-85748, Garching b. München, Germany}%

\author{Vojt\v{e}ch Nov\'{a}k}
\affiliation{Department of Computer Science, Faculty of Electrical Engineering and Computer Science, VSB-Technical University of Ostrava, Ostrava, Czech Republic}%
\affiliation{IT4Innovations National Supercomputing Center, VSB - Technical University of Ostrava, 708 00 Ostrava, Czech Republic}%
\affiliation{Department of Informatics and Statistics, Marine Research Institute, Klaipeda University, Lithuania}

\author{Bruno Senjean}
\affiliation{Institute Charles Gerhardt Montpellier, Université Montpellier, CNRS, ENSCM,  Route de Mende, 1919, Montpellier, France}%

\author{Martin Beseda}
\affiliation{Dipartimento di Ingegneria e Scienze dell’Informazione e Matematica, Università dell’Aquila, Via Vetoio, I-67010 Coppito, L’Aquila, Italy}%

%\collaboration{CLEO Collaboration}
\noaffiliation

\date{\today}% It is always \today, today,

\begin{abstract}
\noindent This work investigates the performance of numerical optimization algorithms applied to the State-Averaged Orbital-Optimized Variational Quantum Eigensolver for the $H_2$ molecule under various quantum noise conditions. The goal is to assess the stability, accuracy, and computational efficiency of commonly used gradient-based, gradient-free, and global optimization strategies within the Noisy Intermediate-Scale Quantum  regime. We systematically compare six representative optimizers, BFGS, SLSQP, Nelder–Mead, Powell, COBYLA, and iSOMA, under ideal, stochastic, and decoherence noise models, including phase damping, depolarizing, and thermal relaxation channels. Each optimizer was tested over multiple noise intensities and measurement settings to characterize convergence behavior and sensitivity to noise-induced landscape distortions. The results show that BFGS consistently achieves the most accurate energies with minimal evaluations, maintaining robustness even under moderate decoherence. COBYLA performs well for low-cost approximations, while SLSQP exhibits instability in noisy regimes. Global approaches such as iSOMA show potential but are computationally expensive. These findings provide practical guidance for selecting suitable optimizers in variational quantum simulations, highlighting the importance of noise-aware optimization strategies for reliable and efficient quantum chemistry computations on current hardware.
\end{abstract}

%Use showkeys class option if keyword

\maketitle

%\tableofcontents

\newpage
\section*{Introduction}
Quantum computing is rapidly advancing as a promising computational paradigm for tackling problems that remain practically infeasible for classical computers due to \textit{quantum advantage}\cite{daniel1997,harrow2009,daley2022practical}. Its well-known potential applications span quantum chemistry~\cite{cao2019quantum,rajamani2025equiensembledescriptionsystematicallyoutperforms, illesova2025numerical,illesova2025transformation,bezdek2025classical,novak2025reliable}, physics~\cite{di2024quantum,micheletti2021polymer,PhysRevA.111.022437,lamm2020parton,mocz2021toward}, material design~\cite{guo2024harnessing,de2021materials,kang2025quantum,liu20192d,ma2020quantum,bauer2020quantum}, software engineering \cite{trovato2025preliminary,piattini2021quantum,ali2022software,dwivedi2024quantum,mandal2025quantum,de2022software},
benchmarking \cite{lewandowska2025benchmarking,bilek2025experimental,proctor2025benchmarking,lubinski2023application,hashim2025practical,illesova2025qmetric,novak2025optimization}, finance~\cite{yuan2024quantifying,egger2020quantum,rebentrost2018quantum,herman2023quantum,orus2019quantum,chang2023prospects}, and machine learning~\cite{gupta2022quantum,illesova2025importance,novak2025predicting,biamonte2017quantum,illesova2025classical,aggarwal2024detailed,novak2025quantum,illesova2025complementarity}. Among the available approaches, \glspl{vqa} are particularly attractive for the current \gls{nisq} era, as they combine quantum state measurements with classical optimization, thus aiming for possible speedup, while leaving most computations to the classical infrastructure, avoiding decoherence.

A flagship example of \glspl{vqa} is the \gls{vqe}~\cite{tilly2022variational,fedorov2022vqe}, designed to compute ground-state energies of physical systems. Recent extensions, such as the \gls{sa-oo-vqe}~\cite{yalouz2021state,yalouz2022analytical,beseda2024state}, broaden this scope to include excited states. This makes \gls{sa-oo-vqe} a quantum equivalent of the classical MCSCF method~\cite{werner1981quadratically}, offering a systematic path toward excited-state calculations on quantum hardware.

As a starting point, we restrict our present study to the dihydrogen molecule~$\mathrm{H_2}$, the simplest nontrivial electronic system. Although elementary, $\mathrm{H_2}$ provides an ideal benchmark for algorithm testing or basic numerical analysis due to its simple electronic structure, modest resource requirements, and relevance as the minimal model for chemical bonding. In subsequent work, the methodology developed here will be extended to more complex and chemically relevant molecules.

As the critical aspect of \glspl{vqa} is their reliance on classical numerical optimization \cite{moll2018quantum,qi2024variational}, the efficiency and reliability of these algorithms ultimately hinge on the optimizer’s ability to navigate a high-dimensional, noisy cost function landscape \cite{ravi2023navigating,huembeli2021characterizing,fontana2021evaluating}. This landscape is shaped by both the chosen ansatz and unavoidable imperfections such as sampling noise and different types of decoherence \cite{ge2022optimization,choy2023molecular}. Understanding how different optimization strategies behave in such environments is therefore essential for the practical implementation of not only \glspl{vqa}, but also all algorithms on \gls{nisq} computers.

In this manuscript, we present a systematic numerical study of several optimization methods applied to \gls{sa-oo-vqe} for $\mathrm{H_2}$, spanning gradient-based, gradient-free, and global strategies, directly continuing our previous works, focused both on \gls{sa-oo-vqe} \cite{yalouz2021state,beseda2024state} and the Variational Hamiltonian Ansatz approaches\cite{illesova2025numerical}. By analyzing the optimizers' behavior under idealized conditions and various noise models, we identify their strengths, limitations, and also the specific effects of different types of noise. Eventually, we provide practical, general guidance for selecting optimizers in \gls{nisq} era quantum simulations based on extensive statistical tests of our results, so they also lay the groundwork for scaling to larger molecular systems in the following work.

The researched $\mathrm{H_2}$ geometry, \gls{sa-oo-vqe} method, the adopted optimization approaches, and the selected quantum estimators emulating different types of noise are described in~\Cref{sec:methodology}. It is followed by~\Cref{sec:results}, which contains all the numerical results comparing the efficiency of optimizers with respect to several different types of quantum noise and the subsequent statistical tests. \Cref{sec:software} describes the software implementation and provides details and the link to the replication package. Finally,~\Cref{sec:conclusion} contains the conclusions, providing general advice for the numerical optimization in the environments with quantum noise, based on statistically significant results.

\section{Methodology}\label{sec:methodology}
This section presents the computational setup and methodological framework used in this study. The \gls{sa-oo-vqe} formulation, the tested optimization algorithms, and the noise and estimator configurations used to assess their performance are described here.

\subsection{SA-OO-VQE}
In this work,
we study the convergence of
a \gls{vqe}
towards the ground- and first-excited-state
energies of the $\mathrm{H}_2$ molecule with an internuclear distance of 0.74279~\AA, which directly corresponds to the bond length in the equilibrium geometry.
The electronic structure was treated within the \gls{cas} approximation
using $\text{CAS}(n_{\mathrm{elec}}, n_{\mathrm{orb}}) = \text{CAS}(2,2)$
and the correlation-consistent polarized valence double-$\zeta$ (cc-pVDZ) basis set.
As a standard \gls{vqe}
relies on the Rayleigh--Ritz
variational principle for the ground state only,
we consider its extension to
excited states based
on the generalized variational principle
for an ensemble of states, so-called the
Theophilou--Gross--Oliveira--Kohn
variational principle~\cite{theophilou1979energy,gross1988rayleigh},
thus leading to
the state-average orbital-optimized VQE (SA-OO-VQE)~\cite{nakanishi2019subspace,yalouz2021state,yalouz2022analytical,SAOOVQE,beseda2024state,illesova2025transformation}
which is
the quantum analog of the state-average multi-configurational self-consistent field (SA-MCSCF)
on classical computers~\cite{helgaker2014molecular}.
Within SA-OO-VQE, the cost function reads as follows,
\begin{eqnarray}
    E^{\text{SA-OO-VQE}}& =& \min_{\bmkappa,\bmtheta}  \Big\lbrace \langle\Psi_{\rm A} (\boldsymbol{\theta})|\hat{H} (\boldsymbol{\kappa})|\Psi_{ \rm A} (\boldsymbol{\theta})\rangle  \nonumber \\
      &&+  \langle\Psi_{\rm B} (\boldsymbol{\theta})|\hat{H} (\boldsymbol{\kappa})|\Psi_{ \rm B} (\boldsymbol{\theta})\rangle \Big\rbrace \\
      &\geq & \langle\Psi_0| \hat{H}(\bmkappa^*)  |\Psi_0\rangle
+ \langle{\Psi_1}| \hat{H}(\bmkappa^*) |{\Psi_1}\rangle \nonumber,
\label{eq:SAOOVQE_energy}
\end{eqnarray}
where
$\bmkappa^*$ and $\bmtheta^*$ denote the minimizing sets of parameters, and
\begin{eqnarray}
 |\Psi_{\rm A}(\boldsymbol{\theta})\rangle  = \hat{U}(\boldsymbol{\theta})|\Phi_{\rm A}\rangle \text{ and }
 |\Psi_{\rm B}(\boldsymbol{\theta})\rangle  = \hat{U}(\boldsymbol{\theta})|\Phi_{\rm B}\rangle
\end{eqnarray}
are prepared on the quantum
computer
using the circuit-parametrized ansatz $\hat{U}(\boldsymbol{\theta})$ (being the generalized unitary coupled-cluster ansatz with single and double excitations~\cite{romero2018strategies} and in our case consisting of 3 trainable parameters)
applied on a
set of two orthonormal initial states $\ket{\Phi_{\rm A}}$ and $\ket{\Phi_{\rm B}}$ (with $\langle\Phi_{\rm A}| \Phi_{\rm B}\rangle = 0$).
In this work, the Hartree--Fock state
and the first-excited singlet configuration state function
were used for $\ket{\Phi_A}$ and $\ket{\Phi_B}$, respectively.
The orbital dependence of the Hamiltonian in the active space approximation is
represented by the parameters $\boldsymbol{\kappa}$,
and its derivation is well detailed in Ref.~\cite{yalouz2021state}.
According to Eq.~(\ref{eq:SAOOVQE_energy}) and the variational principle~\cite{theophilou1979energy,gross1988rayleigh}, the SA-OO-VQE energy is lower-bounded by the ensemble energy of the eigensubspace formed by the eigenstates $\ket{\Psi_0}$
and $\ket{\Psi_1}$ of $\hat{H}(\bmkappa^*)$.
Note that $\ket{\Psi_A}$ and
$\ket{\Psi_B}$ can be any state obtained from a rotation between
$\ket{\Psi_0}$ and $\ket{\Psi_1}$.
In order to force $\ket{\Psi_A}$ and
$\ket{\Psi_B}$ to be equal to either $\ket{\Psi_0}$
and $\ket{\Psi_1}$, we perform an additional rotation on the quantum computer according to Refs.~\cite{yalouz2022analytical,illesova2025transformation}.

\subsection{Optimization Methods}\label{sub:om}
For the analysis in this paper, six optimization methods were selected based on their different properties so that a complete analysis could be performed. The gradient-based optimization methods are employed, specifically \gls{bfgs} \cite{liu1989limited} and \gls{slsqp} \cite{kraft1988software}. These gradient-based methods use derivatives to construct the step size and the step direction, and while they are usually efficient on smooth objective functions and fast in local convergence \cite{daoud2023gradient}, they are usually not the best choice in a noisy environment \cite{friedl2014population,haji2021comparison}. Yet, they are used in our calculations to provide an insight into how the specific noise types affect the landscape and the gradient calculations. 
\gls{bfgs} is a quasi-Newton optimization method that uses an approximation of the Hessian matrix in each iteration, thus avoiding the direct calculation of costly second derivatives, and is particularly effective for medium-sized problems \cite{dai2002convergence}.  While in \gls{slsqp} the optimized problem is approximated as a quadratic programming problem, which is solved to update the overall best solution.

In the next subgroup, we have three gradient-free methods, namely \gls{nm} \cite{nelder1965simplex}, \gls{pm} \cite{powell1964efficient}, and \gls{cobyla} \cite{powell1994direct}. This group was chosen to investigate whether, when the need to calculate gradients is removed, better convergence can be achieved in an environment where quantum decoherence is present. 
The \gls{nm} method in $ n$-dimensional search space uses an $ n+1$-dimensional simplex, which is iteratively updated using reflection, expansion, contraction, and shrinkage operations to search for the optimum. It is simple and efficient in lower dimensions, but convergence may worsen when the problem's dimensionality increases \cite{lagarias1998convergence}. On the other hand, \gls{pm} performs sequential line minimizations along a set of directions. These directions are later updated to improve conjugacy, thus speeding up the convergence. Usually, it works best on smooth problems and can stagnate in non-ideal conditions \cite{poljak1978nonlinear}. The last method of this set is \gls{cobyla}, a trust-region method that builds an approximation of both the objective function and constraints. Updates are performed within a region around the current point, which shrinks or expands adaptively \cite{pellow2021comparison}. 

The last investigated method is an example from global optimizers, specifically the \gls{isoma}~\cite{zelinka2016soma,zelinka2023isoma}. Global methods oftentimes use stochastic approaches and a population of candidate solutions to search for global optima. \gls{isoma}'s candidate solutions migrate toward the best-performing solution with adaptive step lengths, balancing exploration and exploitation, making it effective for global optimization of complex, multimodal landscapes \cite{klein2024optimizing}. Further details for the optimization methods are described in \Cref{app:opt_settings}.
\subsection{Estimators}\label{sub:est}
To examine the performance under noisy conditions, we first needed a benchmark; for this, a noiseless  estimator was configured, so that we could test the optimization method without the influence of inherent stochasticity or any decoherence. Meaning that the performance of the optimization method was tested on the base problem, without any influence of decoherence, which stems from quantum mechanics. The next step was to add the stochasticity. For this, we used four estimators, where the effect of finite sampling was included; the number of measurements for these estimators was set to four levels,
\begin{equation}
    \text{n}_{m} \in \lbrace 256,\, 512, \,1024,\, 6144 \rbrace.
\end{equation}
This setting will allow us to quantify how the number of measurements influences the optimization process and the convergence.

Beyond sampling noise, we constructed a variety of realistic noise models using Qiskit Aer’s \texttt{Estimator}~\footnote{\url{https://qiskit.github.io/qiskit-aer/stubs/qiskit\_aer.primitives.Estimator.html}}.

% The types of noise models we tested included phase damping, depolarizing channels, and thermal relaxation channels.

Three decoherence channels from Qiskit Aer~\footnote{\url{https://qiskit.github.io/qiskit-aer/stubs/qiskit\_aer.noise.NoiseModel.html}} were used: phase damping, depolarizing, and thermal relaxation. 
The phase-damping channel is defined as
\begin{equation}
\mathcal{E}_{\mathrm{PD}}(\rho)=E_0\rho E_0^\dagger+E_1\rho E_1^\dagger,
\end{equation}
with
\begin{equation}
E_0=\begin{bmatrix}1&0\\0&\sqrt{1-\lambda}\end{bmatrix}, \quad
E_1=\begin{bmatrix}0&0\\0&\sqrt{\lambda}\end{bmatrix},
\end{equation}
where $\lambda$ is the dephasing probability~\footnote{\url{https://qiskit.github.io/qiskit-aer/stubs/qiskit\_aer.noise.PhaseDampingError.html}}. 
The depolarizing channel is given by
\begin{equation}
\mathcal{E}_{\mathrm{Depol}}(\rho)=(1-p)\rho+\frac{p}{d}I,
\end{equation}
where $p$ is the depolarizing probability and $d$ is the Hilbert-space dimension~\footnote{\url{https://qiskit.github.io/qiskit-aer/stubs/qiskit\_aer.noise.DepolarizingError.html}}. 
Finally, the thermal relaxation channel is expressed as
\begin{equation}
\mathcal{E}_{\mathrm{TR}}(\rho)=\sum_{i=0}^{2}E_i\rho E_i^\dagger,
\end{equation}
where $E_i$ depend on the gate duration $t_g$ and relaxation times $T_1$ and $T_2$~\footnote{\url{https://qiskit.github.io/qiskit-aer/stubs/qiskit\_aer.noise.ThermalRelaxationError.html}}.

All noise types were tested with multiple levels of error rates to encompass near-ideal, realistic, and really noisy conditions. For these calculations, the number of measurements was set to $6144$ to minimize the role of stochasticity, as that was investigated independently.

Phase damping was applied with the following rates
\begin{equation}
    \text{r}_{PD} \in \{ 1 \% , \,5\%, \,10\%,\, 20\%\},
\end{equation}
which affected the $R_z$ rotation gates. The depolarizing channel affecting the following single-qubit, $x,y,z,h,r_x,r_y,r_z$, and two-qubit $cx$ gates, was simulated with the rates equaling to
\begin{equation}
    \text{r}_{Depol} \in \{ 1 \% , \,5\%, \,10\%,\, 20\%\}.
\end{equation}
As we also wanted to test thermal relaxation conditions, two distinct sets were investigated. The first one, which mimics realistic conditions, has relatively long $T_1$ and $T_2$ times. Here we varied the $T_2$ time as follows
\begin{equation}
    T_2 \in \{70, 80, 180, 380\}~\mu \text{s},
\end{equation}
and setting $T_1 = T_2 + 20 \mu \text{s}$. The intention of the second set was to look at shorter relaxation times, where 
\begin{equation}
    T_2 \in \{50, 100, 200, 300\}~\text{ns},
\end{equation}
and $T_1 = T_2$. For both of the sets examining thermal relaxation, realistic gate durations of 50~ns for single-qubit and 150~ns for two-qubit operations were set. Together, these two types of models capture both extended coherence and rapid decay conditions, allowing us to examine the optimization methods' abilities under varying conditions.

\section{Results}\label{sec:results}

To evaluate the effect of different types of noise described in \Cref{sub:est}, on the performance of \gls{sa-oo-vqe}, we focus on the six methods described in \Cref{sub:om}. 

\subsection{Accuracy of Results}
\begin{figure*}
\centering

% Row 1: shared legend strip
\onlyTopThirty[width=0.95\textwidth]{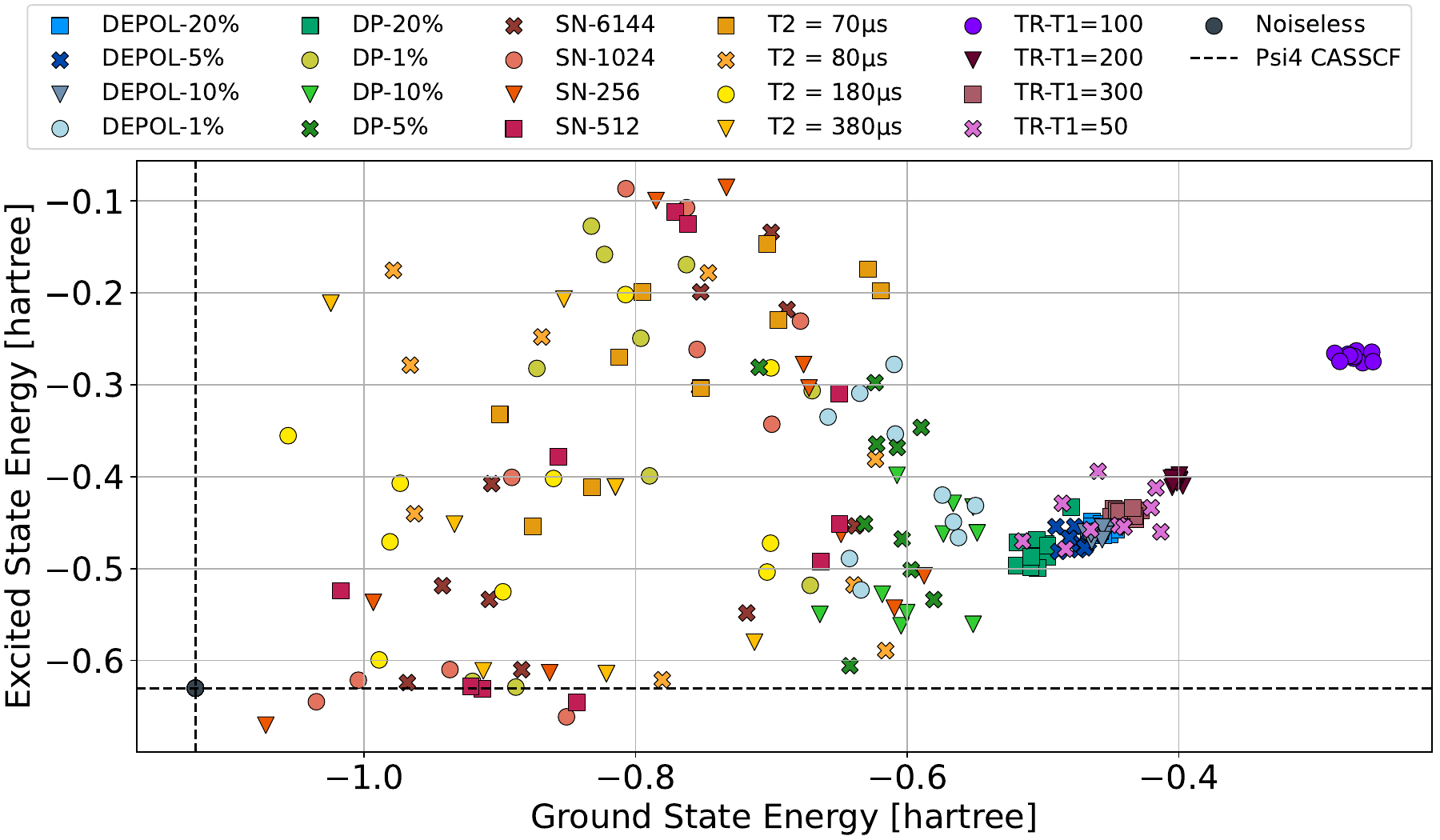}

\vspace{0.8em}

% Row 2
\subfigure[\gls{slsqp}\label{fig:slsqp_scatter}]{
  \cropTopThirty[width=0.48\textwidth]{figs/slsqp/slsqp_scatter_slsqp.pdf}
}\hfill
\subfigure[\gls{isoma}\label{fig:isoma_scatter}]{
  \cropTopThirty[width=0.48\textwidth]{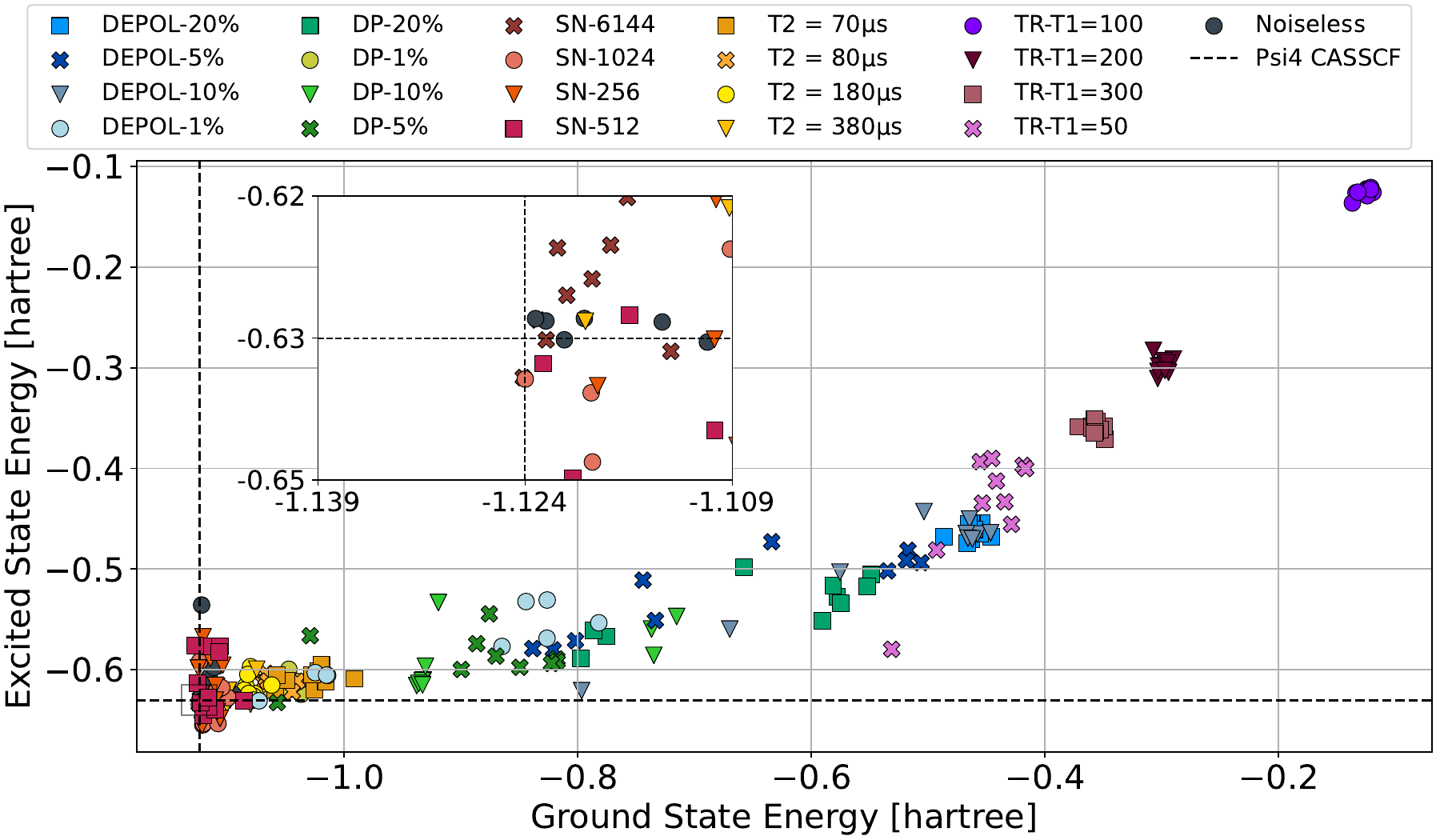}

}

\vspace{0.9em}

% Row 3
\subfigure[\gls{pm}\label{fig:powell_scatter}]{
  \cropTopThirty[width=0.48\textwidth]{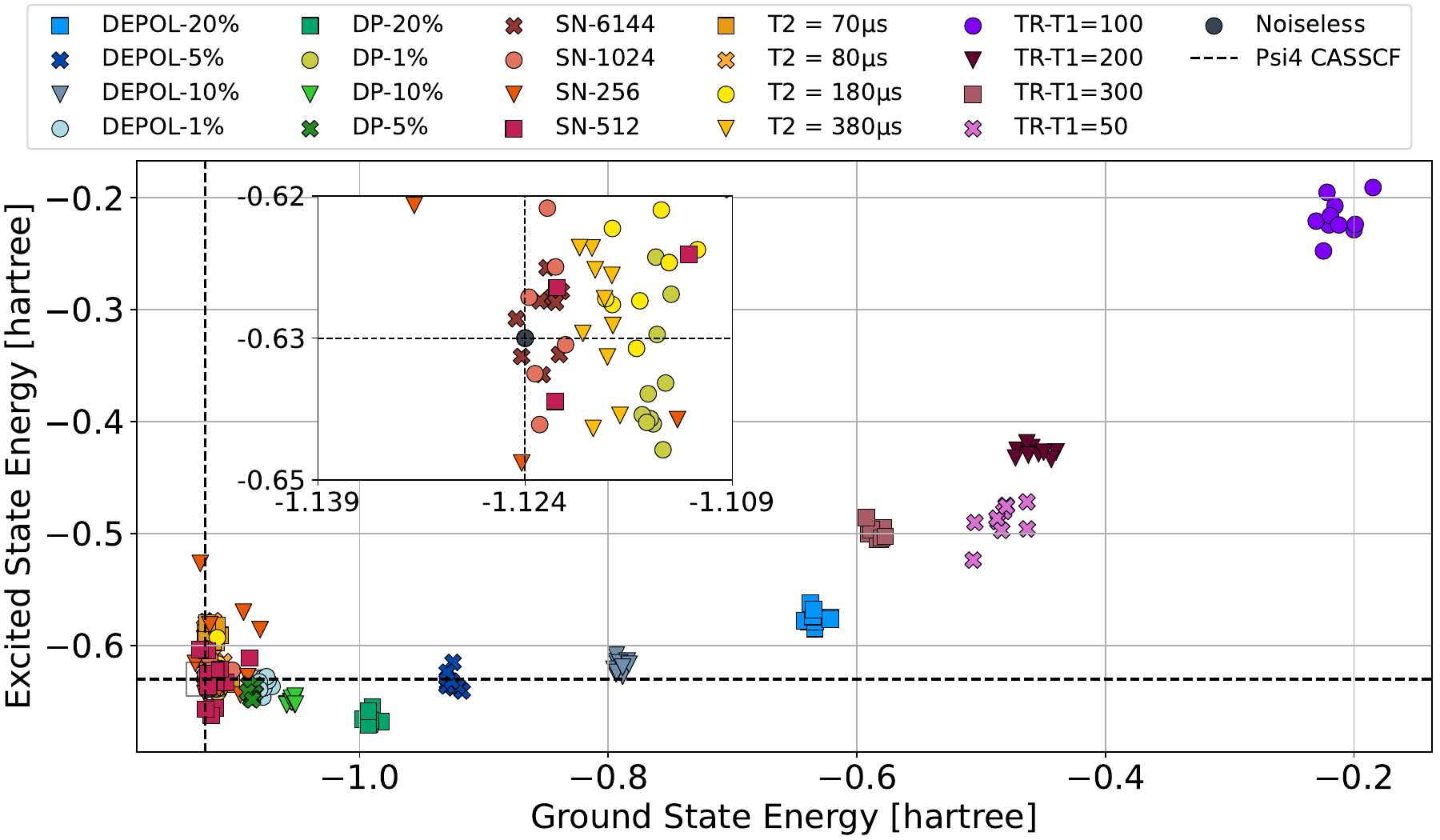}
}\hfill
\subfigure[\gls{nm}\label{fig:nelder-mead_scatter}]{
  \cropTopThirty[width=0.48\textwidth]{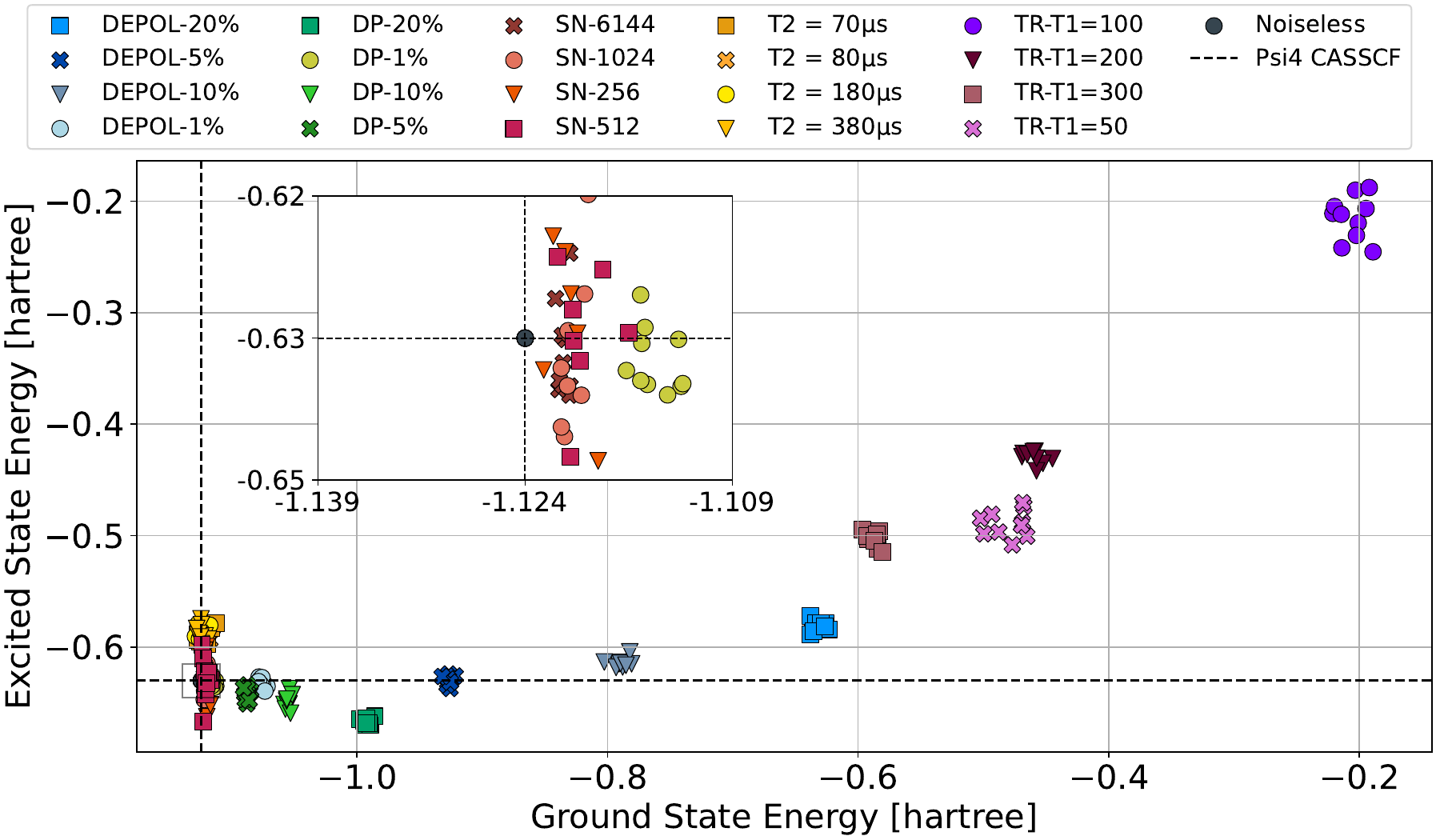}
}

\vspace{0.9em}
\subfigure[\gls{cobyla}\label{fig:cobyla_scatter}]{
  \cropTopThirty[width=0.48\textwidth]{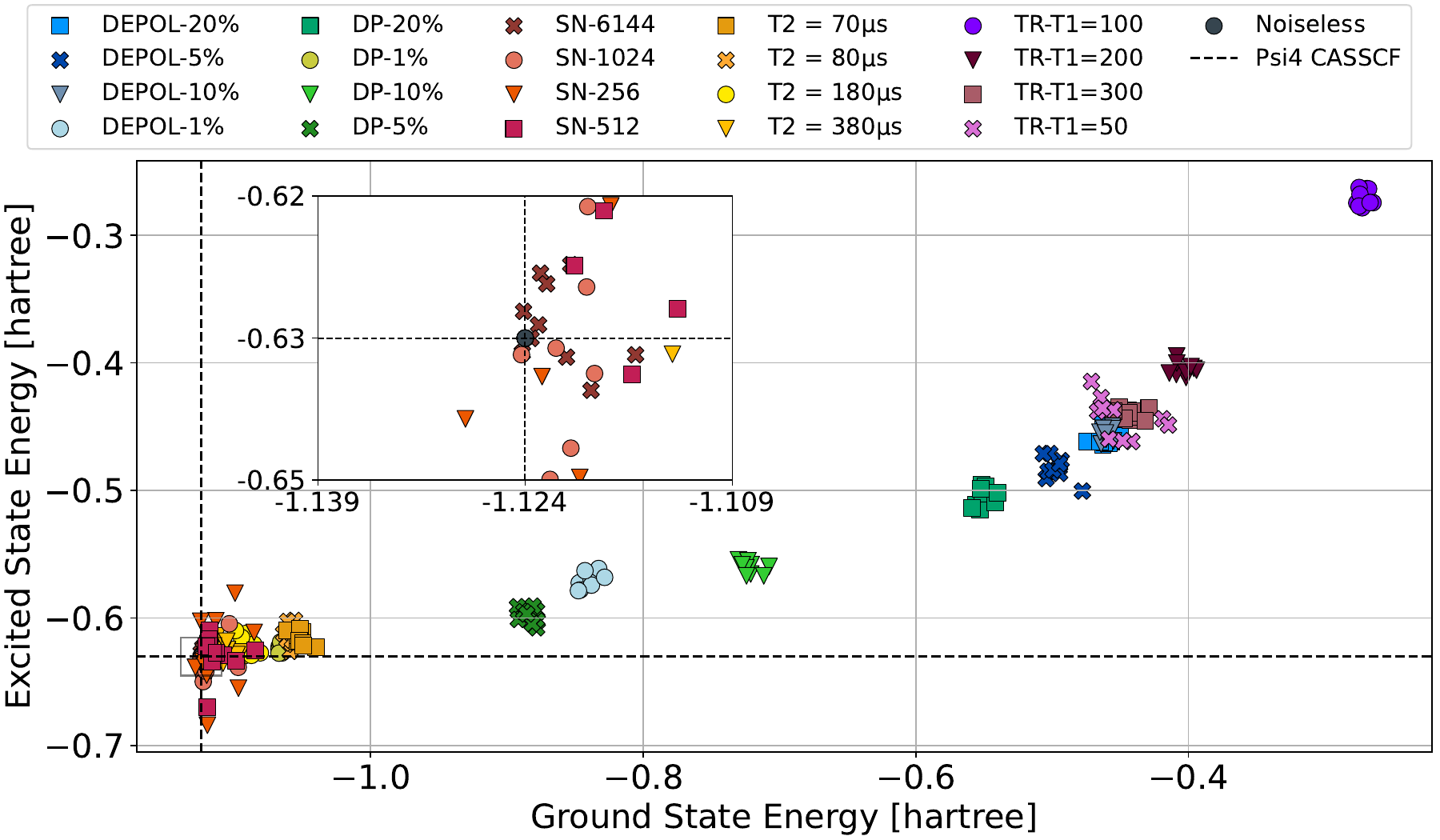}
}\hfill
\subfigure[\gls{bfgs}\label{fig:bfgs_scatter}]{
  \cropTopThirty[width=0.48\textwidth]{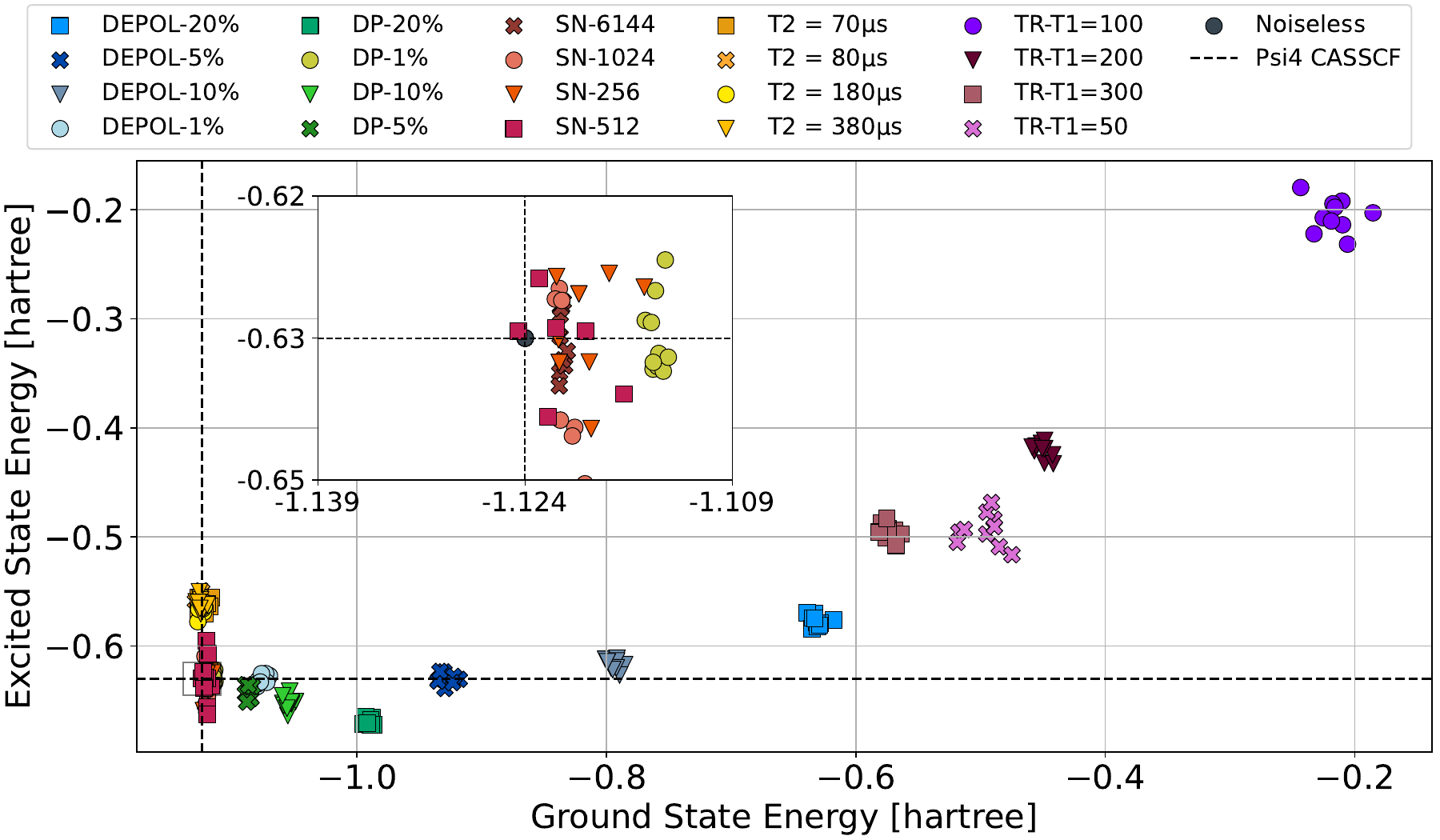}
}
% Row 4

\caption{Scatter plots of ground state energy (x-axis) and excited state energy (y-axis) for each optimization method. Dashed lines indicate the ideal values. All noise models are shown. The shared legend is displayed once at the top and removed from individual panels.}
\label{fig:combined_scatter}
\end{figure*}

At first, the accuracy was investigated. As \gls{sa-oo-vqe} is able to calculate both ground and excited states, the results are visualized in \Cref{fig:combined_scatter}.
In these scatter plots, the x-axis represents the energy of the ground state, the y-axis corresponds to the excited state energy, and the dashed lines show the optimal solution. Different noise types are represented by distinct markers.

It can be seen that \gls{slsqp}, visualized in \Cref{fig:slsqp_scatter}, fails to converge correctly even with just stochastic noise, and the problems only get worse when different noise types are considered. The spread of outcomes of different runs with the same noise is also large. \gls{isoma}, shown in \Cref{fig:isoma_scatter}, also shows some degree of spread, for it is significantly smaller than in the \gls{slsqp} case. The solutions for stochastic noise are clustered around the optimized value; this is, in a way, an expected behaviour as \gls{isoma} is a global optimizer, and some kind of variance, around the optimal value is expected.

If we have a look at \gls{pm}, \Cref{fig:powell_scatter}, \gls{nm}, \Cref{fig:nelder-mead_scatter}, and \gls{cobyla}, \Cref{fig:cobyla_scatter} we can observe similar behaviour for these three optimizers. They tend to cluster relatively well, for each type of noise, and also, we can see that if the rate of noise is increased for the same noise type, the results are worse, which agrees with the expected behaviour. The best results, at least from the point of optimized energies, were obtained via \gls{bfgs}, \Cref{fig:bfgs_scatter}. Here we see well-defined clusters, with expected behaviour.

To establish a baseline for all noise models, let us have a look at ideal, noiseless conditions first. Under these idealized conditions, the results shown in \Cref{tab:ideal_summary}, tell us an expected results, as all local optimizer managed to converge to the energy value $\mu_{\text{Final}} -1.124053$, which is the optimal state-averaged energy, whose ground and excited state components correspond to the reference values obtained via Psi4 \cite{psi4_1.4}. The notable difference is in the mean number of evaluations, $\mu_{\text{Evals}}$, that were necessary to achieve convergence. Here we can see that both \gls{bfgs} and \gls{slsqp} are the best performing optimizers, taking tens of evaluations to converge, thus being the most efficient optimizers, with \gls{bfgs} taking a slight lead. It is also important to note that the $\sigma_{\text{Evals}} < 15$, which is the standard deviation,  for both of the optimizers, making sure that even in the worst-case, the optimizer did not require a significantly larger amount of function evaluations. The other local optimizers needed hundreds of function evaluations to converge, and the $\sigma_{\text{Evals}}$ was also proportionally larger. The third best optimization method for ideal conditions is \gls{nm}, followed by \gls{pm} and then \gls{cobyla}. What was expected is that the global optimizer, \gls{isoma}, would have the hardest time reaching the minimum, ending with over a thousand function evaluations and unable to, on average, achieve the global optimum. The high number of function evaluations is due to the population-based core of the algorithm, and the non-optimal value of energy shows us that in the ideal condition and in the case of a small system like $H_2$, it is unnecessary to employ a sophisticated global optimizer as the search space's dimensionality is low, and there are no distortions that would require such method.

With the noiseless baseline established, the focus turns to the first kind of noise, which is inherent for all calculations performed on quantum computers, and that is the sampling noise, which stems from the inherent stochasticity of quantum mechanics. The presence of sampling noise, data shown in \Cref{tab:sn_summary}, in the optimization process does not prevent most optimizers from attaining the ideal state-average energy, but the accuracy is lower than in the case of ideal conditions. Gradient-based \gls{bfgs} consistently reaches within $10^{-3}$ of the desired value, while keeping the number of iterations in the hundreds. This number also decreases with the increase of measurements, showing a trend that we can achieve faster convergence with a slightly higher number of measurements. This makes \gls{bfgs} highly efficient. Deterministic derivative-free methods, \gls{nm} and \gls{pm}, also show the ability to achieve near-perfect final energy, but the number of function evaluations is notably higher than for \gls{bfgs}. For \gls{isoma} and \gls{cobyla}, the results are quite similar, but \gls{cobyla} requires about ten times fewer function evaluations, making it the optimizer with the fewest evaluations required. The one method that failed in all cases is \gls{slsqp}, as it was unable to achieve the optimum and still required a high number of evaluations.

So overall, \gls{bfgs} proves to be a quick and reliable choice, followed by \gls{cobyla} if we need good enough results and low computational cost, or \gls{pm} and if we want better accuracy, which comes with $3-4$ times higher computational cost. While \gls{nm} achieves good results in terms of accuracy, the number of evaluations it requires lets it be outperformed by other methods. Even the global optimizer, \gls{isoma}, needs about half the iterations to grant comparable results.

Moving on to the models containing decoherence. The first results depicting different levels of dephasing noise are shown in \Cref{tab:dp_summary}. The introduction of dephasing noise into our model leads to systematic worsening of our results. For the $1\%$ level of noise, \gls{bfgs}, \gls{nm}, and \gls{pm} perform relatively well with \gls{cobyla} and \gls{isoma} coming in the second tier, whereas \gls{slsqp} has significant convergence problems even in the lowest noise levels, rendering it unusable. 
With the increasing dephasing noise levels, the performance of the optimizers worsens, while the number of function evaluations remains similar, telling us that the dephasing noise distorts the landscape evenly without creating any new difficulties. What is interesting is the fact that for \gls{nm},
higher levels of noise lead to a decrease in evaluations that are necessary while having results comparable to \gls{bfgs} and \gls{pm}. This tells us that for \gls{nm}, the distortions of the landscape made by dephasing noise lead to faster convergence, probably because the inherent changes make the optimizer stay longer with larger differences in chosen coordinates.
Overall, we see that fine results are achieved under $1\%$ noise, and the absolute error is increasing linearly with the noise.

In the case of depolarizing noise, we can observe higher levels of deviations than in the case of dephasing noise. This is visualized in \Cref{tab:depol_summary}. The best overall performer is \gls{bfgs}, leading both in the best energy and a fairly low number of iterations as opposed to \gls{nm} and \gls{pm}, where the number of evaluations is in thousands. In this type of noise, we see \gls{cobyla} failing, and along with \gls{slsqp} being the most affected by the noise. \gls{isoma} is showing similar behavior as in previous cases, that is, having trouble finding the optimal values. 

Moving on to the case of thermal relaxation under feasible conditions that are similar to those on current quantum computers. From \Cref{tab:t2_summary} we observe the fact that \gls{bfgs} is consistently the top performer. This shows that in small search spaces, this method is resistant to thermal relaxation, and also is the fastest converging method. On the other hand, \gls{slsqp} is consistently the worst, performance-based, as it has trouble even in the longest relaxation time. The other two methods we can say are not a viable choice are \gls{nm} and \gls{isoma}, because the number of evaluations is significantly higher, and thus this makes them inefficient. The last examined method is \gls{pm}, where the results did not linearly correlate with the thermal relaxation level; thus, it is a choice that is unreliable, as with the scaling of the thermal relaxation, we do not get improvement in the expected direction.

The last set of estimators, with unrealistic times of thermal relaxation, \cite{rigetti2012superconducting} is displayed in \Cref{tab:trt1_summary}. These numbers display several things, firstly, that under these noise levels, there is no way for any algorithm to converge, as with these times, the information encoded into the ansatz vanishes, this is the effect known as noise limit, as even with the largest relaxation time, only the last few gates that are applied have any reasonable chance to affect the calculations. While there still remains variance in terms of the number of function evaluations, the expected behavior is displayed, as the results have a tendency to cluster in the same region, around similar values. This shows that these levels of noise are the only dominant factor, and that the choice of the optimization method itself is not significant, as it has a minor influence, and the obtained results are unusable.

\begin{table}[htbp]
\centering
\small
\begin{tabular}{lrrrr}
\toprule
Optimizer & $\mu_{\text{Final}}$ & $\sigma_{\text{Final}}$ & $\mu_{\text{Evals}}$ & $\sigma_{\text{Evals}}$ \\
\midrule
\gls{bfgs}   & -1.124053 & 0.000000 & 37.10  & 9.80  \\
\gls{cobyla} & -1.124051 & 0.000004 & 433.60 & 70.52 \\
\gls{isoma}  & -1.115658 & 0.008464 & 1135.80& 366.53 \\
\gls{nm}     & -1.124053 & 0.000000 & 295.20 & 25.27 \\
\gls{pm}     & -1.124053 & 0.000000 & 396.00 & 33.38 \\
\gls{slsqp}  & -1.124053 & 0.000000 & 52.00  & 12.65 \\
\bottomrule
\end{tabular}
\caption{Summary of optimization results under ideal conditions.}
\label{tab:ideal_summary}
\end{table}

\begin{table}[htbp]
\centering
\small
\begin{tabular}{lrrrr}
\toprule
Optimizer & $\mu_{\text{Final}}$ & $\sigma_{\text{Final}}$ & $\mu_{\text{Evals}}$ & $\sigma_{\text{Evals}}$ \\
\midrule
\multicolumn{5}{l}{\textbf{SN-256}} \\
\gls{bfgs}       & -1.120850 & 0.001351 & 363.10  & 190.07 \\
\gls{cobyla}     & -1.112210 & 0.021526 & 286.70  & 54.63  \\
\gls{isoma}      & -1.113026 & 0.010187 & 2040.40 & 393.25 \\
\gls{nm}         & -1.121963 & 0.001194 & 4787.10 & 827.51 \\
\gls{pm}         & -1.104752 & 0.025028 & 988.40  & 172.07 \\
\gls{slsqp}      & -0.886073 & 0.157799 & 1608.10 & 325.35 \\
\midrule
\multicolumn{5}{l}{\textbf{SN-512}} \\
\gls{bfgs}       & -1.121398 & 0.000759 & 365.50  & 134.29 \\
\gls{cobyla}     & -1.118548 & 0.008037 & 248.30  & 35.34  \\
\gls{isoma}      & -1.114753 & 0.009960 & 2215.50 & 607.06 \\
\gls{nm}         & -1.120785 & 0.000954 & 5082.40 & 1251.06 \\
\gls{pm}         & -1.116395 & 0.009086 & 1027.10 & 212.34 \\
\gls{slsqp}      & -0.793686 & 0.173310 & 1527.30 & 336.02 \\
\midrule
\multicolumn{5}{l}{\textbf{SN-1024}} \\
\gls{bfgs}       & -1.121119 & 0.000820 & 295.10  & 67.72  \\
\gls{cobyla}     & -1.114944 & 0.007365 & 270.60  & 50.13  \\
\gls{isoma}      & -1.113874 & 0.009434 & 1739.00 & 500.15 \\
\gls{nm}         & -1.120932 & 0.000928 & 5179.90 & 569.66 \\
\gls{pm}         & -1.120962 & 0.004637 & 837.00  & 182.30 \\
\gls{slsqp}      & -0.903307 & 0.172749 & 1674.30 & 280.76 \\
\midrule
\multicolumn{5}{l}{\textbf{SN-6144}} \\
\gls{bfgs}       & -1.121203 & 0.000382 & 296.50  & 58.39  \\
\gls{cobyla}     & -1.122400 & 0.001345 & 254.40  & 32.49  \\
\gls{isoma}      & -1.116929 & 0.011109 & 2037.50 & 749.76 \\
\gls{nm}         & -1.121183 & 0.000428 & 5176.50 & 880.26 \\
\gls{pm}         & -1.122546 & 0.002113 & 964.70  & 264.61 \\
\gls{slsqp}      & -0.782598 & 0.149267 & 1283.10 & 225.45 \\
\bottomrule
\end{tabular}
\caption{Summary of optimization results under sampling noise.}
\label{tab:sn_summary}
\end{table}

\begin{table}[htbp]
\centering
\small
\begin{tabular}{lrrrr}
\toprule
Optimizer & $\mu_{\text{Final}}$ & $\sigma_{\text{Final}}$ & $\mu_{\text{Evals}}$ & $\sigma_{\text{Evals}}$ \\
\midrule
\multicolumn{5}{l}{\textbf{DP-1\%}} \\
\gls{bfgs}        & -1.114730 & 0.000635 & 335.70  & 59.57  \\
\gls{cobyla}      & -1.066868 & 0.002831 & 274.40  & 41.16  \\
\gls{isoma}       & -1.043178 & 0.012105 & 2356.70 & 1096.66 \\
\gls{nm}          & -1.114112 & 0.001037 & 4090.20 & 1887.48 \\
\gls{pm}          & -1.114574 & 0.001005 & 1620.30 & 800.79 \\
\gls{slsqp}       & -0.807131 & 0.121553 & 1679.20 & 412.94 \\
\midrule
\multicolumn{5}{l}{\textbf{DP-5\%}} \\
\gls{bfgs}        & -1.087991 & 0.001559 & 347.80  & 85.84  \\
\gls{cobyla}      & -0.882381 & 0.007467 & 264.10  & 48.66  \\
\gls{isoma}       & -0.890502 & 0.082490 & 2648.50 & 1176.03 \\
\gls{nm}          & -1.086391 & 0.001657 & 2943.00 & 742.09  \\
\gls{pm}          & -1.087121 & 0.001709 & 1305.00 & 384.04  \\
\gls{slsqp}       & -0.638770 & 0.047198 & 1315.20 & 231.53  \\
\midrule
\multicolumn{5}{l}{\textbf{DP-10\%}} \\
\gls{bfgs}        & -1.053551 & 0.002349 & 296.70  & 60.82  \\
\gls{cobyla}      & -0.714788 & 0.010197 & 260.60  & 39.63  \\
\gls{isoma}       & -0.869096 & 0.100150 & 2081.40 & 645.04  \\
\gls{nm}          & -1.054817 & 0.002169 & 3041.80 & 1059.16 \\
\gls{pm}          & -1.054651 & 0.002056 & 1460.80 & 559.85  \\
\gls{slsqp}       & -0.585455 & 0.046439 & 1433.30 & 455.71  \\
\midrule
\multicolumn{5}{l}{\textbf{DP-20\%}} \\
\gls{bfgs}        & -0.993124 & 0.003935 & 334.10  & 83.85  \\
\gls{cobyla}      & -0.552636 & 0.008491 & 284.30  & 18.92  \\
\gls{isoma}       & -0.642074 & 0.102140 & 1808.40 & 380.16  \\
\gls{nm}          & -0.992108 & 0.002812 & 2737.40 & 829.93  \\
\gls{pm}          & -0.993125 & 0.004564 & 1470.50 & 417.57  \\
\gls{slsqp}       & -0.500512 & 0.007260 & 1609.90 & 521.27  \\
\bottomrule
\end{tabular}
\caption{Summary of optimization results under dephasing noise models.}
\label{tab:dp_summary}
\end{table}

\begin{table}[htbp]
\centering
\small
\begin{tabular}{lrrrr}
\toprule
Optimizer & $\mu_{\text{Final}}$ & $\sigma_{\text{Final}}$ & $\mu_{\text{Evals}}$ & $\sigma_{\text{Evals}}$ \\
\midrule
\multicolumn{5}{l}{\textbf{DEPOL-1\%}} \\
\gls{bfgs}        & -1.075892 & 0.001902 & 304.70  & 64.05  \\
\gls{cobyla}      & -0.842345 & 0.007302 & 285.20  & 65.42  \\
\gls{isoma}       & -0.910764 & 0.107383 & 2103.40 & 639.58 \\
\gls{nm}          & -1.075821 & 0.002196 & 2636.80 & 702.12 \\
\gls{pm}          & -1.076834 & 0.002692 & 1435.10 & 305.30 \\
\gls{slsqp}       & -0.644954 & 0.044021 & 1274.10 & 183.04 \\
\midrule
\multicolumn{5}{l}{\textbf{DEPOL-5\%}} \\
\gls{bfgs}        & -0.926838 & 0.003819 & 415.70  & 123.57 \\
\gls{cobyla}      & -0.497159 & 0.006110 & 247.40  & 26.25  \\
\gls{isoma}       & -0.662891 & 0.139207 & 2432.30 & 987.37 \\
\gls{nm}          & -0.926605 & 0.004067 & 2640.40 & 701.85 \\
\gls{pm}          & -0.927095 & 0.002709 & 1725.00 & 520.84 \\
\gls{slsqp}       & -0.478193 & 0.008645 & 1266.20 & 230.18 \\
\midrule
\multicolumn{5}{l}{\textbf{DEPOL-10\%}} \\
\gls{bfgs}        & -0.794687 & 0.005111 & 337.00  & 81.42  \\
\gls{cobyla}      & -0.463104 & 0.003334 & 236.00  & 23.67  \\
\gls{isoma}       & -0.529579 & 0.116088 & 2281.30 & 572.57 \\
\gls{nm}          & -0.791182 & 0.003995 & 2843.70 & 1138.18 \\
\gls{pm}          & -0.792864 & 0.005558 & 1823.60 & 657.97 \\
\gls{slsqp}       & -0.461766 & 0.005858 & 1304.40 & 222.70 \\
\midrule
\multicolumn{5}{l}{\textbf{DEPOL-20\%}} \\
\gls{bfgs}        & -0.630411 & 0.001677 & 316.40  & 58.63  \\
\gls{cobyla}      & -0.460570 & 0.004747 & 282.00  & 43.91  \\
\gls{isoma}       & -0.464260 & 0.006281 & 2110.40 & 529.50 \\
\gls{nm}          & -0.633490 & 0.004929 & 2943.10 & 739.55 \\
\gls{pm}          & -0.629442 & 0.005980 & 1369.70 & 374.21 \\
\gls{slsqp}       & -0.460514 & 0.007062 & 1370.10 & 229.13 \\
\bottomrule
\end{tabular}
\caption{Summary of optimization results under depolarizing noise models.}
\label{tab:depol_summary}
\end{table}

\begin{table}[htbp]
\centering
\small
\begin{tabular}{lrrrr}
\toprule
Optimizer & $\mu_{\text{Final}}$ & $\sigma_{\text{Final}}$ & $\mu_{\text{Evals}}$ & $\sigma_{\text{Evals}}$ \\
\midrule
\multicolumn{5}{l}{\textbf{T2 = 70$\mu$s}} \\
\gls{bfgs}   & -1.122428 & 0.002691 & 293.70  & 54.20  \\
\gls{cobyla} & -1.055532 & 0.003830 & 445.50  & 56.68  \\
\gls{isoma}  & -1.029268 & 0.018616 & 1953.80 & 399.60 \\
\gls{nm}     & -1.120829 & 0.003170 & 3047.40 & 1063.07 \\
\gls{pm}     & -1.117958 & 0.003686 & 1367.40 & 344.32 \\
\gls{slsqp}  & -0.762712 & 0.098260 & 751.80  & 246.60 \\
\midrule
\multicolumn{5}{l}{\textbf{T2 = 80$\mu$s}} \\
\gls{bfgs}   & -1.122921 & 0.003397 & 324.90  & 84.45  \\
\gls{cobyla} & -1.059595 & 0.005007 & 422.10  & 63.62  \\
\gls{isoma}  & -1.061441 & 0.028576 & 1965.60 & 548.02 \\
\gls{nm}     & -1.121465 & 0.002912 & 2743.90 & 1066.15 \\
\gls{pm}     & -1.115830 & 0.004948 & 1085.30 & 224.42 \\
\gls{slsqp}  & -0.790968 & 0.144783 & 645.20  & 175.81 \\
\midrule
\multicolumn{5}{l}{\textbf{T2 = 180$\mu$s}} \\
\gls{bfgs}   & -1.123770 & 0.003853 & 369.50  & 105.66 \\
\gls{cobyla} & -1.091851 & 0.005649 & 387.80  & 69.27  \\
\gls{isoma}  & -1.079421 & 0.005654 & 1888.70 & 398.09 \\
\gls{nm}     & -1.122740 & 0.003461 & 3047.70 & 1062.96 \\
\gls{pm}     & -1.084546 & 0.101001 & 765.40  & 337.73 \\
\gls{slsqp}  & -0.865482 & 0.135434 & 797.40  & 208.41 \\
\midrule
\multicolumn{5}{l}{\textbf{T2 = 380$\mu$s}} \\ 
\gls{bfgs}   & -1.125712 & 0.005031 & 329.50  & 86.37  \\
\gls{cobyla} & -1.103739 & 0.009152 & 299.20  & 34.36  \\
\gls{isoma}  & -1.098443 & 0.013342 & 2586.10 & 818.29 \\
\gls{nm}     & -1.124246 & 0.003378 & 2745.90 & 828.78  \\
\gls{pm}     & -1.118999 & 0.000687 & 351.70  & 212.30 \\
\gls{slsqp}  & -0.808040 & 0.195677 & 794.50  & 255.29 \\
\bottomrule
\end{tabular}
\caption{Summary of optimization results under $T_2$ noise models mimicking realistic conditions.}
\label{tab:t2_summary}
\end{table}

\begin{table}[htbp]
\centering
\small
\begin{tabular}{lrrrr}
\toprule
Optimizer & $\mu_{\text{Final}}$ & $\sigma_{\text{Final}}$ & $\mu_{\text{Evals}}$ & $\sigma_{\text{Evals}}$ \\
\midrule
\multicolumn{5}{l}{\textbf{TR-T1 = 50}} \\
\gls{bfgs}   & -0.491281 & 0.015304 & 392.70  & 104.28 \\
\gls{cobyla} & -0.438886 & 0.014340 & 433.60  & 114.71 \\
\gls{isoma}  & -0.452341 & 0.056327 & 2336.40 & 581.44 \\
\gls{nm}     & -0.493719 & 0.009967 & 3079.60 & 1270.86 \\
\gls{pm}     & -0.505781 & 0.010593 & 1456.10 & 386.38 \\
\gls{slsqp}  & -0.437642 & 0.025683 & 1821.00 & 313.57 \\
\midrule
\multicolumn{5}{l}{\textbf{TR-T1 = 100}} \\
\gls{bfgs}   & -0.211467 & 0.017875 & 473.90  & 140.48 \\
\gls{cobyla} & -0.267755 & 0.005440 & 285.30  & 56.76  \\
\gls{isoma}  & -0.127668 & 0.007830 & 1916.10 & 548.77 \\
\gls{nm}     & -0.216100 & 0.015541 & 3008.40 & 900.78  \\
\gls{pm}     & -0.220340 & 0.012304 & 1989.30 & 886.94  \\
\gls{slsqp}  & -0.267370 & 0.003906 & 1865.70 & 426.46  \\
\midrule
\multicolumn{5}{l}{\textbf{TR-T1 = 200}} \\
\gls{bfgs}   & -0.446706 & 0.007617 & 354.20  & 71.42  \\
\gls{cobyla} & -0.395097 & 0.007213 & 301.70  & 44.36  \\
\gls{isoma}  & -0.298103 & 0.005032 & 2139.70 & 524.19 \\
\gls{nm}     & -0.460963 & 0.006728 & 2863.50 & 797.29  \\
\gls{pm}     & -0.462090 & 0.004899 & 1753.80 & 325.92  \\
\gls{slsqp}  & -0.400840 & 0.004440 & 1335.89 & 292.10  \\
\midrule
\multicolumn{5}{l}{\textbf{TR-T1 = 300}} \\
\gls{bfgs}   & -0.571573 & 0.004647 & 402.40  & 111.41 \\
\gls{cobyla} & -0.442324 & 0.008752 & 282.40  & 77.91  \\
\gls{isoma}  & -0.354515 & 0.006189 & 2207.70 & 792.10 \\
\gls{nm}     & -0.587468 & 0.005057 & 2852.80 & 926.76  \\
\gls{pm}     & -0.586655 & 0.007705 & 1696.00 & 264.08  \\
\gls{slsqp}  & -0.435964 & 0.006146 & 744.40  & 251.22  \\
% \midrule
% \multicolumn{5}{l}{\textbf{Other TR-T1 cases (BFGS only)}} \\
% TR-T1 = 52  & -0.397180 & 0.010944 & 317.50 & 2.12   \\
% TR-T1 = 55  & -0.378869 & 0.002566 & 379.00 & 101.82 \\
% TR-T1 = 57  & -0.363914 & 0.007986 & 373.00 & 100.41 \\
% TR-T1 = 60  & -0.144880 & 0.003289 & 524.50 & 92.63  \\
% TR-T1 = 70  & -0.173418 & 0.006439 & 370.00 & 91.92  \\
% TR-T1 = 80  & -0.172533 & 0.005834 & 383.50 & 101.12 \\
% TR-T1 = 90  & -0.193128 & 0.013141 & 376.00 & 111.72 \\
\bottomrule
\end{tabular}
\caption{Summary of optimization results under \textbf{TR-T1} noise models to test unrealistic conditions.}
\label{tab:trt1_summary}
\end{table}

\begin{figure}
    \centering
    \includegraphics[width=0.99\linewidth]{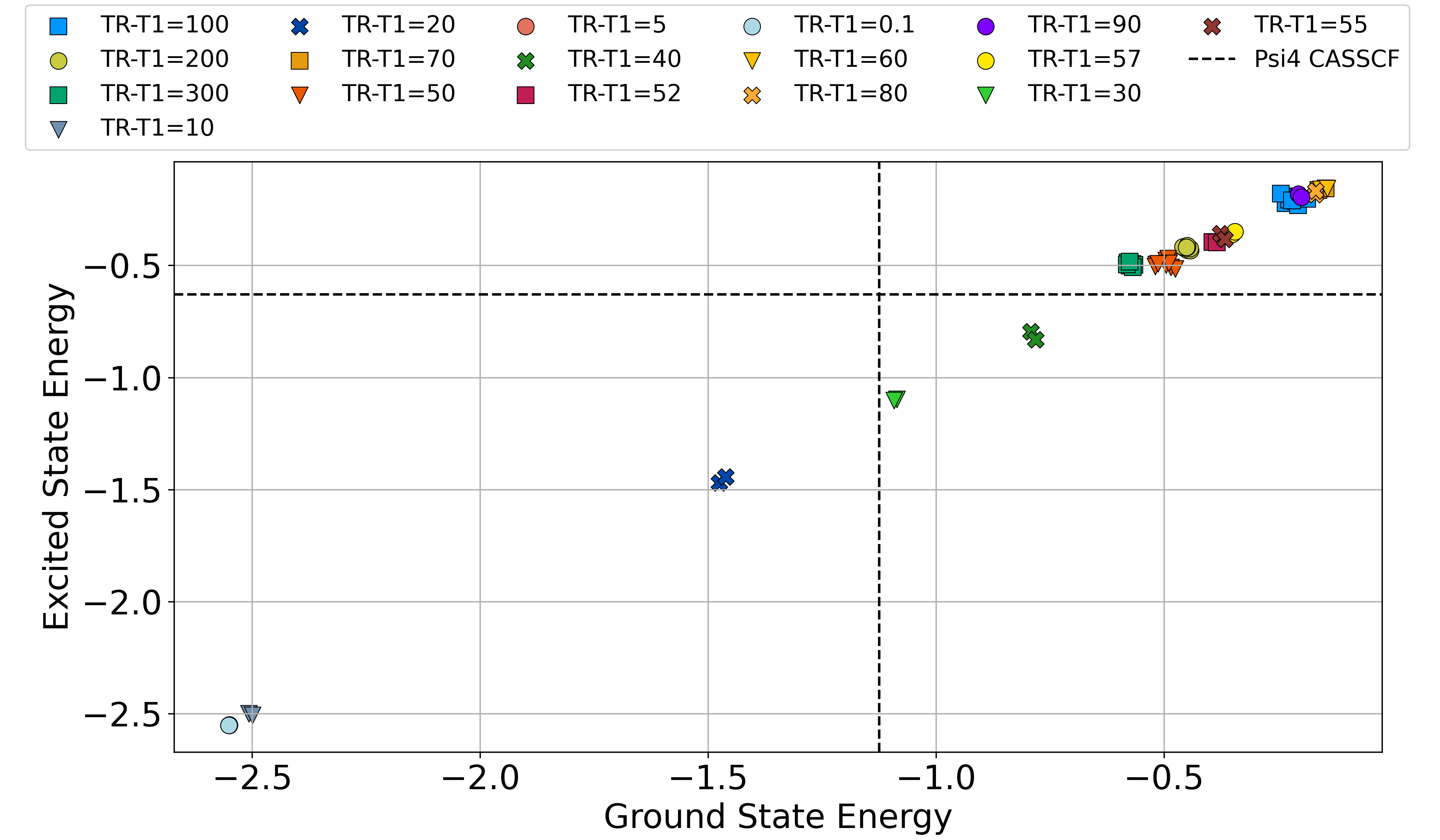}
    \caption{Further analysis and scaling of unrealistic $T_1$ and $T_2$ values.}
    \label{fig:bfgs-analysis}
\end{figure}

Also, an additional analysis was done with respect to the unrealistic $T_2$ and $T_1$ values, to better understand how the method behaves near its noise limit. The values themselves were chosen to reflect the range all the way to $0.1$ ns. It is shown in \Cref{fig:bfgs-analysis}. This was done only with \gls{bfgs} optimization method and the values were scaled all the way to $0.1$. This shows that with large enough levels of noise, the results are converging to a value, which is mainly influenced by the Hamiltonian of the problem.

\subsection{Statistical Analysis}
To broaden the scope of the obtained results and to obtain the general findings as much as possible, an encompassing statistical analysis was done next. First, the analysis of the spread, i.e., how large the clusters are, and differences of the clusters created by different estimator settings was performed to see if there are indeed statistical differences between the individual clusters. This analysis was performed for every optimizer independently to examine how the different types of decoherence affect the results of the optimization process. 

Next, we focused on the comparison of the optimizers themselves, first in a setting-by-setting manner, where the comparison is between distinct optimization methods, while keeping the conditions the same. Eventually, we investigated at the overall ranking of the methods, with respect to the chemical accuracy, as that is the most important aspect when running chemical calculations. A brief additional analysis regarding the number of cost function evaluations was also performed, further supporting the results and strengthening the overall conclusions of this study.

In the first parts of the statistical analysis, the goal was to determine whether the clusters for different estimators are significantly different in the position of centroids or not, to better judge if all noises affect the optimization process distinctly. For this, we have chosen a per optimization method approach, so that the effect of the decoherence could be studied independently.

At first, a \gls{manova} \cite{smith1962multivariate, o1985manova} was chosen to determine if there are statistically significant differences  with respect to the position of centroids between the clusters of data all at once. 

The suitability of \gls{manova} was evaluated through a series of assumption checks, the full details of which are reported in \Cref{app:manova}. In brief, tests of multivariate normality and homogeneity of covariance structures (including Mardia’s test, Box’s $M$, Levene’s, and Brown--Forsythe procedures) consistently indicated significant deviations from the required conditions. In particular, the equality of covariance matrices and variances across noise settings was systematically violated, demonstrating that the data exhibit heterogeneous dispersion and non-Gaussian structure. These results indicate that the parametric assumptions underlying \gls{manova} are not satisfied for the present dataset.

So instead, we adopted the \gls{permanova} \cite{anderson2014permutational} to help us figure out if there are statistically significant differences in the individual clusters, as \gls{permanova} is a non-parametric method that partitions variation in a distance matrix among groups and assesses significance using permutations. It does not assume multivariate normality but relies on the assumption of independent observations.

The assumption of independent observations is met by the design of how these results were obtained. But let us have a look at the multivariate dispersions, so that we can interpret the \gls{permanova} results more clearly, when we present them. 

For this, the \gls{permdisp} \cite{anderson2006distance} approach was chosen. In this procedure, the distance of each observation to its group centroid in multivariate space is computed, and these distances are then subjected to a one-way \gls{anova} \cite{st1989analysis}. The details of this approach and the analysis results themselves are discussed in detail in \Cref{app:permdisp}.

The detailed analysis is described in \Cref{app:permanova}, but we can say that overall, the pairwise post hoc analyses show a pretty consistent pattern across all of the optimization methods. For most of the pairs, there is a significant difference in both the location of the centroid and in the dispersion. The pairs that are statistically not different mostly consist of sampling noise, low levels of dephasing channel, and some categories of $T2$.

\subsection{Bootstrapping}
Next, we wanted to gain more insight about the results of the optimization methods, so we turned to Bootstrapping $95\%$ predictions for all optimizers in the ground vs excited energy plane, with the goal to be able to concretely predict the optimization outcomes for all methods and settings. This means that out of 100 samples, 95 of them will be in our defined area.

We visualize each family with a 95\% predictive ellipse in the $(x,y)$ plane, defined as the set of points whose squared Mahalanobis distance \cite{de2000mahalanobis} to the group mean $\boldsymbol{\mu}$ (with covariance $\boldsymbol{\Sigma}$) does not exceed a bootstrap-estimated cutoff $d_{0.95}^2$,
\begin{equation}
\mathcal{E}_{0.95}=\left\{\mathbf{x}\in\mathbb{R}^2:\;(\mathbf{x}-\boldsymbol{\mu})^\top \boldsymbol{\Sigma}^{-1}(\mathbf{x}-\boldsymbol{\mu})\le d_{0.95}^2\right\}.
\end{equation}

For each family, $\boldsymbol{\mu}$ and $\boldsymbol{\Sigma}$ are computed from its points, then $d_{0.95}^2$ is obtained by bootstrap resampling of the group, taking within-bootstrap 95th percentiles of Mahalanobis distances and using their median across bootstraps. The ellipse center indicates location (group mean), its area reflects within-group dispersion, and its tilt encodes covariance structure. The separation of centers suggests location differences, while differing sizes or overlaps highlight dispersion heterogeneity, complementing the formal \gls{permanova} and \gls{permdisp} results.

\begin{figure*}
\centering

% -------- Row 1 : legend strip --------
% (use any one ellipse plot if they share an identical legend)
\onlyTopThirty[width=0.95\textwidth]{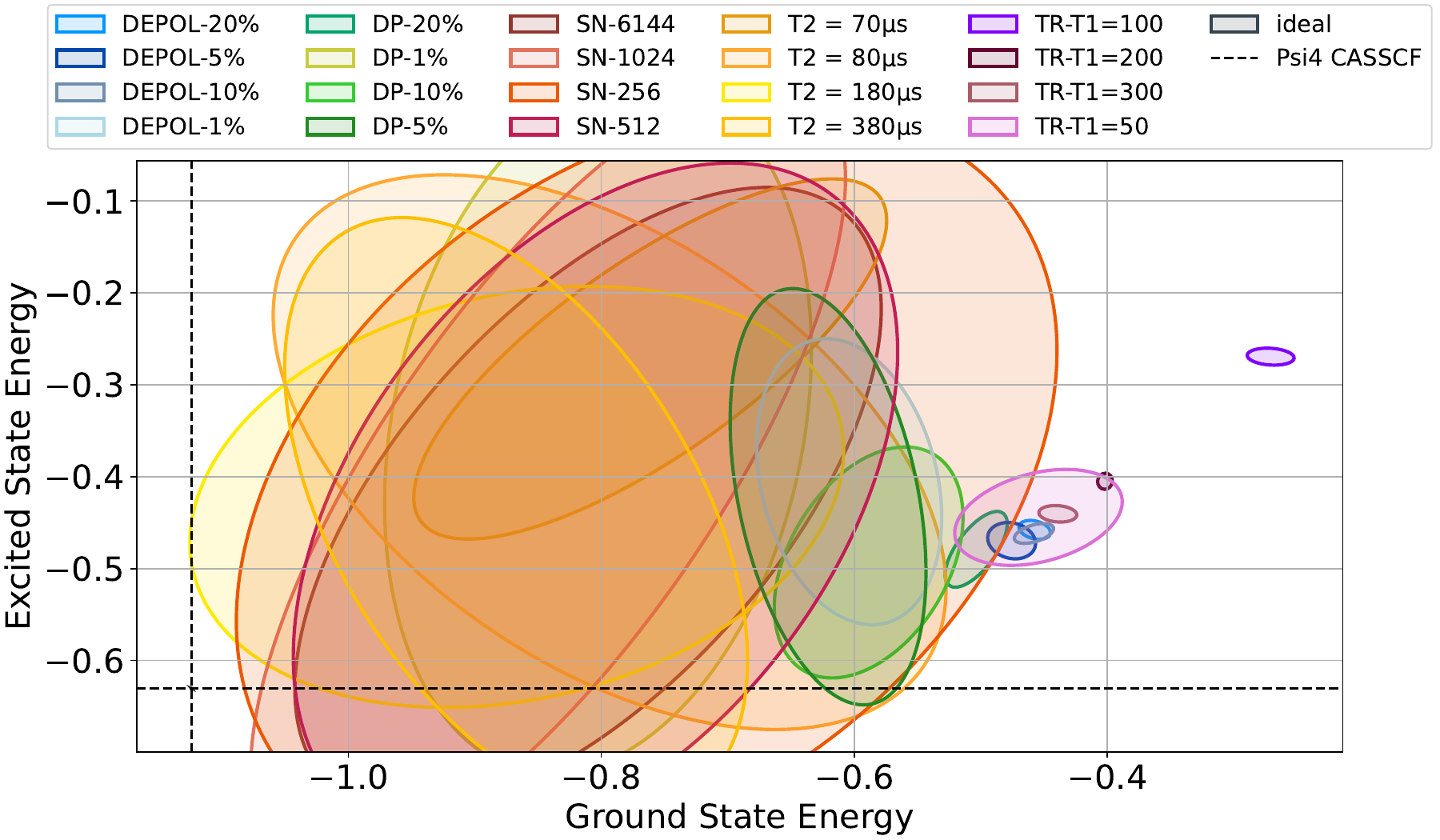}

\vspace{0.8em}

% -------- Row 2 --------

\subfigure[\gls{slsqp}\label{fig:slsqp_ellipses}]{
  \cropTopThirty[width=0.48\textwidth]{figs/stats/slsqp_ellipses_slsqp.pdf}
}\hfill
\subfigure[\gls{isoma}\label{fig:isoma_ellipses}]{
  \cropTopThirty[width=0.48\textwidth]{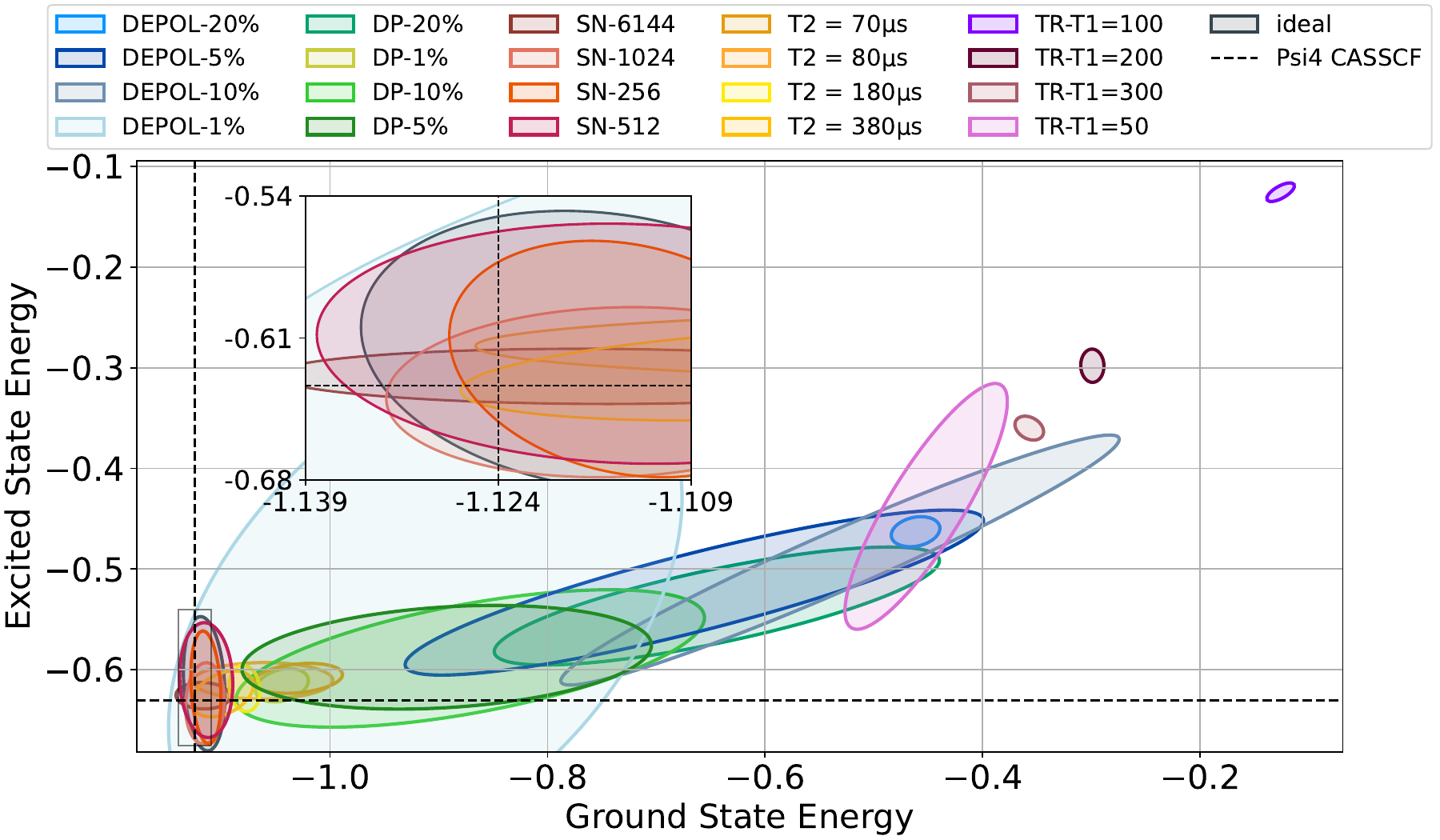}
}

\vspace{0.9em}

% -------- Row 3 --------
\subfigure[\gls{pm}\label{fig:pm_ellipses}]{
  \cropTopThirty[width=0.48\textwidth]{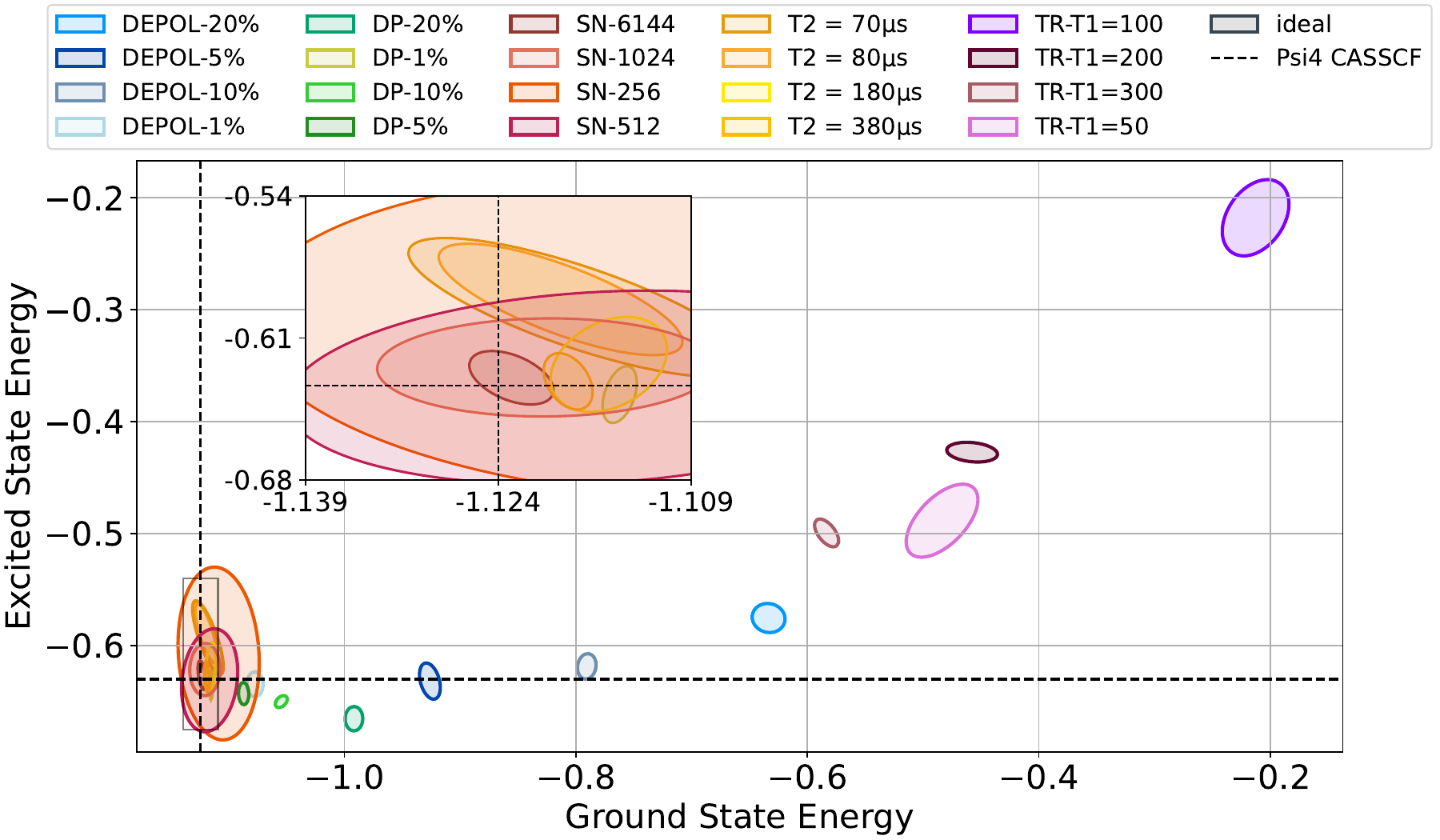}
}\hfill
\subfigure[\gls{nm}\label{fig:nm_ellipses}]{
  \cropTopThirty[width=0.48\textwidth]{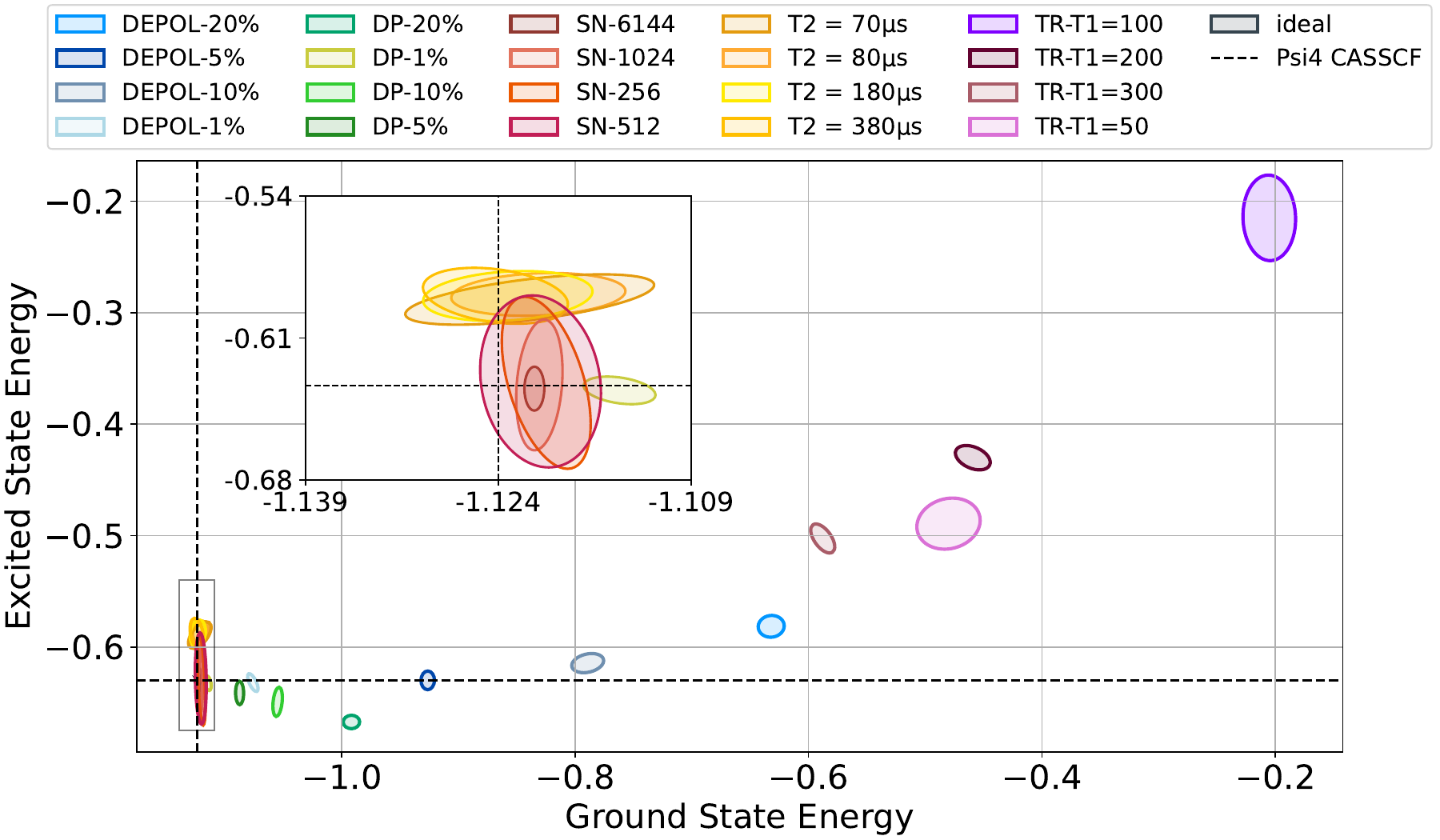}
}

\vspace{0.9em}

% -------- Row 4 --------
\subfigure[\gls{cobyla}\label{fig:cobyla_ellipses}]{
  \cropTopThirty[width=0.48\textwidth]{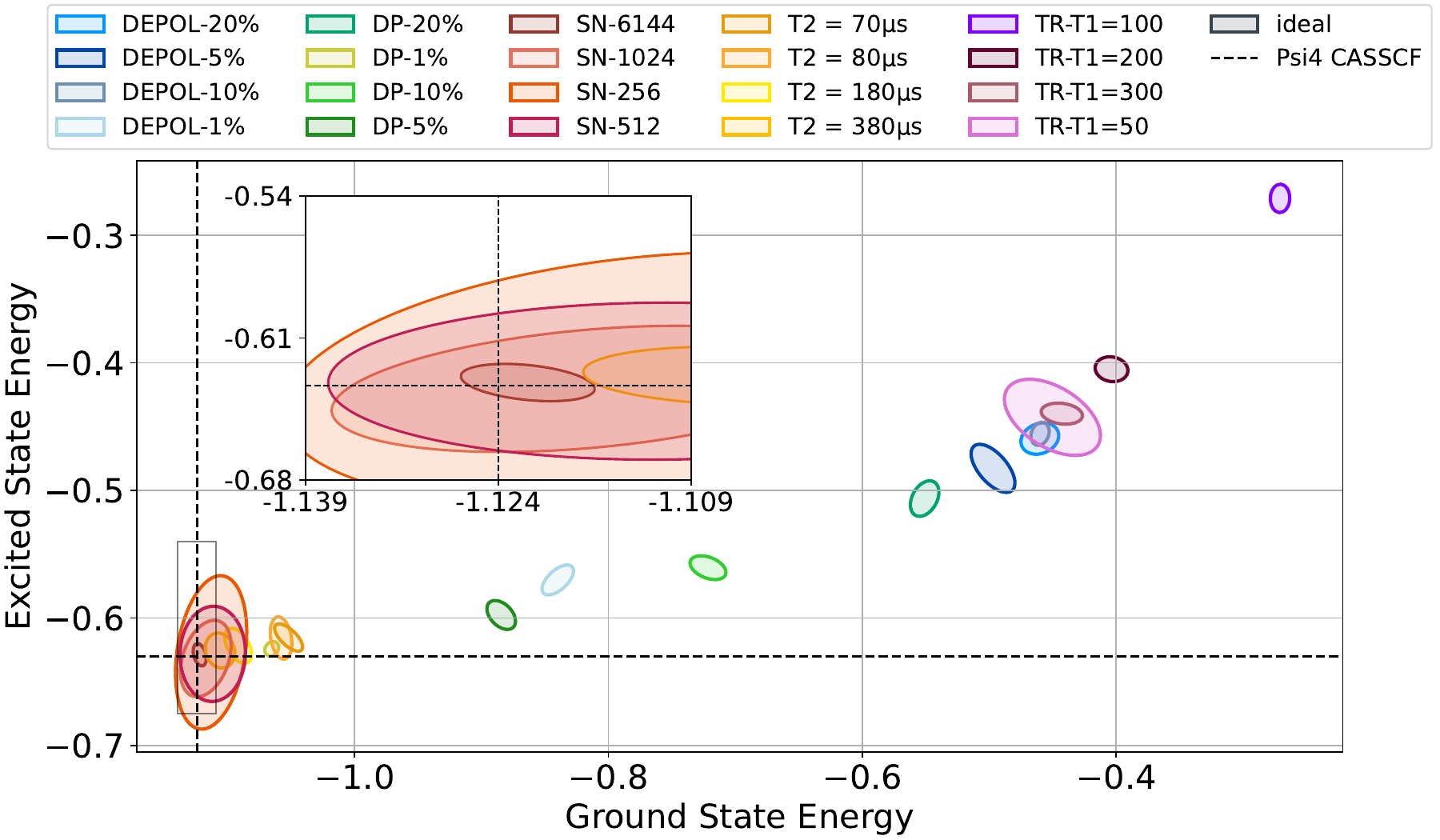}
}\hfill
\subfigure[\gls{bfgs}\label{fig:bfgs_ellipses}]{
  \cropTopThirty[width=0.48\textwidth]{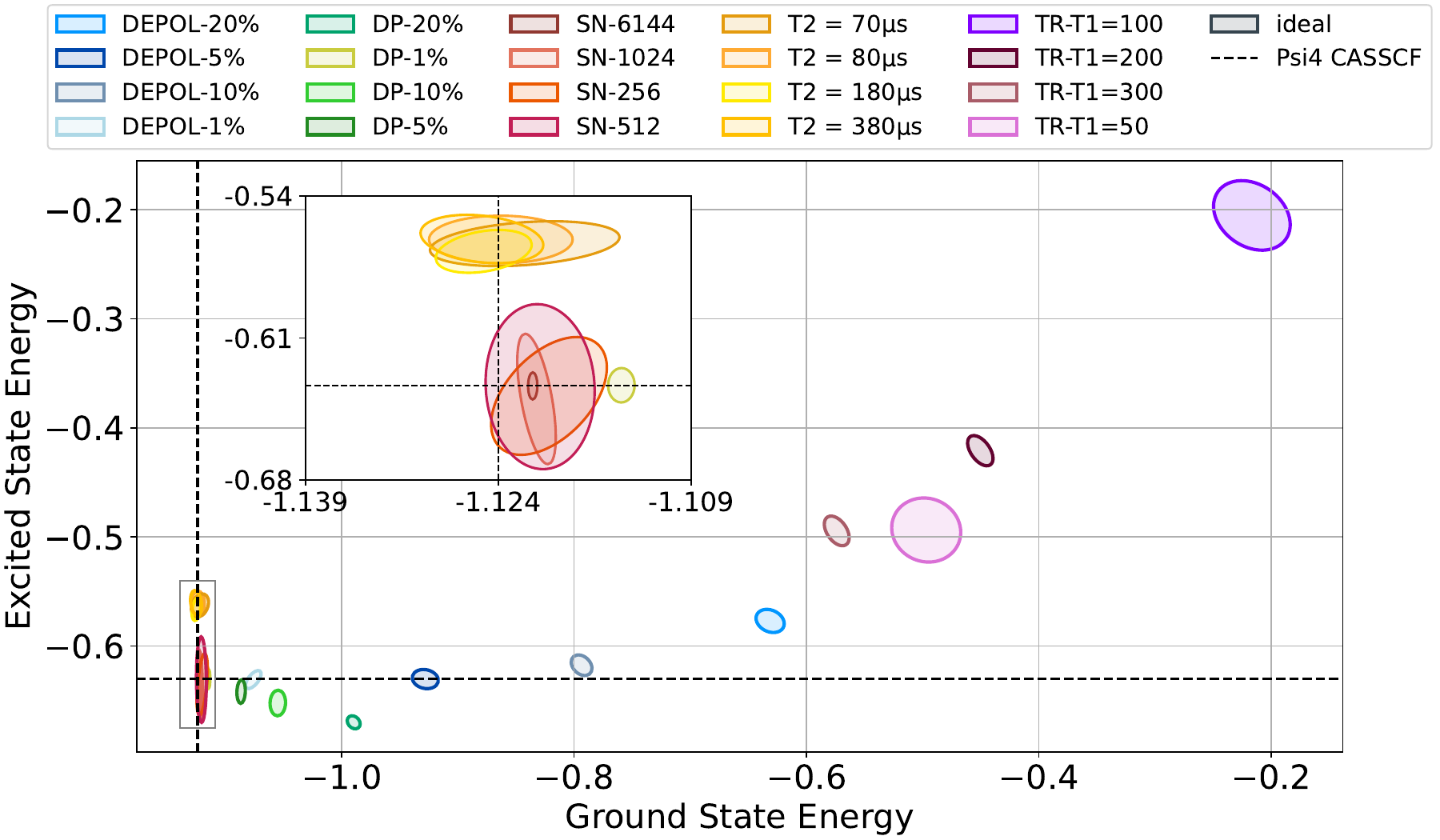}

}

\caption{Bootstrapped 95\% prediction ellipses in the ground versus excited state energy plane for all optimization methods. Ellipses indicate the variability and orientation of results within families. Dashed lines mark the reference energies from Psi4 CASSCF.}
\label{fig:all_ellipses}
\end{figure*}

The ellipses are visualized in \Cref{fig:all_ellipses}. In these figures, we observe several things, firstly, that for reasonable noise levels \gls{slsqp}, \Cref{fig:slsqp_ellipses}, is widely spread, showing that the obtained results will not be clustered, so this method is not really a good choice if we desire consistent results. The global optimizer \gls{isoma}, \Cref{fig:isoma_ellipses}, is compared to \gls{slsqp}, which is more clustered, but due to its heuristic nature, it is still outperformed by \gls{pm}, \Cref{fig:pm_ellipses}, \gls{nm}, \Cref{fig:nm_ellipses}, \gls{bfgs}, \Cref{fig:bfgs_ellipses}, and \gls{cobyla}, \Cref{fig:cobyla_ellipses}. 

For the last four mentioned optimizers, we see tighter clusters, pointing to the fact that these methods are under the same conditions clustering to the same value, showing that we are getting consistent values, and that each of the runs is well representative of the behavior of the optimizer.

This concludes the intra-optimizer analysis of the runs, and we can now move to the comparison of different optimization methods' performance.

\subsection{Overall Comparison}
To begin with, the overall performance of the optimizers was compared. This was done by comparing their relative accuracy with respect to the Psi4 reference values, obtained by running \gls{casscf}. This is shown in \Cref{tab:optimizer_summary}, where the overall mean and root-mean-square (RMS) distance which is calculated as
\begin{equation}
\mathrm{RMS} = \sqrt{\frac{1}{n}\sum_{i=1}^{n}\left(x_i - x_i^{\mathrm{ref}}\right)^{2}}
\end{equation}
where $x_i$ denotes the computed value for the $i$-th data point, $x_i^{\mathrm{ref}}$ is the corresponding reference value obtained from \gls{casscf} using Psi4, and $n$ is the total number of data points. The average rank across categories with its standard deviation (SD), and the number of categories and total data points used in the calculation, are shown. There we can observe that \gls{pm} achieved the lowest overall mean error equaling to $0.209$, and it was closely followed by \gls{nm} at $0.211$, and \gls{bfgs} with a value of $0.216$, while the worst performance is associated with \gls{slsqp}, with a value of $0.523$, and \gls{isoma} with \gls{cobyla} were constant under-performers with respect to the best three methods.

\begin{table}[t]
\centering
\small
\caption{Overall optimizer performance. Lower values of mean and RMS distance indicate better agreement with the Psi4 reference. Ranks are average $\pm$ SD across 21 categories.}
\label{tab:optimizer_summary}
\begin{tabular*}{\linewidth}{@{\extracolsep{\fill}}lcccccc}
\toprule
Optimizer & Mean & RMS & Avg.~Rank & SD & Cat. & Points \\
\midrule
\gls{pm}      & 0.209 & 0.355 & 2.36 & 1.22 & 21 & 210 \\
\gls{nm} & 0.211 & 0.356 & 2.10 & 1.09 & 21 & 210 \\
\gls{bfgs}        & 0.216 & 0.357 & 2.64 & 1.33 & 21 & 210 \\
\gls{isoma}       & 0.328 & 0.473 & 4.57 & 0.93 & 21 & 210 \\
\gls{cobyla}      & 0.329 & 0.460 & 3.90 & 1.41 & 21 & 210 \\
\gls{slsqp}       & 0.523 & 0.559 & 5.43 & 1.25 & 21 & 210 \\
\bottomrule
\end{tabular*}
\end{table}

The global Friedman's test \cite{pereira2015overview},
with the null hypothesis defined as
\begin{equation}
H_0:\ \overline{R}_1=\overline{R}_2=\cdots=\overline{R}_k
\end{equation}
where $\overline{R}_j$ is the mean rank of optimizer $j$ across $n$ categories and $k$ is the number of optimizers, confirmed that optimizer choice had a strong effect, with the values displayed in \Cref{tab:friedman}, and are indicative of a moderate-to-strong effect across the 21 categories.  To explain further, the Friedman test evaluates whether the rankings of optimizers differ across blocks (categories), while Kendall’s $W$ \cite{kendall1939problem, friedman1940comparison}, with the null hypothesis equaling to
\begin{equation}
H_0:\ W=0
\end{equation}
where $W\in[0,1]$ is Kendall’s coefficient of concordance over the rank matrices, provides an effect size, ranging from 0, meaning no agreement in rankings, to 1, signaling perfect agreement. 
Kendall’s coefficient of concordance $W$ is computed as 
\begin{equation}
W = \frac{12S}{k^{2}\left(n^{3}-n\right)}
\end{equation}
where $S = \sum_{j=1}^{k}\left(\overline{R}_j - \frac{n+1}{2}\right)^{2}$, $k$ is the number of optimizers, $n$ is the number of categories (blocks), and $\overline{R}_j$ is the sum of ranks assigned to optimizer $j$ across all categories.

The observed value 
of $W = 0.518$ therefore reflects a moderate-to-strong agreement among categories. The Friedman's test is a valid choice here as it is suitable for repeated measurements of data, and the categories are treated as blocks in the test. Also, there is no assumption of normality, and we only need ordinal data and balanced block sizes, which we both have.

\begin{table}[t]
\centering
\small
\caption{Global Friedman test of optimizer differences. The test was conducted across 21 categories common to all six optimizers.}
\label{tab:friedman}
\begin{tabular*}{\linewidth}{@{\extracolsep{\fill}}lcccc}
\toprule
$k$ (methods) & $n$ (blocks) & $\chi^2$ & $p$-value & Kendall’s $W$ \\
\midrule
6 & 21 & 54.35 & $1.8\times 10^{-10}$ & 0.518 \\
\bottomrule
\end{tabular*}
\end{table}

\begin{figure}[t]
    \centering
    \includegraphics[width=\linewidth]{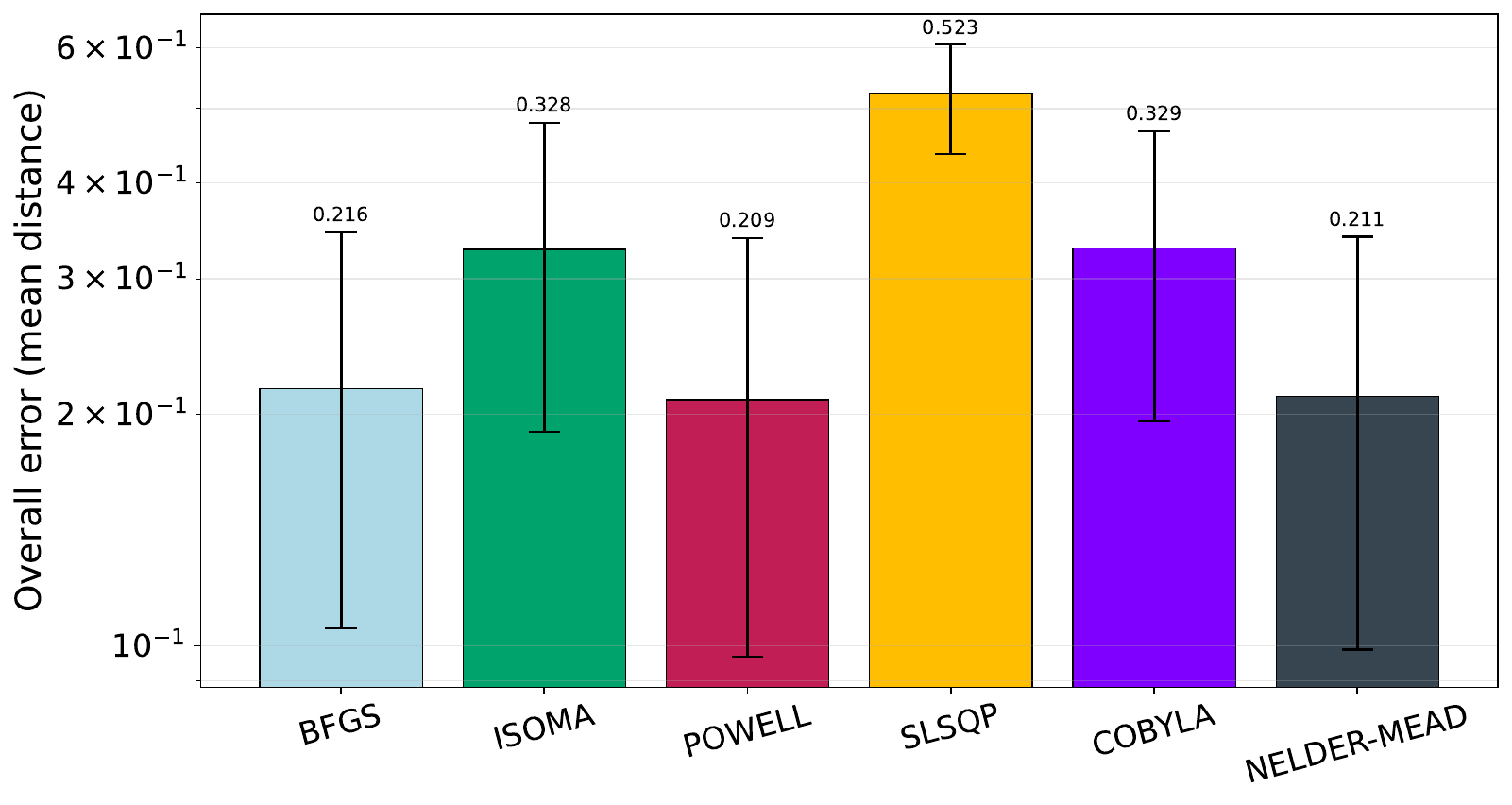}
    \caption{%
    Overall optimizer performance measured as mean distance to the Psi4 reference (bars $\pm$95\% bootstrap confidence interval on a log scale). Lower values indicate better agreement with the reference.}
    \label{fig:overall_error}
\end{figure}

Next, we have a look at the distribution of the mean distances from the Psi4 reference values and their associated $95\%$ confidence intervals, which are illustrated in \Cref{fig:overall_error}. This type of visualization shows that \gls{pm}, \gls{nm}, and \gls{bfgs} form a distinct group with similar error rates, while \gls{isoma} and \gls{cobyla} follow behind with \gls{slsqp} being the worst one overall. The logarithmic scale is used to emphasize the magnitude of the errors and accurately visualize the difference between the best and the worst performer.

Next, we want to further investigate if there are any significant differences in the optimizers, so we turn to the pairwise Wilcoxon signed-rank test \cite{woolson2007wilcoxon}, 
with the null hypothesis for comparing two optimizers defined as
\begin{equation}
H_0:\ \tilde{\Delta}=0
\end{equation}
where $\tilde{\Delta}$ is the median of paired differences between two optimizers.

The Holm correction \cite{holm1979simple} was also applied to control the family-wise error rate across all pairwise comparisons, as it provides stricter Type I error control than the Benjamini–Hochberg procedure, which is more suitable for exploratory analyses \cite{bender2001adjusting}.

This test was performed on all 21 categories, and as it relies on paired comparisons between optimizers, it only assumes that the distribution of paired differences is symmetric around the median, and unlike parametric alternatives, it does not require normality of the data, so it is suitable for our case. Holm correction was chosen to control the family-wise error rate across all 
pairwise comparisons.

The results for this test are presented in \Cref{tab:wilcoxon}, and they reveal that \gls{slsqp} was significantly worse than all other optimization methods, with $p-$values below the significance threshold of $0.05$. In a similar manner, \gls{isoma} and \gls{cobyla} were consistently outperformed by the other three methods, among which no significant difference was found any significant difference.

\begin{table}[t]
\centering
\small
\caption{Significant pairwise comparisons between optimizers based on Wilcoxon signed-rank tests across 21 categories. Reported are median paired differences (method\(_a\) – method\(_b\)) and Holm-adjusted $p$-values. Negative differences indicate that method\(_a\) performed better.}
\label{tab:wilcoxon}
\begin{tabular*}{\linewidth}{@{\extracolsep{\fill}}lccc}
\toprule
Comparison & Median diff. & $W$ & $p_{\text{Holm}}$ \\
\midrule
\gls{pm} vs \gls{slsqp}       & -0.314 & 0.0  & $2.5\times 10^{-5}$ \\
\gls{nm} vs \gls{slsqp}       & -0.312 & 0.0  & $2.5\times 10^{-5}$ \\
\gls{bfgs} vs \gls{slsqp}     & -0.307 & 0.0  & $2.5\times 10^{-5}$ \\
\gls{cobyla} vs \gls{slsqp}   & -0.195 & 0.0  & $4.2\times 10^{-4}$ \\
\gls{isoma} vs \gls{slsqp}    & -0.195 & 0.0  & $4.2\times 10^{-4}$ \\
\gls{pm} vs \gls{isoma}       & -0.119 & 12.0 & $0.014$ \\
\gls{pm} vs \gls{cobyla}      & -0.118 & 12.0 & $0.014$ \\
\gls{nm} vs \gls{isoma}       & -0.117 & 12.0 & $0.014$ \\
\gls{nm} vs \gls{cobyla}      & -0.116 & 12.0 & $0.014$ \\
\gls{bfgs} vs \gls{isoma}     & -0.111 & 12.0 & $0.014$ \\
\gls{bfgs} vs \gls{cobyla}    & -0.110 & 12.0 & $0.014$ \\
\bottomrule
\end{tabular*}
\end{table}

To focus on the problem with finer granularity and examine the differences within the individual noise categories. This is done because different types of noise may favor different optimization methods, and the good optimizer choice with respect to the known noise is crucial for obtaining the desired results.

\begin{figure*}[t]
    \centering
    \includegraphics[width=\linewidth]{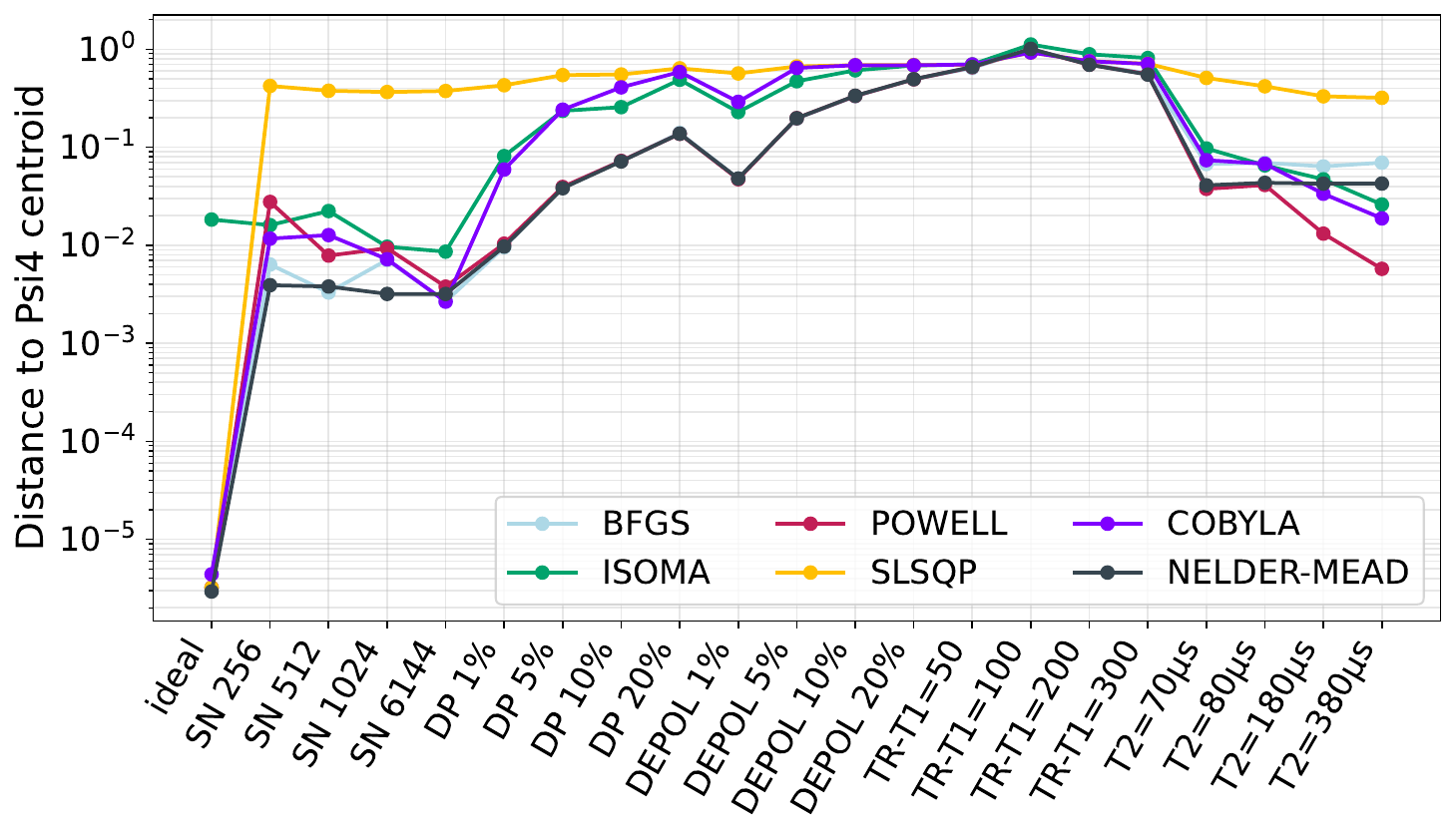}
    \caption{%
%    \bruno{isn't it fully redundant with Fig. 9 ? I prefer this one, easier to read in my opinion.}
    Mean distance to the Psi4 reference for each optimizer across all noise categories. Lower values indicate better agreement with the reference.}
    \label{fig:distance_by_category}
\end{figure*}

For this, we plotted the grouped error for each optimization method, which is visible in \Cref{fig:distance_by_category}, and defined as the centroid distance to the Psi4 reference, which was computed as
\begin{equation}
D_{jc} = \sqrt{\left(\bar E_{g,jc}-E_g^{\mathrm{ref}}\right)^2 + \left(\bar E_{e,jc}-E_e^{\mathrm{ref}}\right)^2}
\end{equation}
where $D_{jc}$ denotes the centroid distance for optimizer $j$ and noise type $c$, $\bar E_{g,jc}$ and $\bar E_{e,jc}$ are the mean ground and excited energies obtained for that optimizer–noise combination, and $E_g^{\mathrm{ref}}$ and $E_e^{\mathrm{ref}}$ are the corresponding Psi4 reference values.

There we see a direct comparison of the mean errors, for each configuration, where, as expected, for all methods except the \gls{slsqp}, the errors are significantly smaller in the ideal case, as \gls{slsqp} had convergence troubles even in the optimal case. Furthermore, we can observe how the error evolves with different levels of noise.
%For clarity, we present these data in \Cref{fig:distance_by_category}, where we can see more clearly the difference for the same optimizer when the noise levels increase.

% \begin{figure}[t]
%     \centering
%     \includegraphics[width=\linewidth]{figs/stats/interopt_grouped_error_by_category.pdf}
%     \caption{\bruno{don't hesitate to use widespread so that it takes the space of the two columns. Here it is a bit small.}
%     Per-category optimizer performance. Bars represent the mean distance to the Psi4 reference. Lower values indicate better agreement.}
%     \label{fig:grouped_error}
% \end{figure}

Next, across individual noise categories, tied-rank groupings derived from Friedman tests with Wilcoxon–Holm post-hoc comparisons showed consistent structure. The tied rankings were obtained by first applying the Friedman test within each noise setting to assess overall optimizer differences, followed by pairwise Wilcoxon signed-rank tests with Holm correction to identify significant contrasts. Optimizers that did not differ significantly at $\alpha=0.05$ were grouped together by rank groups. This approach ensures that the reported ranks reflect statistically validated groupings rather than raw performance differences.

\begin{figure}[t]
    \centering
    \includegraphics[width=\linewidth]{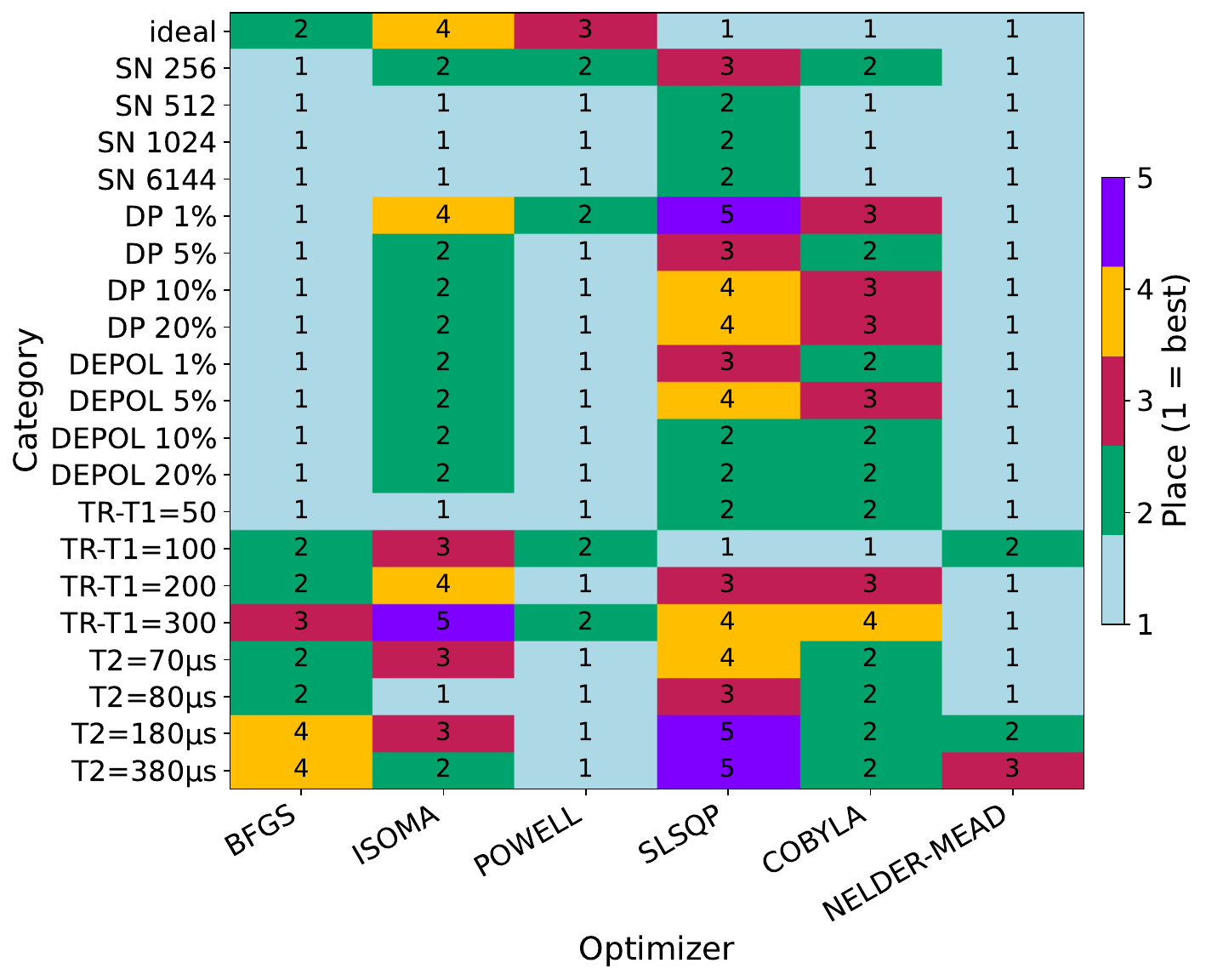}
    \caption{%
    Heatmap of optimizer ranks across noise categories. Darker shades indicate better ranks (1 = best). Each row corresponds to a category, and each column to an optimizer.}
    \label{fig:place_heatmap}
\end{figure}

As displayed in the heatmap in \Cref{fig:place_heatmap}, in most categories, \gls{pm}, \gls{nm}, and \gls{bfgs} occupied the top positions, often forming a statistically indistinguishable leading group, while \gls{slsqp} nearly always fell into the worst group. For example, in the SN-256 setting, \gls{bfgs} and \gls{nm} tied for best, with \gls{slsqp} uniquely last. In DP 1\%, \gls{nm} and \gls{bfgs} dominated, with \gls{slsqp} again trailing. Under T\textsubscript{2}, \gls{pm} emerged as the most stable method, taking sole first place at $180\mu$s and $380\mu$s, whereas \gls{bfgs} dropped markedly in these regimes. These tied rankings underline the heterogeneity across categories but also reinforce the general trend as the three quasi-Newton/simplex–type methods (\gls{pm}, \gls{nm}, \gls{bfgs}) consistently populate the best statistical groups, \gls{isoma} and \gls{cobyla} remain mid-field, and \gls{slsqp} is systematically the worst.

The optimizers were ranked by taking into account each pairwise rating and combining them into the ranked list. It was possible for multiple optimizers to share a rank, so the placement at the first placement meant that there had not been any statistical difference between the optimizers that rank in the same place, and all other optimizers were significantly worse. This process was done with all optimizers until all of them were assigned a place.
For further overview, the average place was calculated along with the number of placements in the best place, and this is all displayed in \Cref{tab:rank_summary}.

\begin{table}[t]
\centering
\small
\caption{Per-category rank summary of optimizers. Reported are the average place (1 = best), standard deviation (SD) of the rank across 21 categories, and the number of first-place finishes (\#wins). Lower average place and more wins 
indicate more robust performance.}
\label{tab:rank_summary}
\begin{tabular*}{\linewidth}{@{\extracolsep{\fill}}lccc}
\toprule
Optimizer & Avg.~Place & SD & \#Wins \\
\midrule
\gls{nm}      & 2.10 & 1.09 & 7 \\
\gls{pm}      & 2.36 & 1.22 & 6 \\
\gls{bfgs}    & 2.64 & 1.33 & 5 \\
\gls{cobyla}  & 3.90 & 1.41 & 2 \\
\gls{isoma}   & 4.57 & 0.93 & 1 \\
\gls{slsqp}   & 5.43 & 1.25 & 0 \\
\bottomrule
\end{tabular*}
\end{table}

Ranking analyses corroborated these conclusions. The Friedman test revealed a significant overall effect ($\chi^2=54.35$, $p=1.8\times10^{-10}$, Kendall’s $W=0.518$). Pairwise Wilcoxon tests with Holm correction further showed that BFGS, NM, and PM are statistically indistinguishable in performance, whereas \gls{slsqp} is significantly worse than all others ($p_{\text{Holm}}\leq 4.2\times10^{-4}$). \gls{cobyla} and \gls{isoma} also underperformed relative to the top three optimizers, although their deficits were less severe.

The optimizers were quantitatively separated by the aggregated distance-to-reference metric. While \gls{cobyla} ($0.329$) and \gls{isoma} ($0.328$) showed intermediate deviations from the exact reference, \gls{pm} ($0.209$), \gls{nm} ($0.211$), and \gls{bfgs} ($0.216$) formed a high-performing cluster with small deviations. With the biggest disparity ($0.523$), \gls{slsqp} stood out and demonstrated its consistently subpar performance in every scenario that was tested.

% \section{Software}\label{sec:software}
% All calculations in this work were obtained via the \gls{sa-oo-vqe} software package \cite{beseda2024state}, which is developed in Python and openly available at \hyperlink{Gitlab}{https://gitlab.com/MartinBeseda/sa-oo-vqe-qiskit} and \hyperlink{Pypi}{https://pypi.org/project/saoovqe/}. The exact version of the solver is archived at Zenodo \todo{ADD Zenodo} along with a replication package for this project, with a Conda environment.
\section{Software}\label{sec:software}
All calculations in this work were obtained via the \gls{sa-oo-vqe} software package \cite{beseda2024state}, which is developed in Python and openly available at \href{https://gitlab.com/MartinBeseda/sa-oo-vqe-qiskit}{https://gitlab.com/MartinBeseda/sa-oo-vqe-qiskit} and \href{https://pypi.org/project/saoovqe/}{https://pypi.org/project/saoovqe/}. The exact version of the solver used in this study corresponds to Git commit \href{https://gitlab.com/illesova.silvie.scholar/sa-oo-vqe-qiskit/-/commit/504013abfe1b32a043a629295da003138aeecfa2}{504013abfe1b32a043a629295da003138aeecfa2}, which is archived  along with a replication package at Zenodo (DOI: \href{https://doi.org/10.5281/zenodo.17296586}{10.5281/zenodo.17296586}) \cite{illesova2025replication}.

\section{Conclusions}\label{sec:conclusion}
 The ideal conditions served as a benchmark, where all local optimizers converged, and only the global optimization method had trouble obtaining the correct results. The best method in this benchmark was \gls{bfgs}, which showed extremely fast convergence. The global convergence method \gls{isoma} did not provide any advantage, mainly due to the small dimensionality of the studied problem. And the outlook is that this method will become better suited when problems with higher dimensionality and with higher multimodality are studied. 

In the case of sampling noise, where finite number of measurements was employed,  convergence with adequate accuracy, $10^{-3}$, is achievable, but new phenomena appears, and that is the fact that even due to this inherent noise, the variational principle, on which \gls{sa-oo-vqe} method is build does not hold, and the obtained values can be lower that achieved energy. This can be mitigated with a sufficient number of measurements, but it still has to be considered. Also, in these conditions, \gls{bfgs}, \gls{nm}, and \gls{pm} attained similar levels of accuracy, but from a pragmatic point of view, \gls{bfgs} comes out the best as the other two methods required significantly higher numbers of function evaluations, thus being inefficient. One method that proved to be unreliable in all sampling noise levels was \gls{slsqp}, which failed to converge in all cases, and an increase in the number of measurements did not lead to an increase in accuracy.

Models with quantum decoherence showed a stronger influence on the results. When the dephasing channel was introduced, a slight but uniform decrease in accuracy arose, but the evaluation counts remained fairly stable. The same levels of depolarizing noise proved to have greater influence than their dephasing counterparts, but still \gls{bfgs} remained the best, mainly due to its highest accuracy and stable and low number of evaluations. Next, in the case of realistic levels of thermal decoherence, the \gls{bfgs} was the best performing, but the results were far from required accuracy, \gls{nm} and \gls{isoma} were consistently inefficient, \gls{pm} showed non-monotonic sensitivity to the noise, and \gls{slsqp} proved to be the worst choice. In the harsh conditions, where the times of thermal relaxations were by two orders of magnitude shorter than those found in current devices, it became apparent that the choice of the optimization methods becomes irrelevant, as there is nearly a full loss of information that is encoded in the ansatz.

The observed performance trends among the optimizers were further supported by statistical testing. The \gls{permanova} analysis shown in \Cref{tab:permanova_all_methods}) verified that the differences are highly significant ($p < 10^{-4}$) for all optimizers) and not anecdotal. Simultaneously, \gls{permdisp} displayed in \Cref{tab:permdisp_all_methods} showed dispersion heterogeneity, indicating that the observed variability is influenced by both centroid shifts and spread differences. These results show that distributional characteristics must be taken into account in order to fully capture optimizer performance, which cannot be achieved solely by averaging outcomes.

Mardia's test of multivariate normality demonstrated that most optimizer-noise families retained approximate normality, with the notable exception of \gls{slsqp} in the ideal family. Specifically, significant deviations were detected in skewness ($p_{\text{skew}}=0.002$) and kurtosis ($p_{\text{kurt}}=0.041$). This highlights that optimizer instability can manifest even under noiseless conditions, underscoring the poor suitability of \gls{slsqp} for variational quantum eigensolver applications.

Comparing different noise types showed that, at equivalent nominal levels, depolarizing noise deteriorated results more than dephasing noise. For instance, \gls{bfgs} maintained $-1.0880$ under 5\% dephasing but converged to $-0.9268$ under 5\% depolarization (\Cref{tab:depol_summary,tab:dp_summary}). Because sensitivity to noise is optimizer-dependent, this emphasizes how crucial it is to match optimizer strategies with the dominant error channels of particular quantum devices.

Interestingly, a counterintuitive trend was identified for NM under dephasing noise, where the number of required function evaluations decreased as noise increased (from $4090$ at 1\% to $2737$ at 20\%, \Cref{tab:dp_summary}). This suggests that increased stochasticity may in some cases reshape the optimization landscape in ways that simplify convergence.

Finally, at extreme thermal relaxation limits (TR-T1 on the order of tens of ns), all optimizers converged to similarly poor energy values (\Cref{tab:trt1_summary}). This demonstrates that under such conditions, the choice of optimizer becomes largely irrelevant compared to the hardware noise floor, as algorithmic improvements cannot compensate for the noise limit.

Overall, \gls{bfgs} is consistently the best performer for small systems and \gls{sa-oo-vqe} methods, where the combinations of accuracy and efficiency were shown. When there is the need to minimize the number of function evaluations, \gls{cobyla} proves to be the best choice if one is open to a slight decrease in accuracy ranging from $1\%$ errors in stochastic noise to to $<6\%$ in the dephasing time study, but the number of function evaluations stays under $500$ in all the cases. On the other hand, one may look to \gls{nm} or \gls{pm} if the number of function evaluations is not an issue to have guaranteed higher levels of accuracy, as these methods were oftentimes among the best performers. \gls{isoma} proves to be unnecessary for the low-dimensional problems, if the landscape does not exhibit pronounced multimodality. And the last examined method, \gls{slsqp}, constantly exhibits insufficient capabilities, rendering itself an ill choice for the investigated noisy environments.

In follow-up work, we will focus on two goals: one is the testing of different optimization models, specifically some that are designed for a noisy environment, so that better convergence can be attained consistently. Among these tested will be CMAES \cite{hansen2003reducing}, BOBYQA \cite{powell2009bobyqa}, Bayesian optimization \cite{garnett2023bayesian} and ADAM \cite{kingma2014adam}. Although the present study highlights clear differences in optimizer performance, it remains unclear which specific features—such as population-based exploration, adaptive step-size control, or implicit regularization—are primarily responsible for improved noise resilience. Further investigation is required to determine how these properties interact with the structure of quantum noise and with the topology of the underlying energy landscape.

Secondly, the optimization methods must be tested on problems with higher dimensionality, where different molecules will be examined along with increases in active space. Here, it is not yet well understood how optimizer behavior scales with circuit depth, entanglement complexity, or the onset of barren plateaus. A systematic exploration of these factors will be essential for predicting performance in large-scale quantum systems. Lastly, to efficiently test the performance of different optimizers on real-world quantum hardware, error mitigation techniques must be considered. At present, the coupling between specific mitigation schemes and optimizer dynamics is poorly characterized—particularly whether noise reduction strategies like zero-noise extrapolation or measurement mitigation affect convergence speed or bias differently across methods.

The final goal is to construct a guide where the best optimization method can be chosen based on a priori knowledge of the problem and noise structure. This guide will ultimately depend on bridging the remaining gaps in understanding between optimizer mechanisms, landscape geometry, and the stochastic nature of real quantum devices.

\begin{acknowledgments}
Vojtěch Novák is supported by Grant of SGS No. SP2025/072, VSB-Technical University of Ostrava, Czech Republic. Martin Beseda was funded by Italian Government (Ministero dell'Università e della Ricerca, PRIN 2022 PNRR) -- cod.P2022SELA7: ''RECHARGE: monitoRing, tEsting, and CHaracterization of performAnce Regressions`` -- Decreto Direttoriale n. 1205 del 28/7/2023.  This work was supported by the Ministry of Education, Youth and Sports of the Czech Republic through the e-INFRA CZ (ID:90254). This project has received funding from the Research Council of Lithuania (LMTLT), agreement No. P-ITP-24-9.
\end{acknowledgments}

\bibliography{apssamp}% Produces the bibliography via BibTeX.

\appendix

\section{Optimizer Settings}\label{app:opt_settings}

 The overview of the optimization methods is described in \Cref{tab:optim_methods}, while their settings are further described here, to facilitate the reproducibility of the results.
 The maximum number of iterations for each state-averaged \gls{vqe} phase is set to $500$ for all optimizers except \gls{isoma}. The \texttt{ftol} parameter is set to $10^{-8}$  for intended accuracy, for \gls{bfgs} and \gls{slsqp}. \gls{isoma} needed a different set of parameters, and all can be seen in the \Cref{tab:optimizers}.

\begin{table}[htbp]
\centering
\caption{Overview of the optimization methods}
\label{tab:optim_methods}
\begin{tabular}{l l c c c}
\toprule
\textbf{Method} & \textbf{Type} & \textbf{Grad.} & \textbf{Constr.} & \textbf{Scope} \\
\midrule
\gls{bfgs}   & Quasi-Newton         & Yes & No        & Local \\
\gls{slsqp}  & Quadratic programming & Yes & Eq./Ineq. & Local \\
\gls{nm}     & Direct search        & No  & No        & Local \\
\gls{pm}     & Line search          & No  & No        & Local \\
\gls{cobyla} & Trust-region         & No  & Eq./Ineq. & Local \\
\gls{isoma}  & Population-based     & No  & Heuristic & Global \\
\bottomrule
\end{tabular}
\end{table}

\begin{table}[htbp]
\centering
\caption{Optimizer settings used in this study.}
\label{tab:optimizers}
\begin{tabular}{|l|l|}
\hline
\textbf{Optimizer} & \textbf{Settings} \\
\hline
\gls{bfgs} & maxiter = 500, ftol = $10^{-8}$ \\\hline
\gls{cobyla} & maxiter = 500 \\\hline
\gls{slsqp} & maxiter = 500, ftol = $10^{-8}$ \\\hline
\gls{nm} & maxiter = 500 \\\hline
\gls{pm} & maxiter = 500 \\\hline
\gls{isoma} & N\_jump = 10, Step = 0.11, PopSize = 25, \\
      & Max\_Migration = 30, Max\_FEs = 750, \\
      & VarMin = $-2\pi$, VarMax = $2\pi$, $m=15$,\\  & $n=5$, $k=10$ \\
\hline
\end{tabular}
\end{table}

% \section{Bootstrapped 95 \% prediction ellipses}\label{ap:boot}

% \bruno{If you gather everything in a single plot, it could eventually be in the main text. I find it a bit more visual than the rest.}

\section{MANOVA}\label{app:manova}
%begin
The null hypothesis for \gls{manova} in two dimensions, which are represented by the ground and excited energies, is defined as
\begin{equation}
H_0:\ \boldsymbol{\mu}_1=\boldsymbol{\mu}_2=\cdots=\boldsymbol{\mu}_g
\end{equation}
where $\boldsymbol{\mu}_i\in\mathbb{R}^2$ is the mean vector of group $i$ and $g$ is the number of groups. The groups are the individual noise type setting, each consisting of ten runs.

The \gls{manova} has three main assumptions that have to be met to allow for this type of statistical analysis. These are the assumptions of normality, homogeneity, and independence.

For the first assumption, of normality, we have chosen Mardia's test \cite{mardia1970measures} to check for multivariate normality, as our data consists of a tuple of ground and excited state energy. The Mardia's test and its results are described in detail in \cref{ap:mardia}.

Now, let us examine the second assumption for the \gls{manova} approach, i.e., the assumption of homogeneity. This assumption was tested by the following three tests.

Box’s M test \cite{box1953non} compares the equality of covariance matrices of all settings. It is defined as
\begin{equation}
M = \left( N - g \right) \ln \lvert \mathbf{S}_p \rvert \;-\; \sum_{i=1}^g (n_i - 1) \ln \lvert \mathbf{S}_i \rvert,
\end{equation}
where $N = \sum_{i=1}^g n_i$ is the total sample size, $n_i$ is the number of samples in the group $g_i$, $g$ is the number of groups, $\lvert \cdot \rvert$ denotes the determinant, $\mathbf{S}_i$ is the sample covariance matrix of group $i$, and $\mathbf{S}_p$ is the pooled covariance matrix,
\begin{equation}
\mathbf{S}_p = \frac{1}{N-g} \sum_{i=1}^g (n_i - 1)\mathbf{S}_i.
\end{equation}
The statistic $M$ is approximately $\chi^2$ distributed under the null hypothesis that all group covariance matrices are equal, 
$H_0: \mathbf{\Sigma}_1 = \mathbf{\Sigma}_2 = \dots = \mathbf{\Sigma}_g$.

And the null hypothesis for Box’s $M$ test for equality of covariance matrices is defined as
\begin{equation}
H_0:\ \boldsymbol{\Sigma}_1=\boldsymbol{\Sigma}_2=\cdots=\boldsymbol{\Sigma}_g
\end{equation}
where $\boldsymbol{\Sigma}_i\in\mathbb{R}^{2\times 2}$ is the covariance matrix of group $i$.

Here, the results are in \Cref{tab:boxm_all_methods}, and we can observe that the Box’s M test was significant for all optimization methods, with $p < 0.001$ in each case, indicating that the assumption of equality of covariance matrices was violated throughout. That said, the equality of group variances was tested next.

\begin{table}[ht]
\centering
\caption{Box’s M test of equality of covariance matrices across optimization methods. 
$p$-values are shown directly; significance threshold is $p > 0.001$.}
\label{tab:boxm_all_methods}
\begin{tabular}{lrrrr}
\toprule
\textbf{Method} & $M$ & $\chi^2$ & df & $p$ \\
\midrule
\gls{bfgs}   & 1058  & 969.3 & 60 & $<0.001$ \\
\gls{cobyla} & 589.2 & 539.7 & 60 & $<0.001$ \\
\gls{pm}     & 1068  & 977.3 & 60 & $<0.001$ \\
\gls{slsqp}  & 1278  & 1167  & 60 & $<0.001$ \\
\gls{nm}     & 724.7 & 663.8 & 60 & $<0.001$ \\
\gls{isoma}  & 698.7 & 639.9 & 60 & $<0.001$ \\
\bottomrule
\end{tabular}
\end{table}

Levene’s test \cite{levene1960robust} evaluates the equality of group variances. For each observation $Y_{ij}$ in group $i$, the transformed score is
\begin{equation}
Z_{ij} = \lvert Y_{ij} - \bar{Y}_{i\cdot} \rvert,
\end{equation}
where $\bar{Y}_{i\cdot}$ is the group mean. The null hypothesis for equality of variances is defined as
\begin{equation}
H_0:\ \sigma_1^{2}=\sigma_2^{2}=\cdots=\sigma_g^{2}
\end{equation}
where $\sigma_i^2$ is the variance of a single coordinate in group $i$.

The results for Levene's test are visible in \Cref{tab:levene_all_methods}. The columns report the optimization method, the Levene’s test statistic ($F$) for the $x$- and $y$-axis variables ($F_x$, $F_y$), and their corresponding $p$-values ($p_x$, $p_y$). And it was also significant for both the $x$ and $y$ axes across all optimization methods, with $p < 0.05$ in every case, showing that the assumption of homogeneity of variances was not satisfied.   
\begin{table}[ht]
\centering
\caption{Levene’s test of equality of variances across optimization methods. 
$p$-values are shown directly; significance threshold is $p > 0.05$.}
\label{tab:levene_all_methods}

\begin{tabular}{lrrrr}
\toprule
\textbf{Method} & $F_x$ & $p_x$ & $F_y$ & $p_y$ \\
\midrule
\gls{bfgs}   & 8.026  & $1.57\times10^{-16}$ & 4.976  & $9.592\times10^{-10}$ \\
\gls{cobyla} & 5.186  & $3.081\times10^{-10}$ & 7.425  & $2.947\times10^{-15}$ \\
\gls{pm}     & 7.697  & $8.165\times10^{-16}$ & 9.257  & $5.227\times10^{-19}$ \\
\gls{slsqp}  & 13.47  & $1.692\times10^{-26}$ & 13.57  & $1.14\times10^{-26}$ \\
\gls{nm}     & 13.63  & $5.48\times10^{-27}$ & 5.534  & $4.802\times10^{-11}$ \\
\gls{isoma}  & 20.25  & $8.8\times10^{-37}$  & 4.596  & $7.572\times10^{-9}$ \\
\bottomrule
\end{tabular}
\end{table}

The last test done for the second assumption of \gls{manova} was the Brown--Forsythe test \cite{brown1974robust}. It is a median-based modification of Levene’s test, which improves robustness under non-normality and was chosen because of the non-normality in the case of the \textit{ideal} family. The transformed scores are defined as
\begin{equation}
Z_{ij} = \lvert Y_{ij} - \tilde{Y}_{i} \rvert,
\end{equation}
where $\tilde{Y}_{i}$ is the group median. The null hypothesis for the Brown-Forsythe test is defined as
\begin{equation}
H_0:\ \sigma_1^{2}=\sigma_2^{2}=\cdots=\sigma_g^{2}
\end{equation}
where $\sigma_i^2$ is the variance about the group median for group $i$.

We apply both tests because Levene’s test provides a general assessment of variance equality, while the Brown-Forsythe variant offers a more robust check against non-normality, allowing us to confirm that any detected heterogeneity is not driven by outliers or deviations from normality.

The results are shown in \Cref{tab:brownforsythe_all_methods}. And we see that the test was significant for both the $x$ and $y$ axes across all optimization methods, with $p < 0.05$ in every case, indicating heterogeneity of variances even under the more robust median-based approach.

\begin{table}[ht]
\centering
\caption{Brown--Forsythe test of equality of variances across optimization methods. 
$p$-values are shown directly; significance threshold is $p > 0.05$.}
\label{tab:brownforsythe_all_methods}
\begin{tabular}{lrrrr}
\toprule
\textbf{Method} & $F_x$ & $p_x$ & $F_y$ & $p_y$ \\
\midrule
\gls{bfgs}   & 6.294  & $8.947\times10^{-13}$ & 4.135  & $9.467\times10^{-8}$ \\
\gls{cobyla} & 2.731  & $2.073\times10^{-4}$  & 6.604  & $1.817\times10^{-13}$ \\
\gls{pm}     & 5.751  & $1.572\times10^{-11}$ & 5.689  & $2.185\times10^{-11}$ \\
\gls{slsqp}  & 10.46  & $3.212\times10^{-21}$ & 8.993  & $2.167\times10^{-18}$ \\
\gls{nm}     & 7.493  & $2.099\times10^{-15}$ & 4.680  & $4.775\times10^{-9}$ \\
\gls{isoma}  & 5.897  & $7.088\times10^{-12}$ & 2.609  & $3.977\times10^{-4}$ \\
\bottomrule
\end{tabular}
\end{table}

From the three tests described above, we can summarize that the assumption of homogeneity was not met, as there is significant heterogeneity across groups. While we can already observe that the \gls{manova} will not be suitable for our case, for the sake of completeness, let us state that the third assumption of independence holds, as all runs were calculated independently of each other, with different random seeds.

% end

\section{PERMDISP}\label{app:permdisp}
The null hypothesis for the one-way \gls{anova} used inside \gls{permdisp} is defined as
\begin{equation}
H_0:\ \mu_1=\mu_2=\cdots=\mu_g
\end{equation}
where $\mu_i$ is the mean of the distance-to-centroid values for group $i$.

Significance is assessed by permutations of the residuals under a reduced model. The null hypothesis is defined as
\begin{equation}
H_0:\ \delta_1=\delta_2=\cdots=\delta_g
\end{equation}
where $\delta_i=\mathbb{E}\!\left[d(\mathbf{x},\bar{\mathbf{x}}_i)\right]$ is the mean distance of observations to the centroid $\bar{\mathbf{x}}_i$ of group $i$.

This means that groups have equal multivariate dispersions, and if it holds, then the differences detected by \gls{permanova} can be attributed solely to differences in group centroids rather than unequal spread of the data. If the null hypothesis is rejected, then the \gls{permanova} can be conducted, but with the interpretation of the results, the fact that the spread of the data can cause the differences has to be taken into account.

\begin{figure*}
\centering

% -------- Row 1 --------
\subfigure[\gls{slsqp}\label{fig:slsqp_permdisp}]{
  \includegraphics[width=0.45\textwidth]{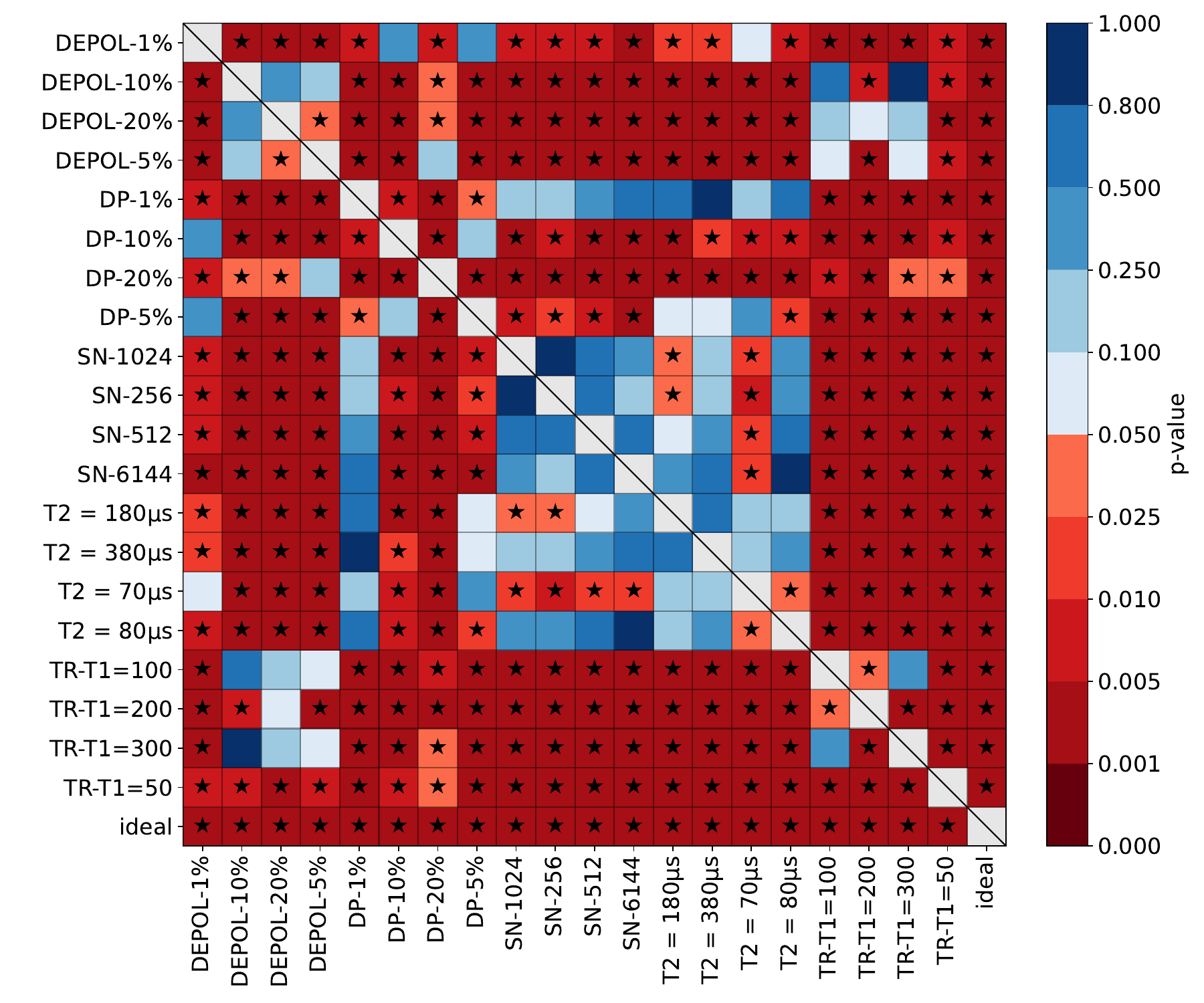}
}\hfill
\subfigure[\gls{isoma}\label{fig:isoma_permdisp}]{
  \includegraphics[width=0.45\textwidth]{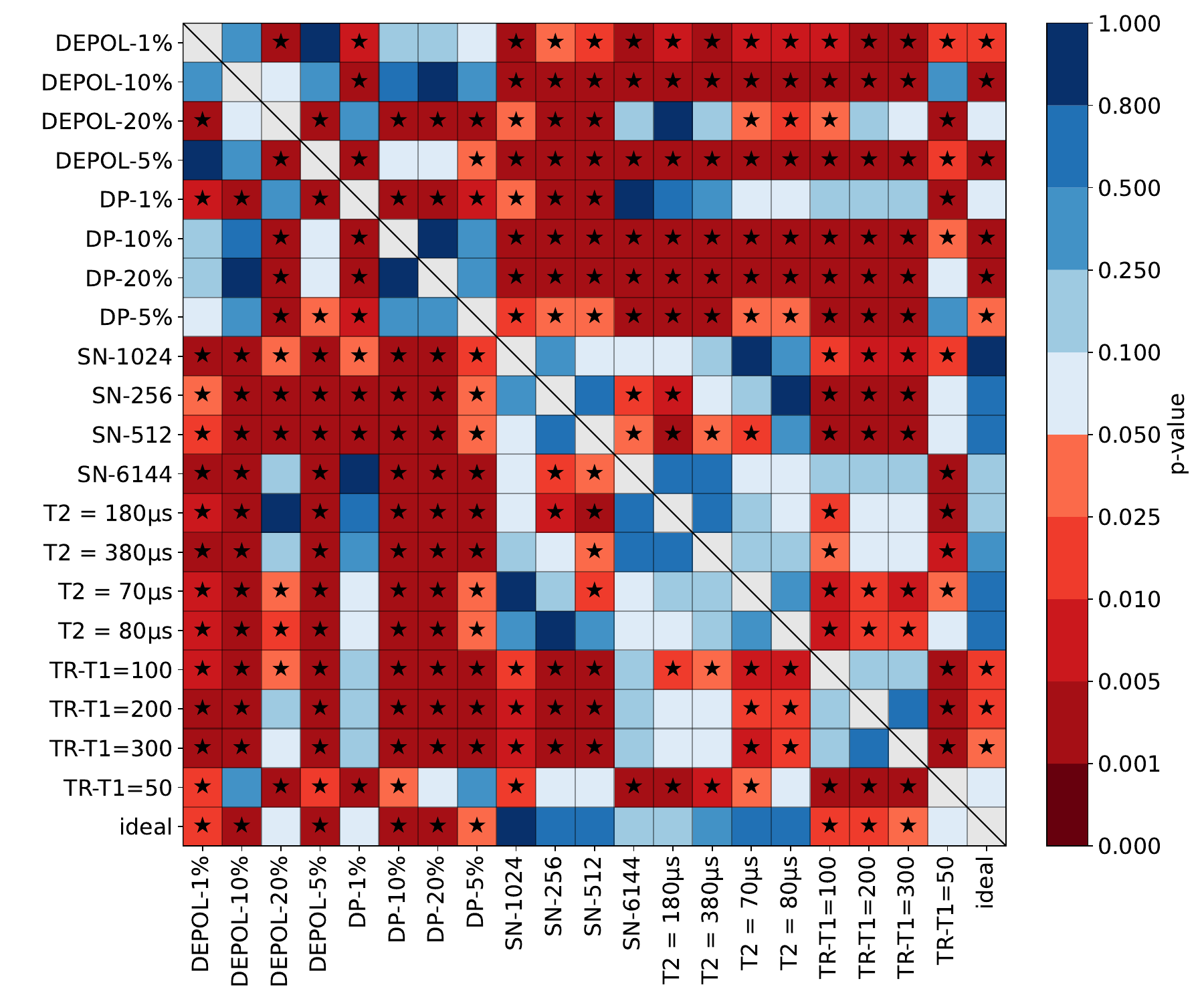}
}

\vspace{0.9em}

% -------- Row 2 --------
\subfigure[\gls{pm}\label{fig:pm_permdisp}]{
  \includegraphics[width=0.45\textwidth]{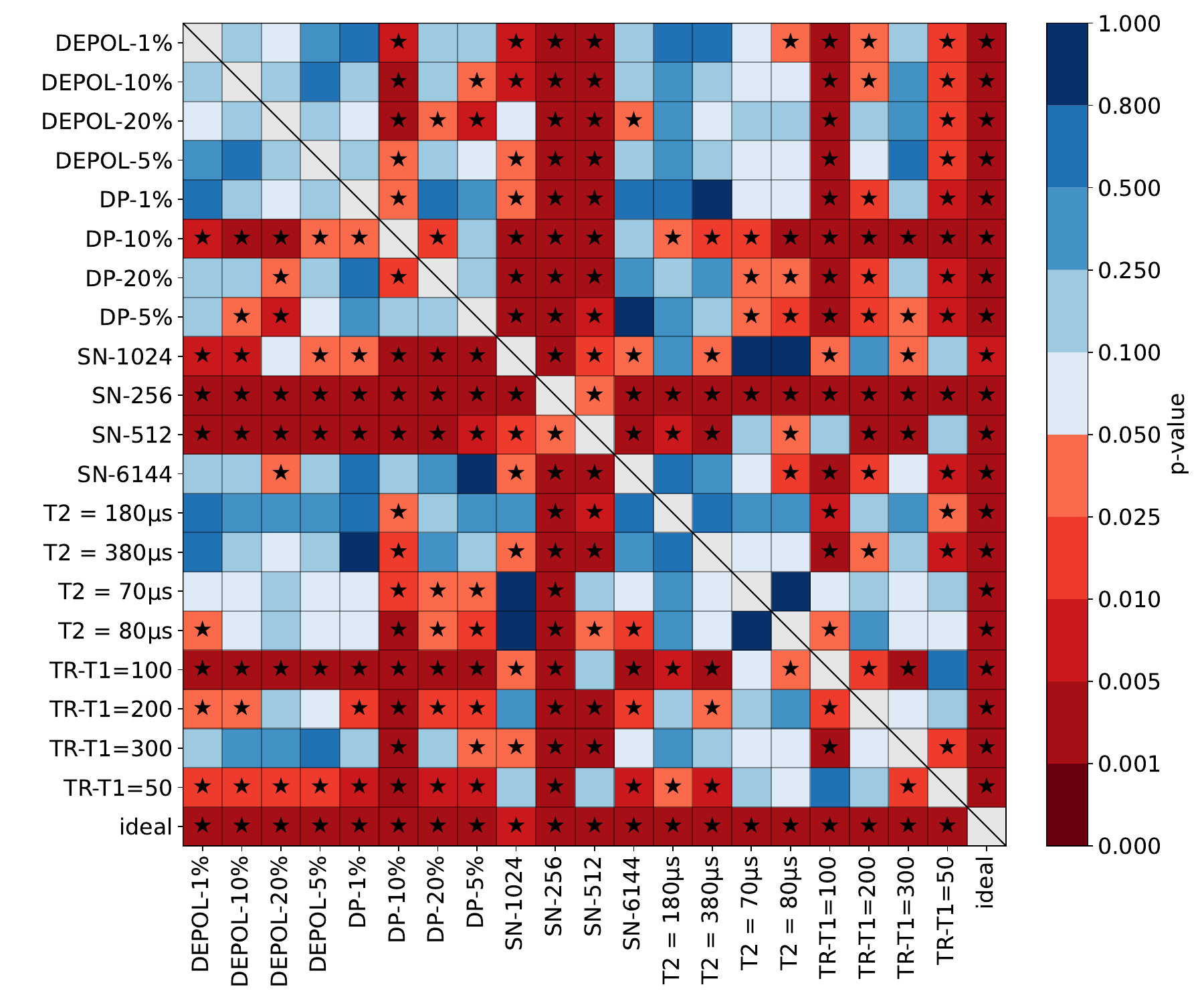}
}\hfill
\subfigure[\gls{nm}\label{fig:nm_permdisp}]{
  \includegraphics[width=0.45\textwidth]{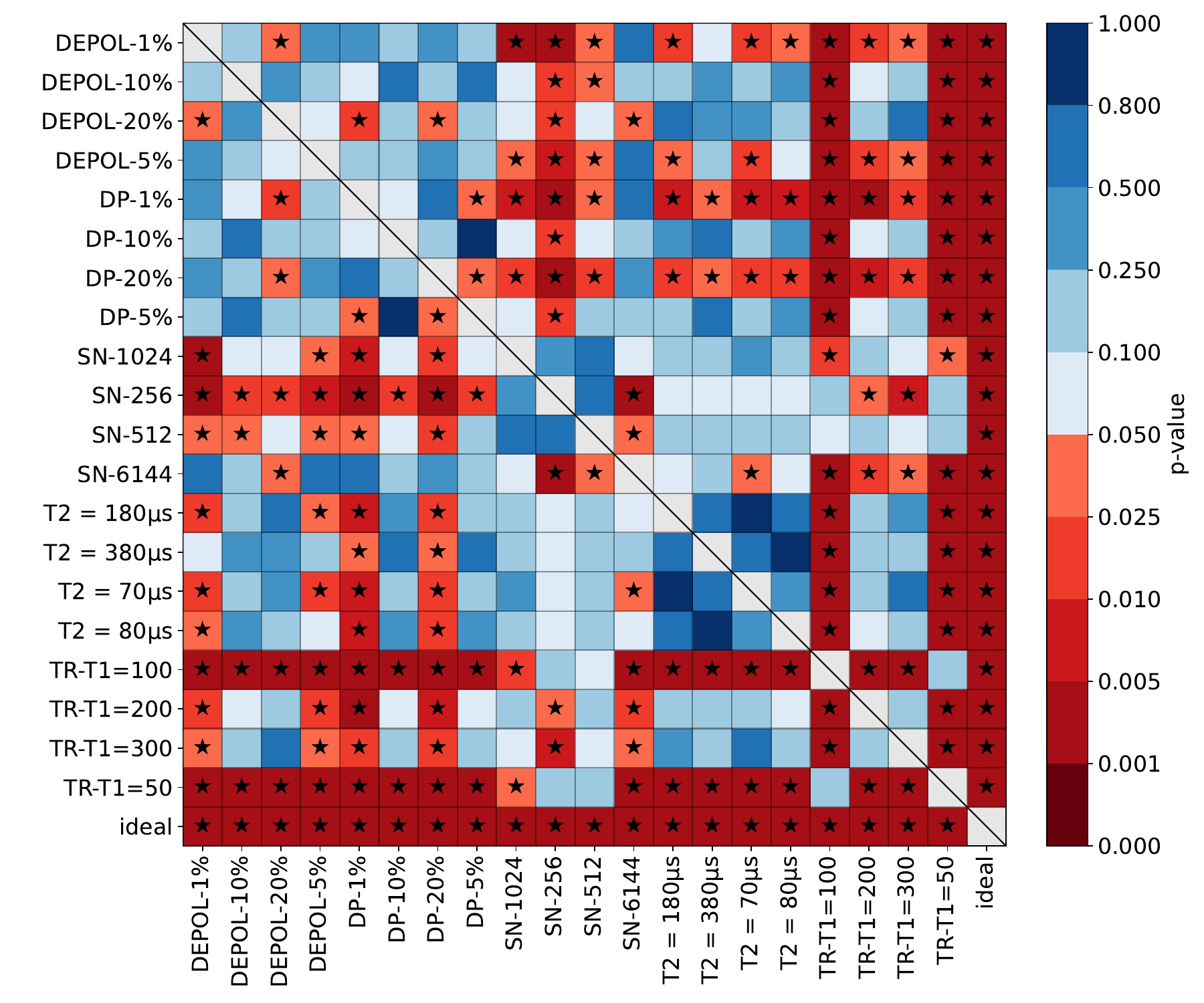}
}

\vspace{0.9em}

% -------- Row 3 --------
\subfigure[\gls{cobyla}\label{fig:cobyla_permdisp}]{
  \includegraphics[width=0.45\textwidth]{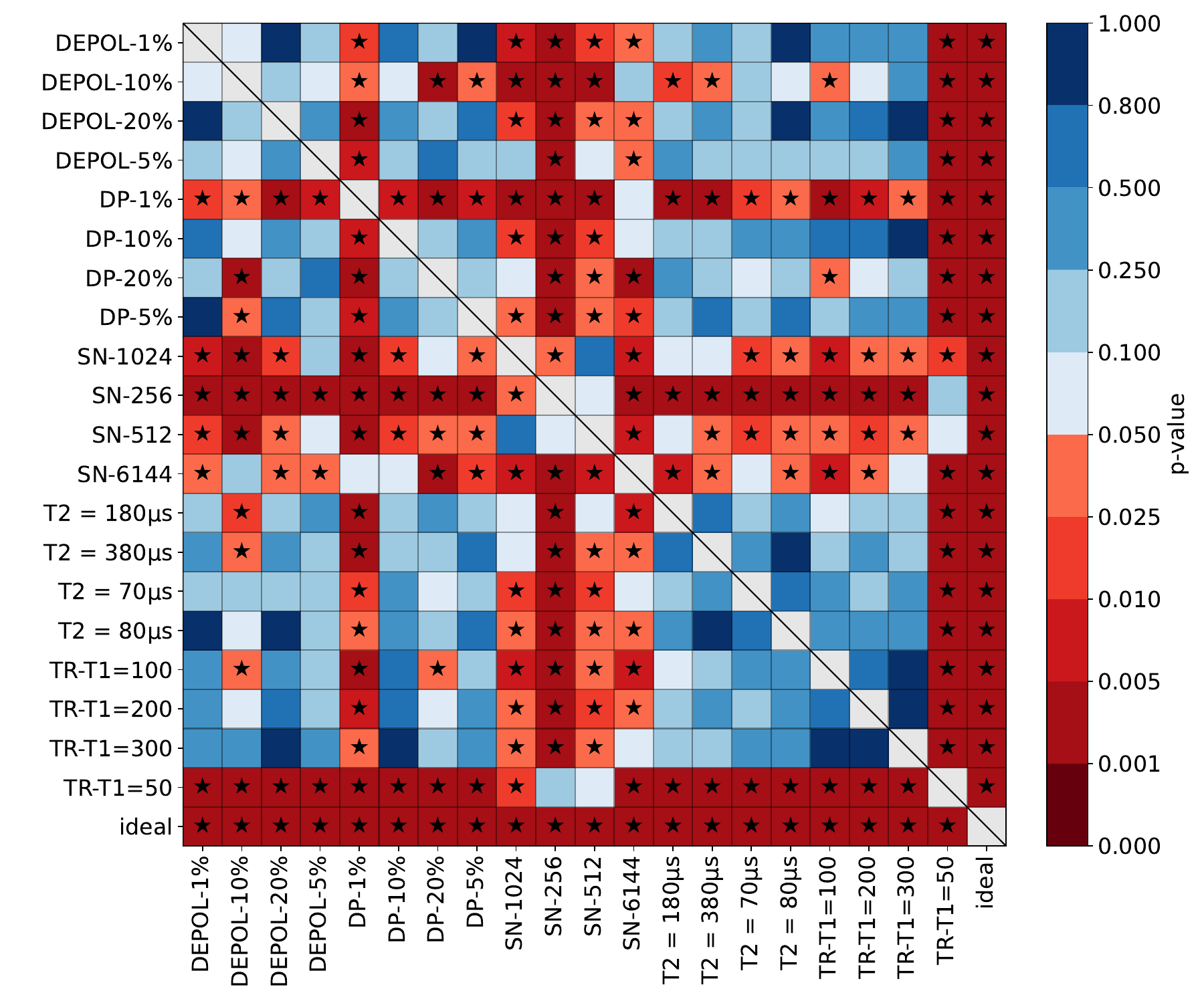}
}\hfill
\subfigure[\gls{bfgs}\label{fig:bfgs_permdisp}]{
  \includegraphics[width=0.45\textwidth]{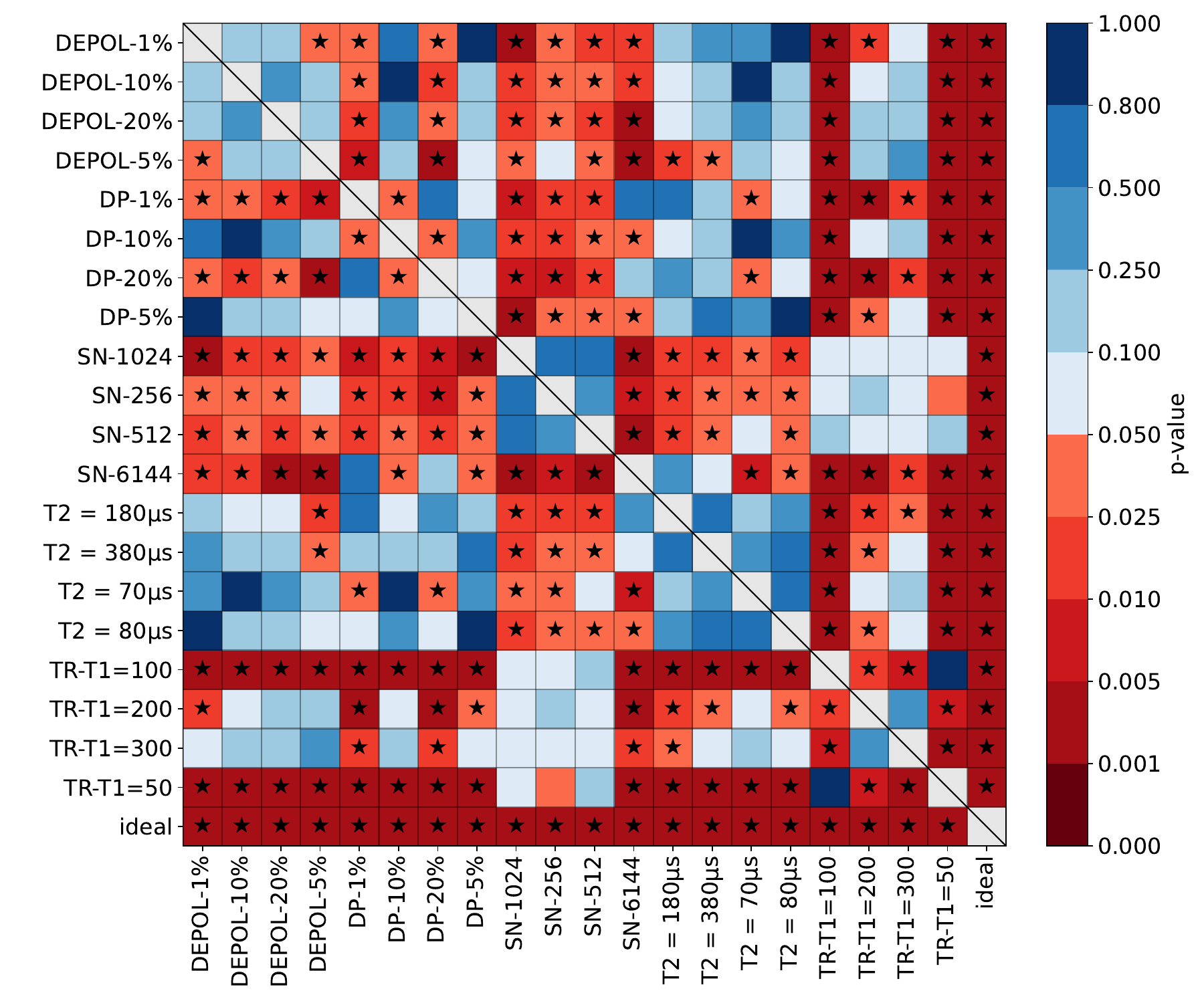}
}

\caption{Pairwise PERMDISP tests for homogeneity of multivariate dispersions across different noise settings for all optimization methods. Cells show $p$-values. Significant values indicate unequal within-family dispersion, which may confound PERMANOVA results. The diagonal is masked.}
\label{fig:all_permdisp}
\end{figure*}

The results of \gls{permdisp} analysis are shown in \Cref{tab:permdisp_all_methods} and in \Cref{fig:all_permdisp}. The columns report the optimization method, the \gls{permdisp} $F$-statistic, the associated degrees of freedom (df), and the permutation-based $p$-value. The test was significant for all optimization methods, with $p < 0.05$ in every case, indicating heterogeneity of multivariate dispersions across groups. This tells us that significant \gls{permanova} results may arise from differences in group dispersions rather than differences in centroid locations.

\begin{table}[ht]
\centering
\caption{PERMDISP test of homogeneity of multivariate dispersions across optimization methods. 
All tests are based on Euclidean distances with 10,000 permutations. 
Significance threshold is $p > 0.05$.}
\label{tab:permdisp_all_methods}

\begin{tabular}{lrrr}
\toprule
\textbf{Method} & $F$ & df & $p$ \\
\midrule
\gls{bfgs}   & 16.66 & (20, 189) & $<0.0001$ \\
\gls{cobyla} & 11.27 & (20, 189) & $<0.0001$ \\
\gls{pm}     & 12.94 & (20, 188) & $<0.0001$ \\
\gls{slsqp}  & 30.20 & (20, 185) & $<0.0001$ \\
\gls{nm}     & 8.11  & (20, 189) & 0.0003    \\
\gls{isoma}  & 16.66 & (20, 189) & $<0.0001$ \\
\bottomrule
\end{tabular}
\end{table}

The figures visualize the distances of individual \gls{sa-oo-vqe} results from their corresponding group centroids in the $(E_{\mathrm{gs}},E_{\mathrm{es}})$ space, thereby illustrating the relative multivariate dispersion under different noise models and optimization methods.
These visualizations are intended to complement the permutation-based PERMDISP statistics reported in the main body and should be interpreted together with those quantitative results.

\section{PERMANOVA}\label{app:permanova}
For the \gls{permanova}, the Euclidean distance defined by 
\begin{equation}
d(x_i, x_j) = \sqrt{\sum_{k=1}^p \left( x_{ik} - x_{jk} \right)^2}
\end{equation}
was chosen, where $x_i$ and $x_j$ are observations in $p$-dimensional space.

The null hypothesis for \gls{permanova} testing equality of centroids is defined as
\begin{equation}
H_0:\ \boldsymbol{\mu}_1=\boldsymbol{\mu}_2=\cdots=\boldsymbol{\mu}_g
\end{equation}
where $\boldsymbol{\mu}_i$ is the multivariate centroid of group $i$ under the Euclidean distance.

The results of the \gls{permanova} are presented in \Cref{tab:permanova_all_methods}. The columns report the optimization method, the \gls{permanova} pseudo-$F$ statistic, the associated degrees of freedom, the permutation-based $p$-value, and the proportion of explained variance ($R^2$). The pseudo-$F$ statistic is defined as the ratio between the mean squares among groups and the mean squares within groups,  
\begin{equation}
F = \frac{SS_\text{between}/(k - 1)}{SS_\text{within}/(N - k)},
\end{equation}
where $SS_\text{between}$ and $SS_\text{within}$ are the sums of squared distances among group centroids and within groups, respectively, $k$ is the number of groups, and $N$ is the total number of observations. These sums of squares are computed directly from the distance matrix rather than from the raw data, which allows the test to be applied to any distance or dissimilarity measure. Larger values of the pseudo-$F$ statistic indicate that the distances between group centroids are greater relative to the dispersion within groups. Statistical significance is determined by randomly permuting group labels to generate the null distribution of $F$, from which the permutation-based $p$-value is obtained.

\begin{figure*}
\centering

% -------- Row 1 --------
\subfigure[\gls{slsqp}\label{fig:slsqp_pBH}]{
  \includegraphics[width=0.45\textwidth]{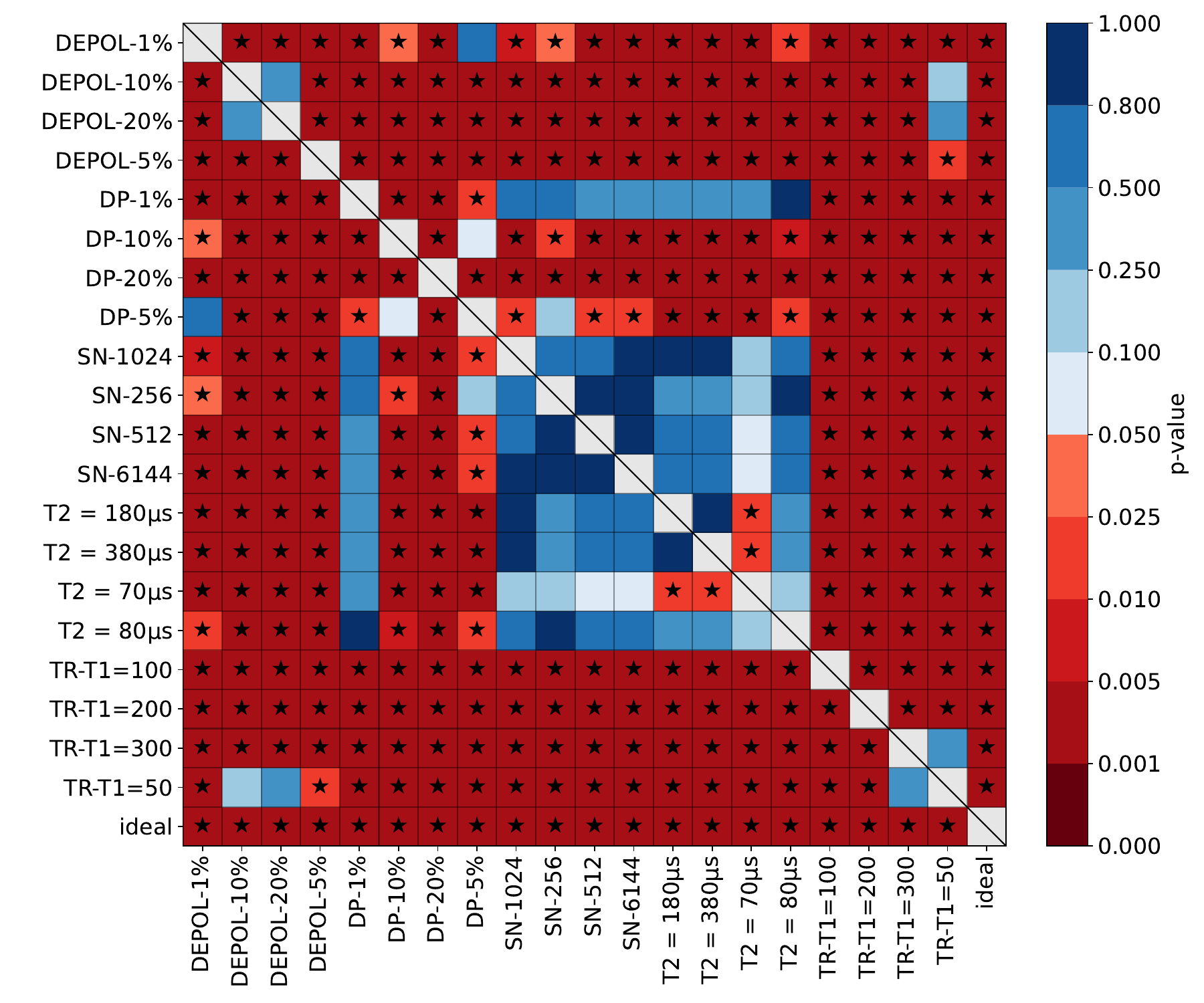}
}\hfill
\subfigure[\gls{isoma}\label{fig:isoma_pBH}]{
  \includegraphics[width=0.45\textwidth]{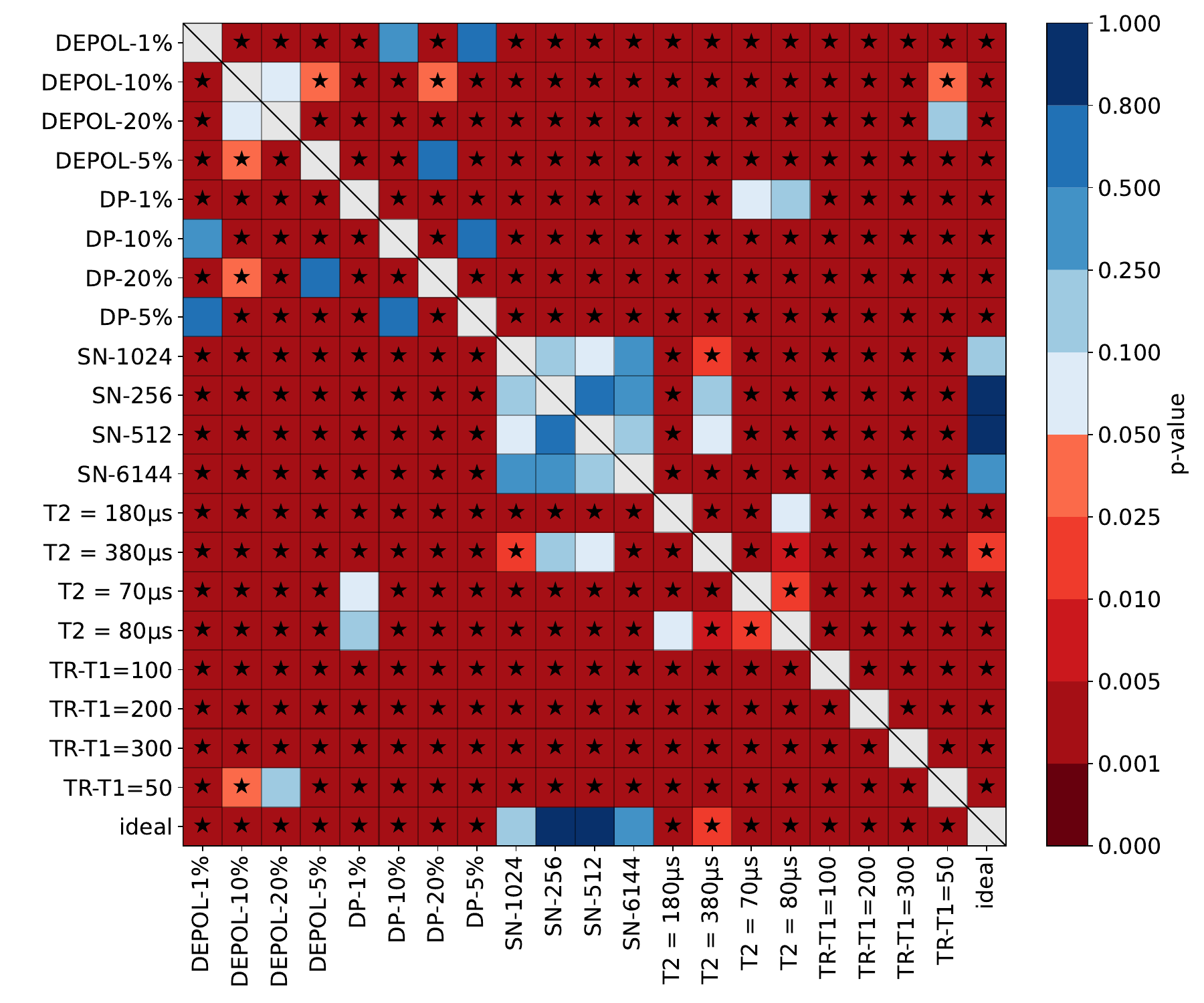}
}

\vspace{0.9em}

% -------- Row 2 --------
\subfigure[\gls{pm}\label{fig:pm_pBH}]{
  \includegraphics[width=0.45\textwidth]{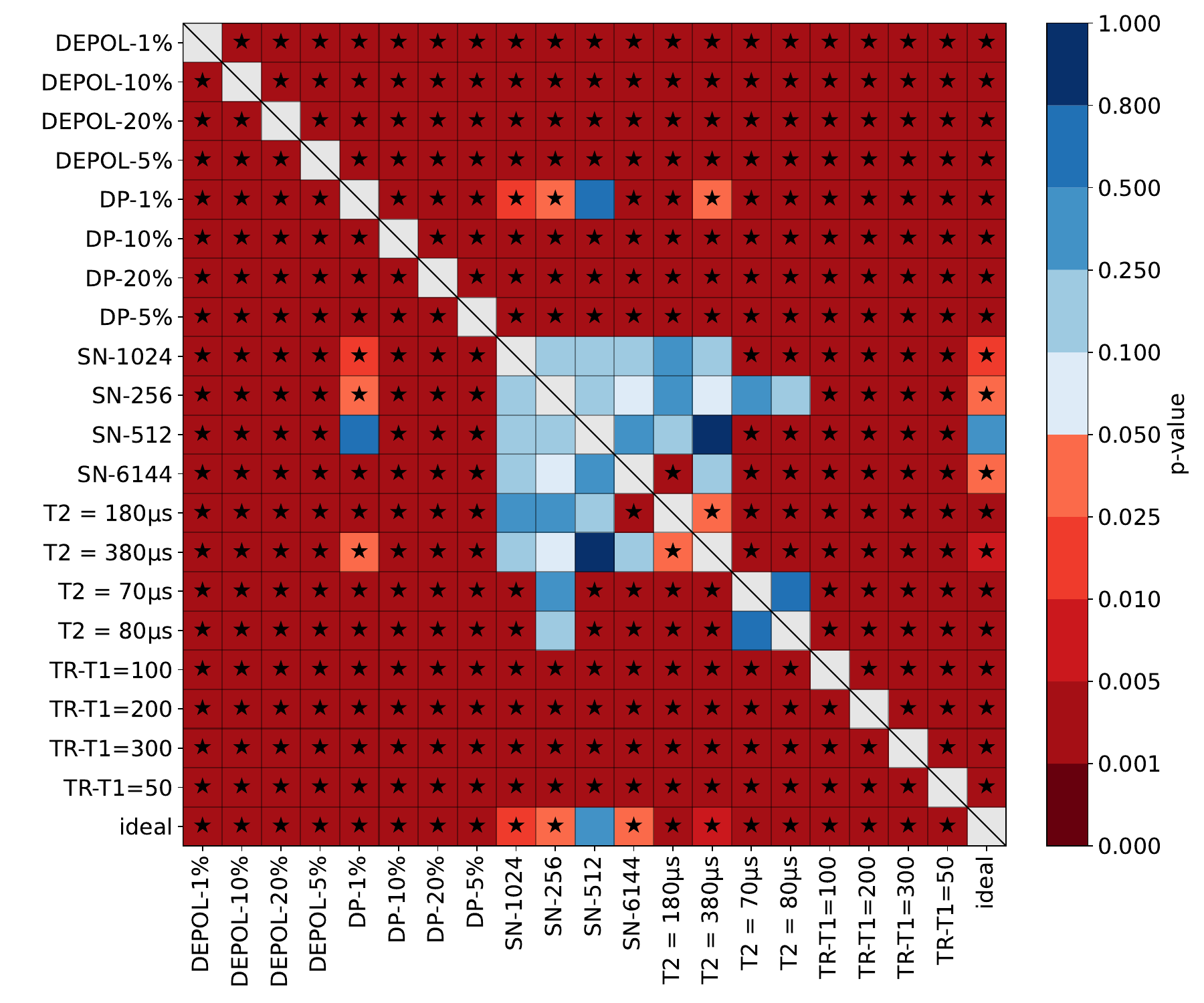}
}\hfill
\subfigure[\gls{nm}\label{fig:nm_pBH}]{
  \includegraphics[width=0.45\textwidth]{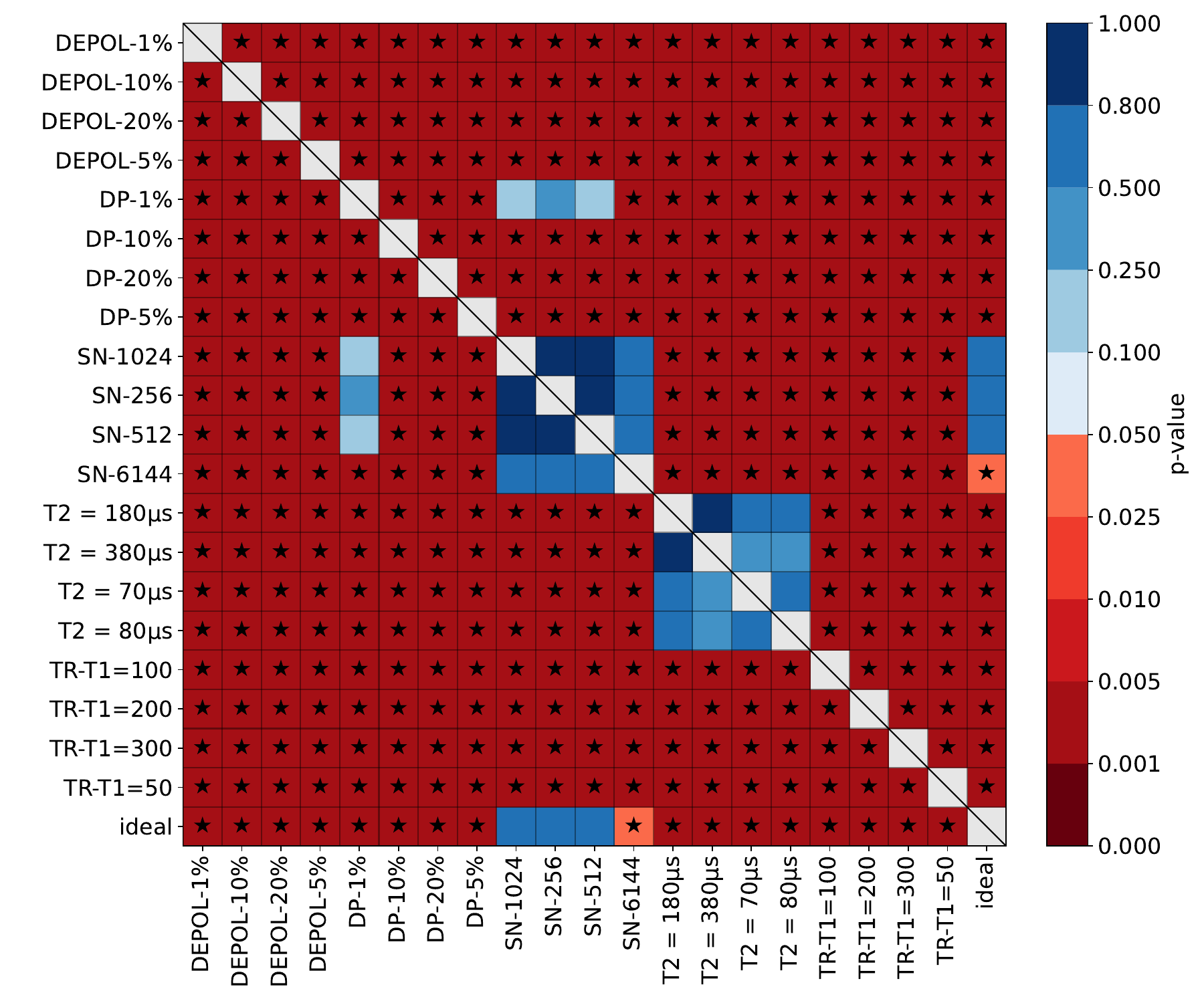}
}

\vspace{0.9em}

% -------- Row 3 --------
\subfigure[\gls{cobyla}\label{fig:cobyla_pBH}]{
  \includegraphics[width=0.45\textwidth]{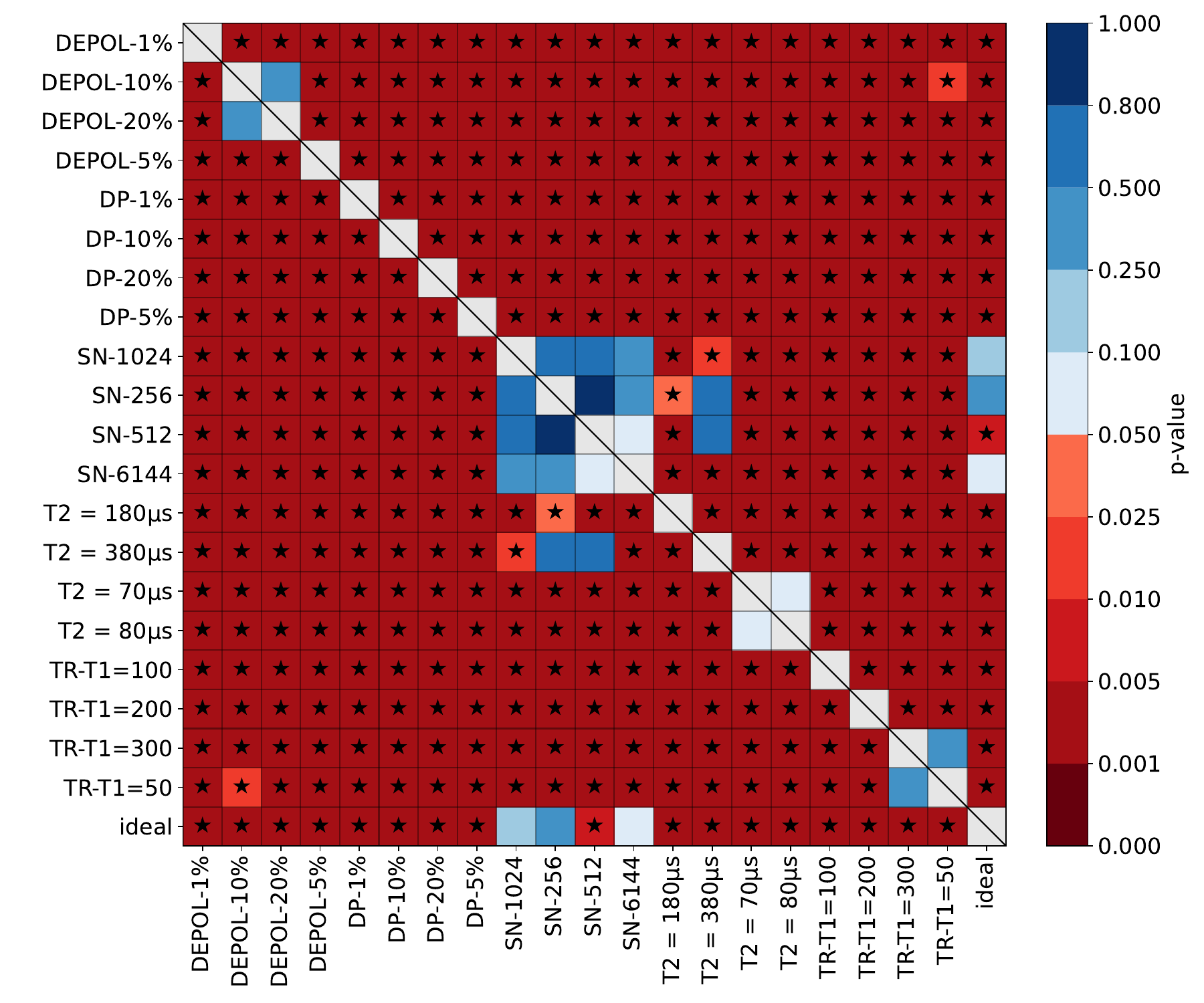}
}\hfill
\subfigure[\gls{bfgs}\label{fig:bfgs_pBH}]{
  \includegraphics[width=0.45\textwidth]{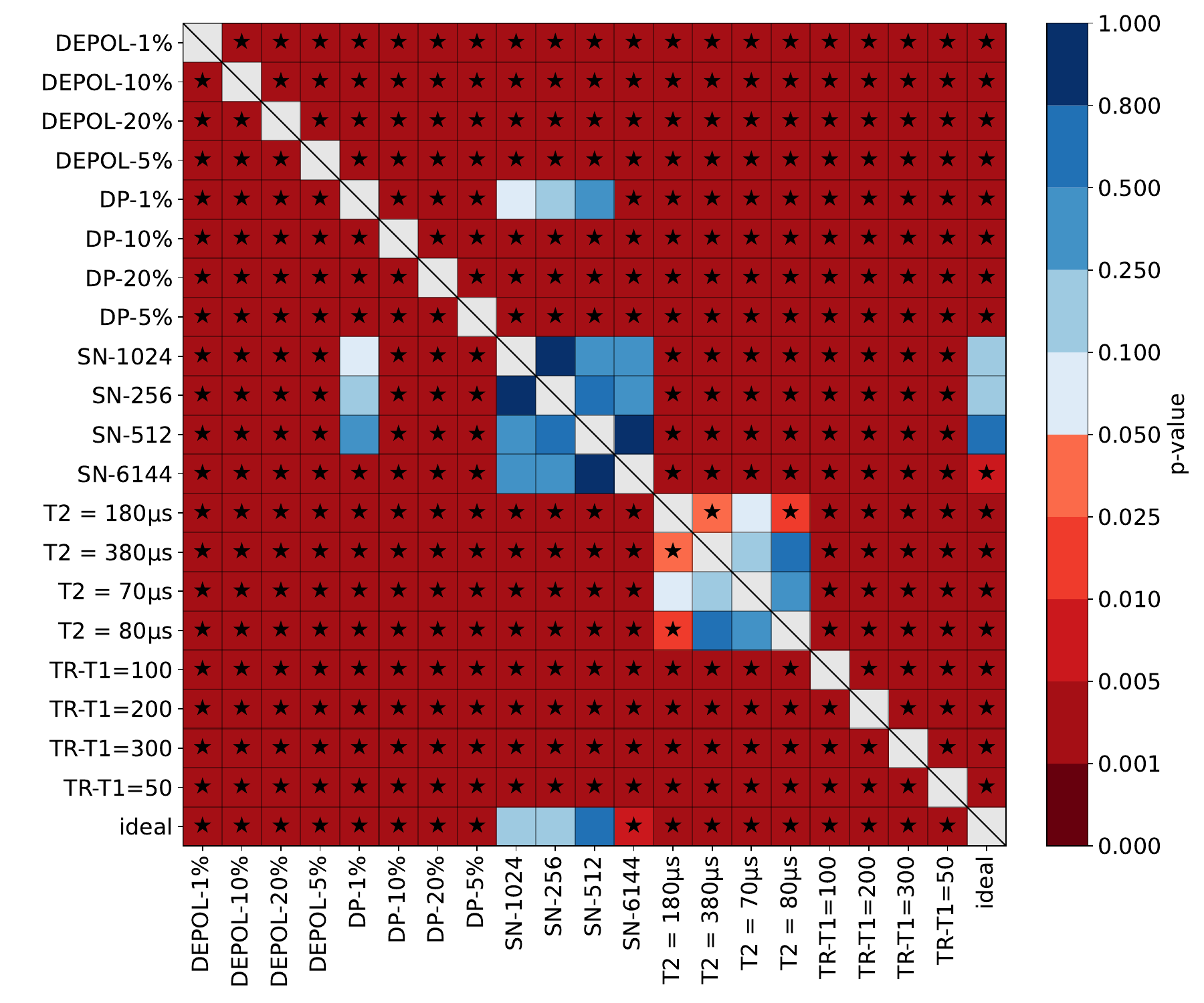}
}

\caption{Pairwise PERMANOVA (Euclidean distance) between noise settings for all optimization methods. Cells show Benjamini–Hochberg FDR-adjusted $p$-values. Warmer colors denote lower $p$-values, indicating stronger evidence of differences in multivariate location. The diagonal is masked.}
\label{fig:all_permanova_pBH}
\end{figure*}

Significant effects were detected by \gls{permanova}, shown in \Cref{fig:all_permanova_pBH}, for all optimization methods as $p$-values were all bellow $0.001$, indicating that the distributions of results obtained with different estimators differ significantly in multivariate space. However, because the \gls{permdisp} test yielded significant results as well as all $p$-values were bellow $0.003$, it is not yet possible to determine whether the observed differences are due to distinct group centroids, location effects, unequal dispersions, or variance heterogeneity.

\begin{table}[ht]
\centering
\caption{\gls{permanova} results across optimization methods. 
$p$-values are obtained by permutation with 10,000 permutations; significance threshold is $p > 0.05$.}
\label{tab:permanova_all_methods}
\begin{tabular}{lrrrr}
\toprule
\textbf{Method} & Pseudo-$F$ & df & $p$ & $R^2$ \\
\midrule
\gls{bfgs}   & 252.4  & (20, 189) & $<0.0001$ & 0.9639 \\
\gls{cobyla} & 6466   & (20, 189) & $<0.0001$ & 0.9985 \\
\gls{pm}     & 3871   & (20, 188) & $<0.0001$ & 0.9976 \\
\gls{slsqp}  & 23.25  & (20, 185) & $<0.0001$ & 0.7154 \\
\gls{nm}     & 7774   & (20, 189) & $<0.0001$ & 0.9988 \\
\gls{isoma}  & 252.4  & (20, 189) & $<0.0001$ & 0.9639 \\
\bottomrule
\end{tabular}
\end{table}

To address these issues, a post hoc tests where performed, in this case in the form of pairwise \gls{permanova} and \gls{permdisp}, so that the results can be compared and summarized. The results were then adapted by Benjamini–Hochberg correction \cite{benjamini1995controlling}, which  was chosen to better capture the real differences and their statistical significance.

The null hypothesis for pairwise \gls{permanova} comparing different settings is defined as
\begin{equation}
H_0:\ \boldsymbol{\mu}_a=\boldsymbol{\mu}_b
\end{equation}
where $\boldsymbol{\mu}_a$ and $\boldsymbol{\mu}_b$ are the multivariate centroids of families $a$ and $b$.

Furthermore, the null hypothesis for pairwise \gls{permdisp} comparing two different settings is defined as
\begin{equation}
H_0:\ \delta_a=\delta_b
\end{equation}
where $\delta_a$ and $\delta_b$ are the mean distances to centroids for families $a$ and $b$.

%\bruno{We go from Figure 6 to Figure 20? Like before, the label should be consistent with the appearance in the main text}
The results for the Post Hoc test for \gls{bfgs} are visualized in \Cref{fig:bfgs_permdisp}, where the pairwise \gls{permdisp} is visualized, \Cref{fig:bfgs_pBH}, where the pairwise \gls{permanova} results are shown in a heatmap with Benjamini-Hochberg corrections applied. From these figures, we can conclude that for the majority of the pairs, there is a significant location displacement, with the exception of the sampling noise settings pairs and the combination of sampling noise settings with ideal setting and dephasing channel with a level of $1\%$. From the second figure, we can see that a lot of pairs have unequal dispersion, so in these cases, the dispersion of the clusters plays a significant role.

The same results, but for \gls{cobyla}, are visualized in \Cref{fig:cobyla_permdisp} for \gls{permdisp} and in \Cref{fig:cobyla_pBH} for the case of \gls{permanova} with Benjamini-Hochberg corrections. The exceptions are similar to the \gls{bfgs} case, particularly among the sampling noise settings and their combinations with the ideal state. Moreover, many pairs also exhibit unequal dispersion, suggesting that group spread differences play an important role in the interpretation of the observed effects under the \gls{cobyla} optimization.

For \gls{isoma} the results for pairwise \gls{permdisp} are in \Cref{fig:isoma_permdisp}, and for pairwise \gls{permanova} in \Cref{fig:isoma_pBH}. These figures demonstrate that most pairs display significant location differences, but there are slightly more pairs that are similar than in the previous two cases.

The results for the post hoc test for \gls{nm} are visualized in \Cref{fig:nm_permdisp}, where the pairwise \gls{permdisp} outcomes are shown, and in \Cref{fig:nm_pBH}, where the pairwise \gls{permanova} results with Benjamini-Hochberg corrections are presented. Similar to the other optimizers, the majority of pairs display significant location differences, except for several comparisons within the sampling noise settings, but a new cluster with similarities emerges in the $T_2$ region.

The results for the post hoc test for \gls{pm} are visualized in \Cref{fig:pm_permdisp}, where the pairwise \gls{permdisp} results are shown, and in \Cref{fig:pm_pBH}, where the pairwise \gls{permanova} outcomes with Benjamini-Hochberg corrections are presented. The cluster of similarities is extended to the combination of sampling noise with a few categories of $T_2$, which is the mean time to dephasing.

The results for the post hoc test for \gls{slsqp} are visualized in \Cref{fig:slsqp_permdisp}, where the pairwise \gls{permdisp} outcomes are shown, and in \Cref{fig:slsqp_pBH}, where the pairwise \gls{permanova} results with Benjamini-Hochberg corrections are presented. Here, because of the overall bad convergence, most pairs are statistically similar.

% ahsgfhadhsjfgdasv
% This appendix gathers the graphical outputs of the PERMANOVA analysis referenced in the main text. The figures display pairwise comparisons between individual noise configurations for each optimization method, evaluated in the two-dimensional $(E_{\mathrm{gs}}, E_{\mathrm{es}})$ space. The visualizations are presented as significance heatmaps derived from permutation testing and summarize the extent to which the distributions of results obtained under different noise models can be considered statistically distinguishable.

% By representing adjusted $p$-values across all tested settings, these plots provide an intuitive overview of how strongly the noise channels alter the location of the solution clusters in multivariate space. Regions of lower $p$-values indicate stronger evidence that two configurations lead to systematically different optimization outcomes, whereas higher values suggest statistically comparable behavior within the resolution of the permutation test.

% These visualizations are intended to complement the PERMANOVA statistics reported in the main body rather than to introduce additional analysis. They serve as an illustrative aid for interpreting the hypothesis-testing results and should be considered together with the quantitative discussion presented in the corresponding section.

\section{Mardia's multivariate normality}\label{ap:mardia}

The Mardia's test looks at the multivariate skewness of the data

\begin{equation}
b_{1,p} = \frac{1}{n^2} \sum_{i=1}^n \sum_{j=1}^n 
\left[ (\mathbf{x}_i - \bar{\mathbf{x}})^\top \mathbf{S}^{-1} (\mathbf{x}_j - \bar{\mathbf{x}}) \right]^3,
\end{equation}

and their kurtosis
\begin{equation}
b_{2,p} = \frac{1}{n} \sum_{i=1}^n 
\left[ (\mathbf{x}_i - \bar{\mathbf{x}})^\top \mathbf{S}^{-1} (\mathbf{x}_i - \bar{\mathbf{x}}) \right]^2.
\end{equation}

Here, $\mathbf{x}_i$ denotes the $i$-th observation vector, $\bar{\mathbf{x}}$ the sample mean, and $\mathbf{S}$ the sample covariance matrix. Here, $n$ is the number of observations (data points) within a given group, and each observation vector $\mathbf{x}_i = (E_{0,i}, E_{1,i})^\mathsf{T}$ contains the ground- and excited-state energies obtained in a single run. The sample mean vector $\bar{\mathbf{x}}$ represents the mean of these energies across all observations, and the sample covariance matrix $\mathbf{S}$ is defined as
\begin{equation}
\mathbf{S} = \frac{1}{n - 1} \sum_{i=1}^{n} (\mathbf{x}_i - \bar{\mathbf{x}})(\mathbf{x}_i - \bar{\mathbf{x}})^\mathsf{T},
\end{equation}
which captures both the variances of $E_0$ and $E_1$ and their covariance.

The skewness statistic $b_{1,p}$ is approximately distributed as
\begin{equation}
n \cdot b_{1,p} \;\sim\; \chi^2_{\tfrac{p(p+1)(p+2)}{6}},
\end{equation}
here, $\chi^2$ denotes the chi-squared distribution, which serves as the reference distribution for the test under the null hypothesis of multivariate normality.

And the null hypothesis for Mardia's multivariate normality based on skewness is defined as
\begin{equation}
H_0:\ b_{1,p}=0
\end{equation}
where $b_{1,p}$ is Mardia's multivariate skewness and $p=2$ is the dimensionality.

While the kurtosis statistic is standardized as
\begin{equation}
z_{\text{kurt}} = \frac{b_{2,p} - p(p+2)}{\sqrt{\tfrac{8p(p+2)}{n}}}
\;\;\sim\; \mathcal{N}(0,1).
\end{equation}

The null hypothesis for Mardia's multivariate normality based on kurtosis is defined as
\begin{equation}
H_0:\ b_{2,p}=p(p+2)
\end{equation}
where $b_{2,p}$ is Mardia's multivariate kurtosis and $p=2$ is the dimensionality.

Non-significant values for both the skewness and kurtosis components ($p > 0.05$) indicate that the data do not significantly deviate from multivariate normality.

The results of this tests are shown in \Cref{tab:mardia_test_bfgs,tab:mardia_test_powell,tab:mardia_test_slsqp,tab:mardia_test_cobyla,tab:mardia_test_neldermead,tab:mardia_test_isoma} where the columns indicate the experimental family, Mardia's skewness statistic ($b_{1,p}$), its chi-square test value ($\chi^2$) with corresponding degrees of freedom (df) and $p$-value ($p_{\text{skew}}$), followed by Mardia's kurtosis statistic ($b_{2,p}$), its standardized $z$-score ($z_{\text{kurt}}$), and the associated $p$-value ($p_{\text{kurt}}$). A higher $\chi^2$ value relative to its degrees of freedom indicates a stronger deviation from multivariate symmetry, meaning the data are less consistent with multivariate normality. These results tell us that almost all of the families do not violate the multivariate normality with the following exceptions: for the \gls{slsqp}, the \textit{ideal} family fails both in skewness and kurtosis with $p_{skew} = 0.002$ and $p_{kurt}=0.041$. The second violation is in the \gls{nm} again for the \textit{ideal} family, where the skewness is violated with $p_{skew} = 0.042$. While these two are the only clear violations of multivariate normality, several other examples came close, especially the \textit{ideal} family in other optimization methods. This shows that the \textit{ideal} family is sensitive and prone to deviations of normality.

\begin{table}[ht]
\centering
\caption{Mardia's multivariate normality test results across all families for \gls{bfgs}.}
\label{tab:mardia_test_bfgs}
\begin{tabular}{lrrrrrrr}
\toprule
family & $b_{1,p}$ & $\chi^2$ & df & $p_{\text{skew}}$ & $b_{2,p}$ & $z_{\text{kurt}}$ & $p_{\text{kurt}}$ \\
\midrule
 DEPOL-1\% & 0.704 & 1.173 & 4 & 0.882 & 4.527 & -1.373 & 0.170 \\
DEPOL-10\% & 0.633 & 1.055 & 4 & 0.901 & 4.840 & -1.249 & 0.212 \\
DEPOL-20\% & 2.420 & 4.034 & 4 & 0.401 & 6.852 & -0.454 & 0.650 \\
 DEPOL-5\% & 0.418 & 0.697 & 4 & 0.952 & 4.398 & -1.424 & 0.155 \\
    DP-1\% & 1.373 & 2.288 & 4 & 0.683 & 5.521 & -0.980 & 0.327 \\
   DP-10\% & 0.200 & 0.333 & 4 & 0.988 & 4.836 & -1.251 & 0.211 \\
   DP-20\% & 1.327 & 2.211 & 4 & 0.697 & 5.524 & -0.979 & 0.328 \\
    DP-5\% & 0.260 & 0.433 & 4 & 0.980 & 4.605 & -1.342 & 0.180 \\
  SN-1024 & 0.947 & 1.579 & 4 & 0.813 & 5.298 & -1.068 & 0.286 \\
   SN-256 & 1.603 & 2.672 & 4 & 0.614 & 5.350 & -1.047 & 0.295 \\
   SN-512 & 0.251 & 0.419 & 4 & 0.981 & 4.495 & -1.385 & 0.166 \\
  SN-6144 & 1.274 & 2.123 & 4 & 0.713 & 4.577 & -1.353 & 0.176 \\
 T2=180$\mu$s & 1.579 & 2.631 & 4 & 0.621 & 6.069 & -0.763 & 0.445 \\
 T2=380$\mu$s & 1.025 & 1.708 & 4 & 0.789 & 6.541 & -0.577 & 0.564 \\
  T2=70$\mu$s & 0.838 & 1.397 & 4 & 0.845 & 5.218 & -1.100 & 0.272 \\
  T2=80$\mu$s & 0.519 & 0.865 & 4 & 0.930 & 5.744 & -0.892 & 0.373 \\
TR-T1=100 & 0.763 & 1.271 & 4 & 0.866 & 6.084 & -0.758 & 0.449 \\
TR-T1=200 & 0.555 & 0.926 & 4 & 0.921 & 4.488 & -1.388 & 0.165 \\
TR-T1=300 & 0.209 & 0.348 & 4 & 0.987 & 5.012 & -1.181 & 0.238 \\
 TR-T1=50 & 1.770 & 2.951 & 4 & 0.566 & 4.928 & -1.214 & 0.225 \\
    ideal & 5.184 & 8.640 & 4 & 0.071 & 6.570 & -0.565 & 0.572 \\
\bottomrule
\end{tabular}
\end{table}

\begin{table}[ht]
\centering
\caption{Mardia's multivariate normality test results across all families for \gls{pm}.}
\label{tab:mardia_test_powell}
\begin{tabular}{lccccccc}
\toprule
family & $b_{1,p}$ & $\chi^2$ & df & $p_{\text{skew}}$ & $b_{2,p}$ & $z_{\text{kurt}}$ & $p_{\text{kurt}}$ \\
\midrule
DEPOL-1\%  & 0.4354 & 0.726  & 4 & 0.948 & 4.823 & -1.256 & 0.209 \\
DEPOL-10\% & 0.5162 & 0.860  & 4 & 0.930 & 4.406 & -1.421 & 0.155 \\
DEPOL-20\% & 1.574  & 2.624  & 4 & 0.623 & 5.262 & -1.082 & 0.279 \\
DEPOL-5\%  & 2.943  & 4.904  & 4 & 0.297 & 6.772 & -0.486 & 0.627 \\
DP-1\%     & 0.6806 & 1.134  & 4 & 0.889 & 5.143 & -1.129 & 0.259 \\
DP-10\%    & 1.506  & 2.510  & 4 & 0.643 & 4.837 & -1.250 & 0.211 \\
DP-20\%    & 1.838  & 3.063  & 4 & 0.547 & 6.659 & -0.530 & 0.596 \\
DP-5\%     & 1.485  & 2.476  & 4 & 0.649 & 5.454 & -1.006 & 0.314 \\
SN-1024    & 5.070  & 8.449  & 4 & 0.076 & 8.171 &  0.067 & 0.946 \\
SN-256     & 0.912  & 1.520  & 4 & 0.823 & 4.795 & -1.267 & 0.205 \\
SN-512     & 2.754  & 4.589  & 4 & 0.332 & 6.556 & -0.620 & 0.535 \\
SN-6144    & 2.384  & 3.973  & 4 & 0.410 & 6.265 & -0.686 & 0.493 \\
T2=180$\mu$s & 2.474 & 3.711 & 4 & 0.447 & 6.178 & -0.683 & 0.494 \\
T2=380$\mu$s & 0.526 & 0.877 & 4 & 0.928 & 4.766 & -1.278 & 0.201 \\
T2=70$\mu$s  & 3.003 & 5.005 & 4 & 0.287 & 7.047 & -0.377 & 0.706 \\
T2=80$\mu$s  & 1.378 & 2.297 & 4 & 0.681 & 6.036 & -0.776 & 0.438 \\
TR-T1=100  & 0.620  & 1.033  & 4 & 0.905 & 5.534 & -0.975 & 0.330 \\
TR-T1=200  & 0.630  & 1.050  & 4 & 0.902 & 5.335 & -1.054 & 0.292 \\
TR-T1=300  & 1.162  & 1.937  & 4 & 0.747 & 5.327 & -1.057 & 0.291 \\
TR-T1=50   & 1.277  & 2.129  & 4 & 0.712 & 6.659 & -0.530 & 0.596 \\
ideal      & 5.184  & 8.640  & 4 & 0.071 & 6.570 & -0.565 & 0.572 \\
\bottomrule
\end{tabular}
\end{table}

\begin{table}[ht]
\centering
\begin{tabular}{lrrrrrrr}
\hline
family & $b_{1,p}$ & $\chi^2$ & df & $p_{\text{skew}}$ & $b_{2,p}$ & $z_{\text{kurt}}$ & $p_{\text{kurt}}$ \\
\hline
DEPOL-1\%   & 0.898 & 1.497 & 4 & 0.827 & 4.070 & -1.553 & 0.120 \\
DEPOL-10\%  & 0.579 & 0.965 & 4 & 0.915 & 4.444 & -1.406 & 0.160 \\
DEPOL-20\%  & 0.586 & 0.977 & 4 & 0.913 & 5.307 & -1.064 & 0.287 \\
DEPOL-5\%   & 0.772 & 1.287 & 4 & 0.864 & 5.284 & -1.074 & 0.283 \\
DP-1\%      & 1.461 & 2.434 & 4 & 0.656 & 4.833 & -1.252 & 0.211 \\
DP-10\%     & 0.398 & 0.664 & 4 & 0.956 & 5.045 & -1.168 & 0.243 \\
DP-20\%     & 1.992 & 3.320 & 4 & 0.506 & 6.500 & -0.593 & 0.553 \\
DP-5\%      & 1.931 & 3.218 & 4 & 0.522 & 6.844 & -0.457 & 0.648 \\
SN-1024     & 0.303 & 0.505 & 4 & 0.973 & 4.070 & -1.554 & 0.120 \\
SN-256      & 1.515 & 2.526 & 4 & 0.640 & 4.368 & -1.436 & 0.151 \\
SN-512      & 0.529 & 0.881 & 4 & 0.927 & 5.300 & -1.065 & 0.287 \\
SN-6144     & 1.060 & 1.767 & 4 & 0.778 & 4.004 & -1.579 & 0.114 \\
T2 = 180$\mu$s  & 0.162 & 0.270 & 4 & 0.992 & 4.394 & -1.425 & 0.154 \\
T2 = 380$\mu$s  & 0.116 & 0.135 & 4 & 0.998 & 3.765 & -1.401 & 0.161 \\
T2 = 70$\mu$s   & 0.539 & 0.899 & 4 & 0.925 & 4.110 & -1.538 & 0.124 \\
T2 = 80$\mu$s   & 0.086 & 0.144 & 4 & 0.998 & 3.852 & -1.639 & 0.101 \\
TR-T1=100   & 0.230 & 0.383 & 4 & 0.984 & 5.112 & -1.142 & 0.254 \\
TR-T1=200   & 0.228 & 0.343 & 4 & 0.987 & 4.615 & -1.269 & 0.204 \\
TR-T1=300   & 0.128 & 0.214 & 4 & 0.995 & 3.918 & -1.613 & 0.107 \\
TR-T1=50    & 0.734 & 1.223 & 4 & 0.874 & 4.855 & -1.243 & 0.214 \\
ideal       & 10.390 & 17.310 & 4 & 0.002 & 13.160 & 2.041 & 0.041 \\
\hline
\end{tabular}
\caption{Mardia's multivariate normality test results across all families for \gls{slsqp}.}
\label{tab:mardia_test_slsqp}
\end{table}

\begin{table}[ht]
\centering
\begin{tabular}{lrrrrrrr}
\hline
family & $b_{1,p}$ & $\chi^2$ & df & $p_{\text{skew}}$ & $b_{2,p}$ & $z_{\text{kurt}}$ & $p_{\text{kurt}}$ \\
\hline
DEPOL-1\%   & 0.704 & 1.173 & 4 & 0.883 & 4.527 & -1.373 & 0.170 \\
DEPOL-10\%  & 0.633 & 1.055 & 4 & 0.901 & 4.840 & -1.249 & 0.212 \\
DEPOL-20\%  & 2.420 & 4.034 & 4 & 0.402 & 6.852 & -0.454 & 0.650 \\
DEPOL-5\%   & 0.418 & 0.697 & 4 & 0.952 & 4.398 & -1.424 & 0.155 \\
DP-1\%      & 1.373 & 2.288 & 4 & 0.683 & 5.521 & -0.980 & 0.327 \\
DP-10\%     & 0.200 & 0.333 & 4 & 0.988 & 4.836 & -1.251 & 0.211 \\
DP-20\%     & 1.327 & 2.211 & 4 & 0.697 & 5.524 & -0.979 & 0.328 \\
DP-5\%      & 0.260 & 0.433 & 4 & 0.980 & 4.605 & -1.342 & 0.180 \\
SN-1024     & 0.947 & 1.579 & 4 & 0.813 & 5.298 & -1.068 & 0.286 \\
SN-256      & 1.603 & 2.672 & 4 & 0.614 & 5.350 & -1.047 & 0.295 \\
SN-512      & 0.251 & 0.419 & 4 & 0.981 & 4.495 & -1.385 & 0.166 \\
SN-6144     & 1.274 & 2.123 & 4 & 0.713 & 4.577 & -1.353 & 0.176 \\
T2=180$\mu$s & 1.579 & 2.631 & 4 & 0.621 & 6.069 & -0.763 & 0.445 \\
T2=380$\mu$s & 1.025 & 1.708 & 4 & 0.789 & 6.541 & -0.577 & 0.564 \\
T2=70$\mu$s  & 0.838 & 1.397 & 4 & 0.845 & 5.218 & -1.100 & 0.272 \\
T2=80$\mu$s  & 0.519 & 0.865 & 4 & 0.930 & 5.744 & -0.892 & 0.373 \\
TR-T1=100   & 0.763 & 1.271 & 4 & 0.866 & 6.084 & -0.758 & 0.449 \\
TR-T1=200   & 0.555 & 0.926 & 4 & 0.921 & 4.488 & -1.388 & 0.165 \\
TR-T1=300   & 0.209 & 0.348 & 4 & 0.987 & 5.012 & -1.181 & 0.238 \\
TR-T1=50    & 1.770 & 2.951 & 4 & 0.566 & 4.928 & -1.214 & 0.225 \\
ideal       & 5.184 & 8.640 & 4 & 0.071 & 6.570 & -0.565 & 0.572 \\
\hline
\end{tabular}
\caption{Mardia's multivariate normality test results across all families for \gls{cobyla}.}
\label{tab:mardia_test_cobyla}
\end{table}

\begin{table}[H]
\centering
\caption{Mardia's multivariate normality test results across all families for \gls{nm}.}
\label{tab:mardia_test_neldermead}
\begin{tabular}{lrrrrrrr}
\toprule
family & $b_{1,p}$ & $\chi^2$ & df & $p_{\text{skew}}$ & $b_{2,p}$ & $z_{\text{kurt}}$ & $p_{\text{kurt}}$ \\
\midrule
DEPOL-1\%  & 0.8959 & 1.493 & 4 & 0.8279 & 5.433 & -1.015 & 0.3103 \\
DEPOL-10\% & 3.5900 & 5.984 & 4 & 0.2004 & 8.629 &  0.249 & 0.8037 \\
DEPOL-20\% & 1.8290 & 3.048 & 4 & 0.5499 & 5.948 & -0.811 & 0.4174 \\
DEPOL-5\%  & 1.4560 & 2.427 & 4 & 0.6577 & 6.254 & -0.690 & 0.4901 \\
DP-1\%     & 0.6382 & 1.064 & 4 & 0.9000 & 4.221 & -1.494 & 0.1352 \\
DP-10\%    & 0.9936 & 1.656 & 4 & 0.7987 & 5.946 & -0.812 & 0.4167 \\
DP-20\%    & 3.3270 & 5.545 & 4 & 0.2358 & 7.403 & -0.236 & 0.8133 \\
DP-5\%     & 0.2449 & 0.408 & 4 & 0.9818 & 3.979 & -1.589 & 0.1120 \\
SN-1024    & 3.4810 & 5.802 & 4 & 0.2144 & 7.349 & -0.258 & 0.7968 \\
SN-256     & 0.5821 & 0.970 & 4 & 0.9143 & 5.503 & -0.987 & 0.3236 \\
SN-512     & 3.0700 & 5.117 & 4 & 0.2755 & 7.192 & -0.319 & 0.7495 \\
SN-6144    & 1.4270 & 2.378 & 4 & 0.6665 & 6.460 & -0.609 & 0.5426 \\
T2 = 180$\mu$s & 0.3522 & 0.587 & 4 & 0.9645 & 4.547 & -1.365 & 0.1723 \\
T2 = 380$\mu$s & 1.0610 & 1.768 & 4 & 0.7783 & 5.285 & -1.073 & 0.2831 \\
T2 = 70$\mu$s  & 1.3240 & 2.207 & 4 & 0.6977 & 5.215 & -1.101 & 0.2709 \\
T2 = 80$\mu$s  & 0.2159 & 0.360 & 4 & 0.9856 & 4.534 & -1.370 & 0.1707 \\
TR-T1=100  & 0.5788 & 0.965 & 4 & 0.9151 & 4.670 & -1.316 & 0.1881 \\
TR-T1=200  & 1.2070 & 2.012 & 4 & 0.7335 & 5.960 & -0.807 & 0.4199 \\
TR-T1=300  & 1.5010 & 2.501 & 4 & 0.6445 & 5.431 & -1.016 & 0.3098 \\
TR-T1=50   & 0.4782 & 0.797 & 4 & 0.9388 & 3.989 & -1.585 & 0.1129 \\
ideal      & 5.9360 & 9.894 & 4 & 0.0423 & 10.030 &  0.801 & 0.4232 \\
\bottomrule
\end{tabular}
\end{table}

\begin{table}[H]
\centering
\caption{Mardia's multivariate normality test results for \gls{isoma}.}
\label{tab:mardia_test_isoma}
\begin{tabular}{lccccccc}
\toprule
family & $b_{1,p}$ & $\chi^2$ & df & $p_{\text{skew}}$ & $b_{2,p}$ & $z_{\text{kurt}}$ & $p_{\text{kurt}}$ \\
\midrule
DEPOL-1\%  & 5.149 & 8.582 & 4 & 0.07245 & 7.764 & -0.09337 & 0.9256 \\
DEPOL-10\% & 3.717 & 6.195 & 4 & 0.1851  & 7.060 & -0.3715  & 0.7103 \\
DEPOL-20\% & 1.037 & 1.729 & 4 & 0.7855  & 5.487 & -0.9933  & 0.3206 \\
DEPOL-5\%  & 1.692 & 2.820 & 4 & 0.5884  & 4.273 & -1.473   & 0.1407 \\
DP-1\%     & 3.509 & 5.849 & 4 & 0.2107  & 8.402 &  0.1588  & 0.8738 \\
DP-10\%    & 3.302 & 5.504 & 4 & 0.2394  & 6.863 & -0.4496  & 0.6530 \\
DP-20\%    & 0.908 & 1.514 & 4 & 0.8242  & 5.000 & -1.186   & 0.2357 \\
DP-5\%     & 2.594 & 4.323 & 4 & 0.3640  & 7.203 & -0.3152  & 0.7526 \\
SN-1024    & 1.558 & 2.597 & 4 & 0.6274  & 5.604 & -0.9472  & 0.3435 \\
SN-256     & 0.266 & 0.443 & 4 & 0.9788  & 4.590 & -1.348   & 0.1777 \\
SN-512     & 1.010 & 1.684 & 4 & 0.7937  & 5.010 & -1.182   & 0.2372 \\
SN-6144    & 4.216 & 7.027 & 4 & 0.1345  & 7.763 & -0.0939  & 0.9252 \\
T2=180$\mu$s   & 2.781 & 4.634 & 4 & 0.3269  & 6.833 & -0.4615  & 0.6445 \\
T2=380$\mu$s   & 2.674 & 4.457 & 4 & 0.3476  & 7.679 & -0.1270  & 0.8990 \\
T2=70$\mu$s    & 0.173 & 0.288 & 4 & 0.9906  & 4.836 & -1.251   & 0.2111 \\
T2=80$\mu$s    & 1.769 & 2.948 & 4 & 0.5666  & 6.639 & -0.5378  & 0.5907 \\
TR-T1=100  & 1.646 & 2.743 & 4 & 0.6018  & 5.731 & -0.8968  & 0.3698 \\
TR-T1=200  & 1.941 & 3.236 & 4 & 0.5192  & 6.705 & -0.5118  & 0.6088 \\
TR-T1=300  & 1.389 & 2.314 & 4 & 0.6782  & 6.106 & -0.7486  & 0.4541 \\
TR-T1=50   & 2.147 & 3.579 & 4 & 0.4660  & 6.139 & -0.7355  & 0.4620 \\
ideal      & 4.799 & 7.998 & 4 & 0.09166 & 8.690 &  0.2729  & 0.7849 \\
\bottomrule
\end{tabular}
\end{table}
\end{document}